\documentclass[12pt,preprint]{aastex}

\usepackage{natbib}

\newcommand{\eps}[1]{\mbox{log~$\epsilon$(#1)}}
\def\eg{\mbox{e.g.}}
\def\ie{\mbox{i.e.}}
\def\logg{\mbox{log~{\it g}}}
\def\Msun{\mbox{$M_{\odot}$}}
\def\vturb{\mbox{$v_{\rm t}$}}
\def\teff{\mbox{$T_{\rm eff}$}}
\def\ncap{\mbox{$n$-capture}}

\slugcomment{Submitted to ApJS}

\shorttitle{The Chemical Compositions of Variable Field Horizontal 
            Branch Stars: RR Lyrae stars}

\shortauthors{B.-Q. For, C. Sneden and G. W. Preston}

\begin{document}

\title{The Chemical Compositions of Variable Field Horizontal Branch Stars: 
RR Lyrae Stars}

\author{Bi-Qing For\altaffilmark{1,2,3} and Christopher Sneden\altaffilmark{1}}
\affil{Department of Astronomy, University of Texas, Austin, 
       TX 78712, USA}			

\author{George W. Preston\altaffilmark{4}}
\affil{Carnegie Observatories, 813 Santa Barbara Street, 
       Pasadena, CA 91101, USA} 

\altaffiltext{2}{John Stocker Fellow, ICRAR, University of Western Australia}
\altaffiltext{3}{Present contact: 
                 35 Stirling Hwy, Crawley, WA 6009, Australia; 
                 biqing.for@uwa.edu.au}

\begin{abstract}

We present a detailed abundance study of 11 RR Lyrae ab-type variables: 
AS Vir, BS Aps, CD Vel, DT Hya, RV Oct, TY Gru, UV Oct, 
V1645 Sgr, WY Ant, XZ Aps, and Z Mic.  
High resolution and high S/N echelle spectra of these variables 
were obtained with 2.5 m du Pont telescope at the Las Campanas Observatory.
We obtained more than 2300 spectra, roughly 200 spectra per star, 
distributed more or less uniformly throughout the pulsational cycles.
A new method has been developed to obtain initial effective temperature 
of our sample stars at a specific pulsational phase. 
We find that the abundance ratios are generally consistent with those of 
similar metallicity field stars in different evolutionary states 
and throughout the pulsational cycles for RR Lyrae stars.
TY Gru remains the only $n$-capture enriched star among the 
RRab in our sample.  
A new relation is found between microturbulence and effective 
temperature among stars of the HB population.
In addition, the variation of microturbulence as a function 
of phase is empirically shown to be similar to the theoretical variation.
Finally, we conclude that the derived \teff\ and \logg\ values of our sample 
stars follow the general trend of a single mass evolutionary track.
\end{abstract}

%%%%%%%%%%%%%%%%%%%%%%%%%%%%%%%%%%%%%%%%%%%%%%%%%%%%%%%%%%%%%%%%%%%%%%%%%%%%%%
\section{INTRODUCTION\label{intro}}
%%%%%%%%%%%%%%%%%%%%%%%%%%%%%%%%%%%%%%%%%%%%%%%%%%%%%%%%%%%%%%%%%%%%%%%%%%%%%%

RR Lyraes (RR Lyr), named after their prototype, are evolved, metal-poor, 
low-mass stars that are fusing helium in their cores and reside in the 
instability strip of the horizontal branch.  
They have long been considered powerful tools to probe many 
fundamental astrophysical problems. 
Due to their distinctive variability and relatively high luminosity, 
they are easily identified even out to large distances. 
Their small dispersion in mean intrinsic luminosity in globular clusters 
suggests that all RR Lyr have similar absolute magnitudes, with a small
correlation in metallicity. 

The distinctive characteristics of RR Lyraes make them good standard 
candles for Galactic and extragalactic populations. 
In the past decades, many studies have been carried out to determine the 
mean absolute magnitudes of RR Lyr and hence their distances. 
The various methods include statistical parallax
(\citealp{Fernley98}; \citealp{GP98}), 
main-sequence fitting in globular clusters \citep{Gratton97}, and the 
Baade-Wesselink technique (\citealp{LJ90}; see \citealp{Gautschy87} for 
a review of this method).
The distance scales are essential in deriving cluster ages, 
which have significant impact for our understanding of stellar structure, 
evolution and ultimately the age of the universe.

The evolutionary states of RR Lyraes also make them ideal tools
for tracing the structure and formation of our Galaxy. 
With ages of $\sim$10 Gyr, they can trace star formation episodes in 
other galaxies (see \eg, \citealp{Clementini10}). 
They also provide evidence of the early merger history of the Milky 
Way \citep{Helmi99} and tidal streams that are associated with the 
formation of the outer halo \citep{Vivas08}. 

Observations of RR Lyr pulsational properties are important in 
constraining both their pulsation models and the physics of their interiors. 
RR Lyr typically have periods of 0.2--1.0 day, with 
magnitude variation of 0.3--2.0 mag.
Most of them pulsate in the radial fundamental mode (RRab stars), 
the radial first overtone (RRc stars) and in some cases, 
in both modes simultaneously (RRd stars). 
Additionally there is a special case, in which the light 
variations of RR Lyraes are modulated with respect to phase and 
amplitude on time scales of days to months, and even years.  
Such modulation is known as the Blazhko effect, named after the Russian 
astronomer who first identified it \citep{Blazhko07}. 
This behavior has been attributed on the one hand to interference of 
radial and non-radial modes of similar frequency (see review by 
\citealp{Preston09, Preston11}), and on the other hand to changes 
in pulsation period induced by changes in envelope structure 
\citep{Stothers06, Stothers10}. 
Vigorous debate about these possibilities is in progress.

The application of RR Lyraes to study the chemical evolution of the
Milky Way disk and halo began with the pioneering 
low-resolution spectroscopic survey by \citet{Preston59}. 
That paper introduced a $\Delta S$ index that describes the relation 
between hydrogen and calcium $K$-line absorption strengths.
The standard $\Delta S$ index is defined near light minimum 
(at phase $\phi$~$\approx$ 0.8).
Early analyses of model stellar spectra \citep{Manduca81} and 
observed high-resolution spectra (\citealp{Preston61}; \citealp{Butler75}) 
showed a correlation between the $\Delta S$ index and metallicity. 
This relation has been calibrated through the studies of metal abundances
in globular clusters (\eg, \citealp{SB78}, 
\citealp{Clementini94, Clementini05}) and presented in various 
forms (see \eg, \citealp{CJ83}). 

While metallicities of RR Lyraes have widely been studied, there are 
only a handful of high-resolution detailed chemical abundance studies of 
field RR Lyraes to date (see \citealp{Clementini95}; 
\citealp{Lambert96}; \citealp{Wallerstein10}; \citealp{Kolenberg10}; 
\citealp{Hansen11}). 
The majority of these investigations concentrated on limited 
pulsational phases near minimum light, because of the relatively slow 
variations in photometric color (hence effective temperature) 
that occur at these phases, and because the minimum light phase is
longer-lived than phases near maximum light.
The exception is the Kolenberg et al. study, in which 
the spectrum analysis was performed around the phase of maximum radius 
($\phi\sim0.35$). 
\citet{Clementini95} deliberately selected RRab 
type variables that have accurate photometric and radial velocity 
data, so that atmospheric parameters could be derived independently of 
excitation and ionization equilibria. 
They obtained 2--6 individual spectra of 10 RR Lyr at pulsational phases 
0.5--0.8, and co-added these spectra to
increase signal-to-noise for chemical composition analysis.
They assumed that lines of most species are formed in conditions 
of local thermodynamic equilibrium (LTE) and that the abundances of RR~Lyr 
share similar patterns to other stars of their metallicity domains.
\citet{Lambert96} gathered spectra of 18 targets; all stars except the 
prototype RR Lyr itself were observed on single occasions at a variety of 
mid-observation phases.
They used photometric information to assist their derivation of iron 
and calcium abundances.  
Recent studies by \citet{Wallerstein10}, \cite{Kolenberg10} 
and \citet{Hansen11} also reported abundances for a few elements
in many RR Lyr stars.

In this paper, we present atmospheric parameters, metallicities,
and detailed chemical compositions of 11 RR Lyr stars which have been
observed intensively throughout multiple pulsational cycles.
On average more than 200 individual spectra were gathered for each target.
These spectra have been described by \citet{For11}, hereafter FPS11,
which discusses the observational data set, and reports the complete set 
of radial velocities and new pulsational ephemerides for the program stars.
In \S2 we briefly summarize the observations and reductions, 
and in \S3 we describe the co-addition of spectra to prepare them
for abundance analysis. 
\S4 discusses the atomic line list and equivalent width measurements, 
\S5 and \S6 describe the initial and derived model atmosphere parameters, 
\S7 describe the optimal phases and \S8 presents the results of 
chemical abundances. Revisiting of the red edge of the RR Lyrae instability 
strip is given in \S9. 
Finally, we describe the evolutionary state of 
these RR Lyr in \S10 and draw a conclusion in \S11.

%%%%%%%%%%%%%%%%%%%%%%%%%%%%%%%%%%%%%%%%%%%%%%%%%%%%%%%%%%%%%%%%%%%%%%%%%%%%%%%
\section{OBSERVATIONS AND DATA REDUCTION\label{obs}}
%%%%%%%%%%%%%%%%%%%%%%%%%%%%%%%%%%%%%%%%%%%%%%%%%%%%%%%%%%%%%%%%%%%%%%%%%%%%%%%

Photometric data from the All Sky Automated Survey (ASAS) and radial 
velocities were presented in FPS11 for a sample of 11 field RRab 
type variable stars, along with their corresponding folded lightcurves 
and radial velocity curves determined from ephemerides derived in that paper.
The RR Lyraes being analyzed here are AS Vir, BS Aps, CD Vel, DT Hya, 
RV Oct, TY Gru, UV Oct, V1645 Sgr, WY Ant, XZ Aps and Z Mic. 
There are no previous detailed chemical abundance studies of 
these stars, except TY Gru \citep{Preston06}. 
We present the basic information about our program stars and the derived 
periods and ephemerides (as shown in Table~1 of FPS11) 
in Table~\ref{targets}. 
We refer the reader to \S3 of FPS11 for details of data reduction. 
Here we summarize the observations. 

The spectroscopic data were obtained with the du Pont 2.5-m telescope at the 
Las Campanas Observatory (LCO), using a cross-dispersed echelle spectrograph. 
We used this instrument with a $1.5\arcsec\times4\arcsec$ entrance slit, 
which gives a resolving power of $R \equiv \lambda/\Delta\lambda \sim 27,000$ 
at the \ion{Mg}{1}~b triplet lines (5180 \AA), and a total wavelength coverage 
of $3500-9000$ \AA.
Integration times ranged from 200--600~s. 
The values of S/N achieved by such integrations can be estimated by 
observations of a star with similar colors to RR Lyr, CS~22175$-$034 
\citep{Preston91}, for which an integration time of 600~s yielded 
S/N$\sim10$ at 4050\AA, 
S/N$\sim15$ at 4300\AA, 
S/N$\sim20$ at 5000\AA, 
S/N$\sim30$ at 6000\AA\ and 
S/N$\sim30$ at 6600\AA.   
We took Thorium-Argon comparison lamp exposures at least once 
per hour at each star position for wavelength calibration. 

The pulsational periods of our program stars tightly cluster around
0.56~days, and so the 600~s maximum integration time corresponds to 
at most $\sim$1.2\% of the period.
The radial velocity excursions over a pulsational cycle are typically
$\sim$65~km~s$^{-1}$.
If we neglect the phase interval 0.85--1.0, in which very rapid velocity
changes occur, then during a 600~s integration the radial velocity
typically changes by only $\sim$0.9 km~s$^{-1}$, much smaller than
a typical absorption line width.
Even during the rapid velocity changes observed in the phase interval
0.85--1.0, the radial velocity changes by only about 5~km~s$^{-1}$
during the maximum integration time; the velocity smearing is
still relatively small in this complex pulsational domain.

%%%%%%%%%%%%%%%%%%%%%%%%%%%%%%%%%%%%%%%%%%%%%%%%%%%%%%%%%%%%%%%%%%%%%%%%%%%%%%
\section{CREATION OF SPECTRA FOR ABUNDANCE ANALYSIS\label{specprep}}
%%%%%%%%%%%%%%%%%%%%%%%%%%%%%%%%%%%%%%%%%%%%%%%%%%%%%%%%%%%%%%%%%%%%%%%%%%%%%%

In this section, we discuss the method of combining spectra for 
Blazhko and non-Blazhko stars. 
Then we describe the scattered light subtraction from 
the combined spectra and the preparation of final spectra 
for equivalent width (EW) measurements and chemical abundance analysis. 

We first shifted individual spectra to rest wavelength by use of the 
IRAF DOPCOR task in the ECHELLE package, having calculated RV$_{\rm obs}$ 
with the FXCOR task. 
The goal is to create as many spectra (or phase bins) as possible 
throughout the pulsational cycle per star. 
However, phase contamination due to rapid changes in the 
atmosphere from phase to phase during a pulsational cycle must be minimized.
A balance between having enough spectra for combining to 
achieve high S/N and avoiding phase contamination is needed. 

We designated a series of phase bins per star. 
Using the phase information in Table~4 of FPS11, we selected about 
10--15 spectra with similar phases for combining, in order to
significantly increase the signal-to-noise for abundance analysis.
For a Blazhko star, we treated the cycles of different RV amplitudes
separately, which resulted in more than one series of phase bins. 
Prior to combination, the individual spectra were examined carefully, 
especially near the H$\alpha$ profile, to guard against any obvious
phase contamination in the averaged spectrum.
The H$\alpha$ profile was chosen because it varied significantly from 
phase-to-phase, and thus any anomalies in its appearance could be 
identified easily. 
The number of spectra for combining was decided on a case-by-case basis
through these inspections of the individual spectra.
We have listed/named the single combined spectrum as the mid-point of 
starting and ending phases (\eg, a spectrum at phase 0.015 
is the combination of spectra that have phases from 0 to 0.03). 
The shapes of metal line profiles of combined XZ Aps and RV Oct spectra 
and their associated H$\alpha$ line profiles (after 
correction for scattered light, see below) are displayed
in Figure~\ref{xzapscomb1_postages}--\ref{rvoctcomb2_postages}. 
The figures show distinctive variations of H$\alpha$ profiles from 
phase to phase.   

Conventional procedures for removal of scattered light from our 
spectra are not feasible because of the short (4 arcsec) entrance 
aperture of the du Pont echelle.  
Therefore, we are obliged to model the scattered light
by the procedures described in \S 3.1 of FPS11.  We proceed as
described below.

To correct for scattered light in the RR Lyr spectra, we first measured 
the peak count of each order of the combined spectrum for each phase. 
This yielded the {\it relative} spectral energy distribution (SED). 
We did the same for the spectra of standard stars (see FPS11) and 
for a family of combinations of their spectra (\eg, one such composite 
contained 50$\%$ of a G6 and 50$\%$ of an A3 spectral type). 
Subsequently, we compared the SEDs of standard stars and their
combination family with the combined RR Lyr spectrum.  
We illustrate SED comparisons between the spectra of standard stars 
and their combination family with RR Lyr spectra in different 
phases in Figure~\ref{seds}.

Once the best match was found (as shown in Figure~\ref{seds}), 
we normalized the combined spectrum with IRAF's CONTINUUM task in 
the ONED package.
We then subtracted the corresponding fractional 
contribution of the inter-order background to the on-order starlight, 
$b_{\lambda}/s_{\lambda}$ 
(corrected by a factor of 5/3 due to different aperture extractions, 
see FPS11), of a particular spectral type from each order. 
The $b_{\lambda}/s_{\lambda}$ values were listed in Table~3 of 
FPS11\footnote{The mean $b_{\lambda}/s_{\lambda}$ of the family 
of spectra combinations are not listed in Table~3 but can be calculated. 
For example, scattered light correction for a 50$\%$ of G6 and 50$\%$ 
of A3 spectral type spectrum would be equal to adding 
50$\%$ $b_{\lambda}/s_{\lambda}$ of G6 and 50$\%$ $b_{\lambda}/s_{\lambda}$ 
of A3 spectral type.} The RR Lyr spectrum corrected for scattered light was then renormalized 
and stitched into 4 long wavelength spectra.
These 4 long wavelength spectra per phase bin were used for the 
abundance analysis.  

To justify that the scattered light correction method we employed 
here was reasonable, we obtained a spectrum of the well-studied 
metal-poor star HD 140283, reduced it and applied the scattered 
light correction in the same manner as we did for our RR Lyr. 
Comparing the EWs of \ion{Fe}{1} lines in the blue and red wavelength 
regions (after scattered light correction) with EWs of \citet{Aoki02}, 
we find: $\Delta$EW(Aoki$-$us)=$-1.3\pm0.4$ m\AA, $\sigma=2.6$ m\AA, 
48 lines, which is good agreement.

%%%%%%%%%%%%%%%%%%%%%%%%%%%%%%%%%%%%%%%%%%%%%%%%%%%%%%%%%%%%%%%%%%%%%%%%%%%%%%
\section{LINE LIST AND EQUIVALENT WIDTH MEASUREMENTS\label{ews}}
%%%%%%%%%%%%%%%%%%%%%%%%%%%%%%%%%%%%%%%%%%%%%%%%%%%%%%%%%%%%%%%%%%%%%%%%%%%%%%

We employed the atomic line list compiled by \citet{For10} 
for our analysis.  
The line wavelengths, excitation potentials (EP) and oscillator strengths 
($\log gf$) and their sources are given in that paper.
For each star, we measured the EWs of unblended atomic absorption 
lines semi-automatically with 
SPECTRE\footnote{An interactive spectrum analysis code \citep{Fitz87}. 
It has been modified to integrate absorption line profiles to 
determine the EW values without manually specifying 
the wavelength.}. 
Each line measurement was visually inspected prior to acceptance of its EW.
Due to the asymmetric line profiles of RR Lyr stars over most of 
their cycle, we adopted the method of integrating over the relative 
absorption across a line profile to determine the EW values. 
Fitting a Gaussian to the line profile was adopted only at the phase 
with sharp (symmetric), non-distorted absorption lines. 
We excluded strong lines, defined as those with reduced widths,  
$\log \rm RW  \equiv \log EW/\lambda \gtrsim -4.0$, because they 
are on the damping portion of the curve-of-growth and thus abundances 
derived from them are sensitive to multiple line formation factors.
Very weak lines ($\log \rm RW$~$<$~$-$5.9) were also excluded because the 
EW measurement errors were too large.

%%%%%%%%%%%%%%%%%%%%%%%%%%%%%%%%%%%%%%%%%%%%%%%%%%%%%%%%%%%%%%%%%%%%%%%%%%%%%%
\section{INITIAL MODEL ATMOSPHERE PARAMETERS\label{modinit}}
%%%%%%%%%%%%%%%%%%%%%%%%%%%%%%%%%%%%%%%%%%%%%%%%%%%%%%%%%%%%%%%%%%%%%%%%%%%%%%

We derived abundances in our RR Lyr stars through EW matching and
spectrum syntheses.
Both methods require model stellar atmospheres that are 
characterized by parameters effective temperature (\teff), surface 
gravity (\logg), metallicity ([M/H]) and microturbulence (\vturb).
We constructed the models by interpolating in Kurucz's 
non-convective-overshooting atmosphere model grid \citep{Castelli97},
using software developed by A. McWilliam and
I. Ivans. 
The elemental abundances were subsequently derived using the latest 
2010 version local thermodynamic equilibrium (LTE), plane-parallel 
atmosphere spectral line synthesis code 
MOOG\footnote{Available at http://www.as.utexas.edu/~chris/moog.html .}
\citep{Sneden73}. 
This code includes treatment of electron scattering contributions
to the near-UV continuum that have been implemented by \cite{Sobeck11}.
Details on estimating initial stellar parameters are given in the 
following subsections.

%%%%%%%%%%%%%%%%%%%%%%%%%%%%%%%%%%%%%%%%%%%%%%%%%%%%%%%%%%%%%%%%%%%%%%%%%%%%%%%
\subsection{Effective Temperature\label{teffinit}}
%%%%%%%%%%%%%%%%%%%%%%%%%%%%%%%%%%%%%%%%%%%%%%%%%%%%%%%%%%%%%%%%%%%%%%%%%%%%%%

Use of spectroscopic constraints alone to determine model atmosphere
parameters can lead to ambiguous results, due to degeneracies in the 
responses of individual EWs changes in various quantities.
This is especially true for \teff\ and \vturb: the lines with lower
EPs are usually those with larger EWs, making it difficult to 
simultaneously solve for \teff\ and \vturb\ unambiguously.
It is important to have a good initial guess at \teff\ from other
data, and the standard method involves photometric color transformations.
Using color-temperature transformations (\eg, \citealp{Alonso96}, 
\citealp{RM05}) it is straightforward to obtain the 
temperatures of the RR Lyr throughout their pulsational cycles.
However, our program stars lack the necessary photometric information.  
Extensive $V$ magnitude data are available for all our stars at the
All-Sky Automated Survey (ASAS) 
website\footnote{http://www.astrouw.edu.pl/asas/} \citep{Pojmanski02}
but $I$ magnitude data have not been gathered.
Therefore  we do not have any color information for our stars and 
development of a new, indirect method to estimate initial
\teff\ values for at individual phases of our RR Lyr stars is needed.

%%%%%%%%%%%%%%%%%%%%%%%%%%%%%%%%%%%%%%%%%%%%%%%%%%%%%%%%%%%%%%%%%%%%%%%%%%%%%%
\subsubsection{Color--Temperature Transformations\label{colortemp}}
%%%%%%%%%%%%%%%%%%%%%%%%%%%%%%%%%%%%%%%%%%%%%%%%%%%%%%%%%%%%%%%%%%%%%%%%%%%%%%

Temperature transformations from photometric indices are generally 
achieved with either a stellar atmosphere model (see \citealp{LJ90}) 
or an empirical color--temperature calibration (see \citealp{Clementini95}). 
The latter method can be problematic because it does not account easily
for metallicity and surface gravity effects. 
Of particular importance is the gravity, which varies about a factor of
ten during the pulsational cycle of an RRab star. 
Ideally, hydrodynamical models would be more suitable to describe 
RR Lyr atmospheres (and thus their \teff\ values at any phase)
but no such models capable of dealing with the fast moving atmospheres
of RR Lyr exist yet. 
Luckily the most dynamical phase (near minimum radius), in which a shock 
wave is produced during the rapid acceleration of an RR Lyr atmosphere, 
only occurs in a very short timescale ($\sim 15$ min). 
\citet{Castor72} found that a
dynamical atmosphere model produces a continuous spectrum that is 
nearly indistinguishable from that of a hydrostatic atmosphere at the same 
temperature and gravity in most of the pulsational cycle. 
A non-linear pulsational model for the prototype star RR Lyr by 
\citet{Kolenberg10} shows that the kinetic energy of its atmosphere 
reaches a minimum at two phases, $\phi$~$\simeq$~0.35 and 0.90 (see 
their Figure~1), for which the dynamical effects are small.  
Accordingly, we assume that the atmospheres of RR Lyrs are in approximate 
quasi-static equilibrium during most of the pulsational phases.

A mirror-image relation between light and radial velocity variations of 
Cepheids has been recognized for more than 80 years \citep{san30}.  
Inspection of the extensive data of \citet{LJ89,LJ90}, hereafter
LJ89 and LJ90, shows that similar mirror image relations also exist 
between the color indices and radial velocities of RR Lyrae stars.  
Because we do not have suitable color data for our RRab stars, 
we decided to use this mirror-image characteristic to estimate 
colors of our stars at the phases of our spectroscopic observations.  
We used the data of \citeauthor{LJ89} to establish relations 
between radial velocity and color indices.  
We then used these relations to estimate colors, and hence 
temperatures from appropriate color-temperature relations.  
This procedure works well: radial velocity is a proxy for color index.

We chose eight RRab stars from LJ89 (SW And, RR Cet, SU Dra, RX Eri, 
RR Leo, TT Lyn, AR Per and TU Uma).
For these stars we first extracted $B-V$, $V-R_{c}$ and $V-I_{c}$ color 
indices\footnote{LJ89 used Johnson-Cousins color system. 
The $V-K$ color index was not chosen because the lack of photometric 
data points for most of the RRab variables in LJ89.} and their RVs 
that correspond to our defined 11 phase bins (\eg, $\phi=$ 0, 0.05, 0.1, 
0.2, 0.3, 0.4, 0.5, 0.6, 0.7, 0.8 and 0.85, 
see Table~\ref{colorteff} for details).
The color index of a phase that most closely matches one of our phase bins
was adopted (\eg, RV at phase 0.8525 in LJ89 was adopted as our RV for 
the defined phase 0.85). 
The published color curves were not corrected for the reddening. 
Thus, we corrected the color indices of $B-V$, $V-R_{c}$ and 
$V-I_{c}$ as follow:
\begin{equation}
c(colors)=(colors)-E(colors),
\end{equation}
where $c(colors)$ is the corrected color index and $E(colors)=kE(B-V)$. 
The values of $k$ and $E(B-V)$ were adopted from Tables~2 $\&$ 3 of LJ90. 
We refer the reader to \S 2b of LJ90 for the extensive discussion of 
their choice of reddening. 

To transform the color indices of LJ89 into \teff\ values, a set of 
synthetic colors computed from model stellar atmosphere grids is needed. 
Calculated colors are given in Table~7 of LJ90, but those are
based on relatively old model atmospheres \citep{Kurucz79}.
Instead, we created grids that correspond to the metallicity of RR Lyr 
in LJ90 with Kurucz's non-convective-overshooting atmosphere
models\footnote{The specific models are under the suffix ODFNEW on 
Kurucz's website: http://kurucz.harvard.edu/grids.html} \citep{Castelli97}.
A surface gravity of $\log g=3.0$ was chosen initially because it is 
a better representation for the mean effective gravity (with only small 
variations) of an RR Lyr star during phases 0--0.8 
(\ie, $3.2 < \log g < 2.8$; see Figure~1 of LJ90). 
However, the effective gravity (which will be described in detail in \S 5.2) 
is an approximation for compensating the dynamical nature of the RR Lyr 
atmospheres, which could be quite different than the actual surface 
gravity in the static model that we applied here.  
Our tests showed that the transformed \teff\ with $\log g=3.0$ 
model was persistently too high to fulfill the spectroscopic constraint 
for all phases of our RR Lyr during the initial spectroscopic analysis. 
We noted that the effective gravity calculated in LJ89 were based on the
Baade-Wesselink (BW) method.
\cite{For10} showed that the \logg\ derived from the BW method by 
others were systematically higher than indicated by the spectroscopic 
method for non-variable horizontal branch stars 
analysis (see Figure~19 of \citealp{For10}). 
Therefore, we employed models with $\log g=2.0$; the new grids are 
presented in Table~\ref{grids}. 

The subsequent color--temperature transformation was carried out 
by employing a linear interpolation scheme:
\begin{equation}
\teff=\teff_{1}+\frac{(\teff_{2}-\teff_{1})}{(c_{2}-c_{1})}\times(c_{*}-c_{1}),
\end{equation} 
\noindent where $\teff_{1}$ and $\teff_{2}$ are two effective 
temperatures from the grid, c$_{1}$ and c$_{2}$ are the color indices 
of \teff$_{1}$, \teff$_{2}$, and c$_{*}$ is 
the color index of the star at a particular phase.  

To derive the \teff--phase relations, we employed only the 
$V-I_{c}$ color because the color--temperature transformation 
became less sensitive to metallicity and gravity 
at longer wavelengths. 
We demonstrate the sensitivity of transformed \teff\ as a function 
of metallicity in Figure~\ref{dev}. 
The strong dependence of $B-V$ on metallicity is caused by 
the line blanketing in the $B$ filter.
The calculated \teff\ for a given observed color index was adopted 
at phase 0.3 of RR Cet for different metallicities with fixed \logg.
The difference was taken between the calculated \teff\ at that 
particular [M/H] minus the \teff\ at [M/H]$=-2.5$. 

We summarize the color--temperature transformations of each phase in 
Table~\ref{colorteff}.
In Figures \ref{bmv_teffphi}, \ref{vmr_teffphi}, and \ref{vmi_teffphi}
we show the transformed \teff\ from $B-V$, $V-R_{c}$ and $V-I_{c}$,
respectively, versus phase for eight selected RRab variables, which 
will be called ``calibration stars'' in the following sections.

Subsequently, we fitted 4$^{\rm th}$-order polynomials to \teff\ values
transformed from $V-I_{c}$ vs phase.  
The fitted curves are called ``calibration curves'' for our RR Lyr. 
Phases during the rising branch of RR Lyr (\ie, after phase $\sim$0.85) 
were excluded to avoid any artificial fit to the data. 
We considered the \teff\ at those phases to be close to their 
descending branch (\ie, phase 0.9 equivalent to phase 0.1). 
This assumption is problematic, but we are unaware of a 
better alternative.
The derived 4$^{\rm th}$-order polynomial equations are given in 
Table~\ref{caleqs} and 
Figure~\ref{all_teffphi} shows the fit to the $V-I_{c}$ data. 

To decide which ``calibration curves'' to use for obtaining the 
initial \teff\ throughout the pulsational cycle of our RR Lyr, we 
compared our RV curves to the RV curves of those eight RRab variables 
selected from LJ89. 
An example of such comparison is shown in 
Figure~\ref{rvmrvmin_ph}, where the RV curve of RV Oct matched the 
RV curve of RR Cet but not that of TT Lyn.
We found that comparing the RV curves 
of our Blazhko stars to the RV curves of calibration stars was 
particularly difficult. 
The RV curves of calibration stars represent typical pulsation 
RV amplitudes of non-Blazhko RRab variables. 
In the case of our Blazhko stars, the RV amplitudes vary significantly 
with Blazhko phase and we could not find any close match between 
the RV curves of our Blazhko stars and those of our calibration stars.
Perforce, we selected the most closely matching RV curve of a calibration star
and used its calibration curve to obtain the initial 
\teff\ in those cases.

%%%%%%%%%%%%%%%%%%%%%%%%%%%%%%%%%%%%%%%%%%%%%%%%%%%%%%%%%%%%%%%%%%%%%%%%%%%%%%%
\subsection{Surface Gravity\label{gravinit}} 
%%%%%%%%%%%%%%%%%%%%%%%%%%%%%%%%%%%%%%%%%%%%%%%%%%%%%%%%%%%%%%%%%%%%%%%%%%%%%%

Due to pulsation, the gravity of RR Lyr varies throughout
the pulsational cycle.
Therefore, the observed gravity at a given phase, which we call the 
effective gravity, must include a dynamical acceleration term:
\begin{equation}
g_{\rm eff}=\frac{GM}{R^2}+\frac{d^2R}{dt^2},
\end{equation}
\noindent where $M$ and $R$ are the mass and the radius of the 
star.
The first term represents the mean gravity of the star, which can be
derived from its mass and mean radius.
The second term represents the variation of gravity,
which takes into account the acceleration of the moving atmosphere.
It can be determined by differentiating the radial velocity curve.

The mass and mean radius can be derived via the BW method, for
which photometric information is required.
Since  we do not have lightcurves for our RR Lyr stars, 
we chose a fixed \logg\ = 2.0 as the initial gravity estimate.

%%%%%%%%%%%%%%%%%%%%%%%%%%%%%%%%%%%%%%%%%%%%%%%%%%%%%%%%%%%%%%%%%%%%%%%%%%%%%%%
\subsection{Metallicity and Microturbulence\label{feinit}} 
%%%%%%%%%%%%%%%%%%%%%%%%%%%%%%%%%%%%%%%%%%%%%%%%%%%%%%%%%%%%%%%%%%%%%%%%%%%%%%

We adopted the [Fe/H] values of \cite{Layden94} as listed in 
Table~1 of \citet{Preston09} as our initial metallicity estimates.
There is no previous derived metallicity for DT Hya and CD Vel in the
literature. 
For these stars we employed [M/H] $= -1.5$, which is similar to the
mean [M/H] of our other program stars.

A constant microturbulence is generally assumed throughout the layers of
stellar atmospheres.
Apart from simplicity, there is no evidence to support this
assumption for real stars.
In fact, some studies suggested that non-constant microturbulence is
more appropriate to physically describe a stellar atmosphere
(\eg, see \citealp{Hardorp67}; \citealp{Kolenberg10}).
In addition, the presence of shock waves during the RR Lyr pulsational
cycle makes \vturb\ unlikely to be constant in their atmospheres
(see theoretical work by \citealp{FGC99}).
To perform the spectroscopic analysis, we adopted \vturb~=~3~km~s$^{-1}$ as an 
initial guess and set it as a free parameter.
The variation of microturbulence as a function of phase/\teff\
is discussed in the following sections.

%%%%%%%%%%%%%%%%%%%%%%%%%%%%%%%%%%%%%%%%%%%%%%%%%%%%%%%%%%%%%%%%%%%%%%%%%%%%%%%
\section{ADOPTED MODEL ATMOSPHERE PARAMETERS\label{modfin}}
%%%%%%%%%%%%%%%%%%%%%%%%%%%%%%%%%%%%%%%%%%%%%%%%%%%%%%%%%%%%%%%%%%%%%%%%%%%%%%%

Final model atmosphere parameters were determined by iteration 
through spectroscopic constraints: (1) for  \teff, that the abundances 
of individual \ion{Fe}{1} and \ion{Fe}{2} lines show no trend with EP;
(2) for \vturb, that the abundances of individual \ion{Fe}{1} and
\ion{Fe}{2} lines show no trend with reduced width $\log\,(\rm{RW})$;
(3) for \logg, that ionization equilibrium be achieved by requiring
equality between the abundances derived from the \ion{Fe}{1} and 
\ion{Fe}{2} species; and (4) for metallicity [M/H], that its value 
is consistent with the [\ion{Fe}{1}/H] determination.
An example of fulfilling the spectroscopic constraints is
presented in Figure~\ref{cdvel0300_Feout}.
The linear regression lines shown in the figure indicate that
\teff\ and \vturb\ have been determined to within the line-scatter
uncertainties, and the agreement between the mean abundances for the
two Fe species indicates choice of a \logg\ that satisfies the
Saha ionization balance.

We present the derived stellar parameters vs pulsational phase of 
RV Oct and AS Vir as examples for Blazhko and non-Blazhko effect stars, 
respectively, in Figures~\ref{par_phase_rvoct} and \ref{par_phase_asvir}.
The dashed lines represent the mean values. 
The top and second panels show the typical \teff\ and \logg\ changes in the 
atmosphere of RR Lyr during the pulsational cycle. 
The third panel shows the consistency of our derived [M/H]. 
The bottom panel shows the variation of \vturb\ as a function of phase. 
Interpolated model atmospheres, constructed as described in
\S\ref{modinit} with the derived parameters listed in Table~\ref{inputmodels}, 
were used to derive the abundances of each star.

%%%%%%%%%%%%%%%%%%%%%%%%%%%%%%%%%%%%%%%%%%%%%%%%%%%%%%%%%%%%%%%%%%%%%%%%%%%%%%%
\subsection{Parameter Uncertainties\label{parerrors}} 
%%%%%%%%%%%%%%%%%%%%%%%%%%%%%%%%%%%%%%%%%%%%%%%%%%%%%%%%%%%%%%%%%%%%%%%%%%%%%%%

To estimate the effects of uncertainties in our spectroscopically-based
\teff\ values on derived abundances, we varied the derived \teff\ of RV Oct 
(as an example) by raising \teff\ by different amounts for all phases. 
The uncertainty of \teff\ was determined for a particular phase when 
the raised \teff\ produced a large trend of derived 
\eps{Fe} ($\Delta$ \eps{Fe} $> \pm0.1$) with excitation potential. 
This yielded estimated \teff\ errors of 100--300~K throughout the cycle. 
The largest uncertainties generally were encountered during the 
most rapidly-changing parts of the pulsational cycles ($\phi <$~0.3 and
$\phi >$~0.8).
The initial \teff\ values for phase 0.9 onward were assumed 
to be close to their descending branch (as discussed in 
\S\ref{colortemp}), which resulted in larger uncertainty considered 
that the \teff\ versus phase curve was asymmetric.  
In addition, fewer Fe lines are available for EW measurements 
in the hotter phases of the descending and rising branches
than at other (cooler) phases.

We estimated \vturb\ uncertainties in a similar manner, assessing the
trends of \eps{Fe} with $\log\,(\rm{RW})$.
This yielded $v_{t}$ errors of 0.1--0.4~km s$^{-1}$ throughout the cycle.  
Finally, assuming that $\log g$ values based on the neutral/ion 
ionization balance of Fe abundance are correct, then from the dependence
of \ion{Fe}{2} abundances on $\log g$ we estimated the $\log g$ 
uncertainty to be $2\sigma$ of the \ion{Fe}{2} abundance error.
The typical mean error of \logg\ is $\sim 0.2$ dex per star. 
We adopted the internal error ($\sigma$) of \ion{Fe}{1} abundances
as the model [M/H] error.

%%%%%%%%%%%%%%%%%%%%%%%%%%%%%%%%%%%%%%%%%%%%%%%%%%%%%%%%%%%%%%%%%%%%%%%%%%%%%%%
\subsection{Reliability of Derived Stellar Parameters\label{parreliable}}
%%%%%%%%%%%%%%%%%%%%%%%%%%%%%%%%%%%%%%%%%%%%%%%%%%%%%%%%%%%%%%%%%%%%%%%%%%%%%%%

%%%%%%%%%%%%%%%%%%%%%%%%%%%%%%%%%%%%%%%%%%%%%%%%%%%%%%%%%%%%%%%%%%%%%%%%%%%%%%%
\subsubsection{Derived Effective Temperature\label{teffreliable}}
%%%%%%%%%%%%%%%%%%%%%%%%%%%%%%%%%%%%%%%%%%%%%%%%%%%%%%%%%%%%%%%%%%%%%%%%%%%%%%%

We compare our final spectroscopic \teff's with the initial 
values that were derived from the calibration curves, in 
top and bottom panels of Figure~\ref{comp_teff} for non-Blazhko and 
Blazhko stars, respectively. 
The scatter with respect to the unity line for the non-Blazhko stars is
$\Delta$($T_{\rm eff, phot}-T_{\rm eff, spec}$)$~= 4\pm10$~K, 
$\sigma = 92$~K, $N = 87$, and it is somewhat larger for the 
Blazhko stars, 
$\Delta$($T_{\rm eff, phot}-T_{\rm eff, spec}$)$~= 8\pm17$~K, 
$\sigma = 151$~K, $N = 78$. 
Most cases of exact agreement (\ie, $\Delta$\teff = 0) were artificially 
caused by the spectroscopic constraints method that we used. 
Those initial \teff\ values either yielded no trend or small trend 
($\Delta$ \eps{Fe} = 0.05) with EP during first iteration.  
Based on the overall calculated $\Delta$\teff, we conclude that even though 
the RV curves of Blazhko stars might not match the RV curves of calibration 
stars, the initial \teff\ values derived from the calibration curves 
worked reasonably well. 
We also showed in a previous section that the selected initial \teff\ yielded 
consistent stellar parameters throughout the pulsational phase for any 
cycle in Blazhko stars (see Figure~\ref{par_phase_asvir} for example).  

We made another comparison with the study of TY Gru 
\citep{Preston06} that was based on MIKE Magellan spectra.
Their derived stellar parameters near minimum light for TY Gru 
were \teff~= 6250$\pm$150 K, \logg~= 2.3$\pm$0.2 dex, 
[M/H]~= $-2.0\pm0.2$, and \vturb~= 4.1$\pm$0.2 km~s$^{-1}$. 
Our derived stellar parameters at phase 0.8 were 
\teff~= 6360$\pm$150 K, \logg~= 2.05$\pm$0.30 and 
\vturb~= 4.15$\pm$0.4 km~s$^{-1}$, which are
within the uncertainties of results of \citet{Preston06}.

%%%%%%%%%%%%%%%%%%%%%%%%%%%%%%%%%%%%%%%%%%%%%%%%%%%%%%%%%%%%%%%%%%%%%%%%%%%%%%%
\subsubsection{Derived Surface Gravity\label{gravreliable}}
%%%%%%%%%%%%%%%%%%%%%%%%%%%%%%%%%%%%%%%%%%%%%%%%%%%%%%%%%%%%%%%%%%%%%%%%%%%%%%%

The \logg\ derived by use of standard  
spectroscopic constraints, \ie, the ionization balance between 
neutral and ionized species, may be lower than the trigonometric 
\logg\ (see \eg, \citealp{AP99}) if radiative processes act
to ionize neutral species beyond standard Saha collisional values.
This is a known issue and has been demonstrated with studies of 
bright metal-poor stars with well-determined distances such as, 
HD 140283 (as mentioned in \S\ref{specprep}).

We performed a standard spectroscopic analysis of HD 140283. 
A summary of the results of this investigation and comparison with 
other studies is given in Table~\ref{methodlogg}.
The spectroscopic \logg\ values derived in \citet{Hosford09} and 
\citet{Aoki02} are lower than those obtained with other methods, and 
are essentially within errors of our spectroscopic values for HD~140283. 
We also note that the slightly higher \logg\ determined by Aoki et al. than 
by either Hosford et al. or us is due to their use of Ti lines, which 
has been shown to cause a systematic offset in spectroscopic 
derived \logg\ (see \S 5.3 of \citealp{For10}).

Ideally we should compare the derived spectroscopic \logg\ 
with physical or trigonometric \logg\ that can be derived 
from stellar parallaxes.  
However, this is not possible for our our RR~Lyr stars, because 
either the reported parallaxes have large errors, or no parallax data 
are available.
Nevertheless, we may evaluate the physical \logg\ by making assumptions 
for the following equation:
\begin{equation}
\logg = \log(M/\Msun) + 4 \log(T_{\rm eff, spec}) - \log(L/L_{\sun})-10.607,
\end{equation}
\noindent in which the constant was calculated by using the solar 
\teff\ and \logg\ values, $M = 0.68$ \Msun\ as typical mass 
of an HB star and absolute magnitude of $M_{V} = +0.6$ \citep{Castellani05}, 
a value consistent with typical RR Lyr stars \citep{BPS92}.
We note that the absolute magnitude is metallicity-dependent, in that
a lower metallicity would result in brighter absolute magnitude 
(see \eg, \citealp{Gratton98}).   

Comparing our derived \logg\ values throughout the pulsational 
cycles with calculated physical \logg\ values, we found that they are 
systematically lower, 
$\Delta$\logg (calculated$-$us) $= 0.80\pm0.02$ dex, 
$\sigma = 0.28$ dex, $N= 165$. 
The large deviation is partly related to the assumptions 
we have made for stellar mass, absolute magnitude, and treatment of 
gravity as mean gravity instead of effective gravity (as described 
in \S~\ref{gravinit}). 
Significant departures from LTE in the ionization equilibrium also 
could drive our spectroscopic gravities to artificially low values.
For example, see the discussion by \citet{Lambert96} for NLTE 
effects on \ion{Fe}{1} and \ion{Fe}{2} lines in RR Lyr. 
However, \citet{Clementini95} argue that NLTE effects are small
in these kinds of stars.
Resolution of the NLTE question is beyond the scope of our study,
but we urge further work on this point in the future.

Finally, note that despite the lower derived \logg\ values 
for our RR Lyr throughout their pulsational cycles, the trend of 
our derived \logg\ variation (see \eg, Figure~\ref{par_phase_rvoct}) 
is quite similar to the effective gravity variation as shown in 
Figure~1 of LJ90.

%%%%%%%%%%%%%%%%%%%%%%%%%%%%%%%%%%%%%%%%%%%%%%%%%%%%%%%%%%%%%%%%%%%%%%%%%%%%%%%
\subsubsection{Derived Metallicity\label{fereliable}}
%%%%%%%%%%%%%%%%%%%%%%%%%%%%%%%%%%%%%%%%%%%%%%%%%%%%%%%%%%%%%%%%%%%%%%%%%%%%%%%

The metallicities of RR Lyr are commonly derived from $\Delta S$--[Fe/H] 
relations calibrated by abundances derived from high-resolution spectroscopy. 
The initial investigation in this area was made by 
\citet{Preston61b}, who employed the single-layer-atmosphere differential 
abundance formalism of \citet{Greenstein48} and \citet{Aller53}, 
with line identifications taken from \citet{Swensson46} 
and \citet{Greenstein47}.
This calibration was supplanted by subsequent analyses of many 
more RR Lyr by \citet{Butler75}, who also used Greenstein's method, 
and by \citet{Butler82}.  
\citet{Layden94} adopted the latter of these, now 30 years old, 
to establish his widely-used abundance scale for the RR Lyr.

We may compare our derived metallicities with those 
in \citet{Layden94}.
As shown in Figure~\ref{comp_metal}, our [Fe/H] values
are lower by $\sim$0.25 dex than those derived by Layden, who used 
the \citet{Butler82} results. 
The downward shift arises from differences in measured equivalent widths, 
adopted $\log gf$ values, and the use of modern model atmospheres 
and spectrum analysis codes instead of one-layer curve-of-growth 
analysis, universally abandoned long ago.  
We note, finally, that our Fe abundance for TY Gruis is in good 
accord with that derived from Magellan/MIKE spectra \citep{Preston06}.

To further investigate the Fe abundance offset, we refer 
back to the well-studied subgiant HD~140283, for which our EWs of 
Fe lines are in good agreement with \citet{Aoki02}.
In \S\ref{specprep} we used this EW agreement to argue that our 
scattered light corrections are reasonable.
Now using the Aoki et al. measured EWs, their chosen $\log gf$ for
\ion{Fe}{1} and \ion{Fe}{2} lines, and their adopted stellar parameters, 
we reproduce almost exactly their published \eps{Fe} with our
analysis code.
Then performing an independent atmospheric analysis in the manner
employed for our RR Lyr spectra, using Aoki's data set, we derive
\teff\ about 150 K lower than theirs, which in turn yields in a 
slightly lower Fe abundance ($\Delta \sim -0.15$ dex). 
However, a derived \teff\ for HD~140283 via the photometric 
``infrared flux method'' calibration \citep{RM05} is consistent with 
our derived spectroscopic \teff.
This lends indirect support to our general metallicity scale.
In addition, we performed a similar test using RR Cet data from 
\citet{Clementini95}. 
Adopting their stellar parameters resulted in \eps{Fe I} = 5.98 and
\eps{Fe II} = 6.05. 
Clementini et al. derived \eps{Fe}= 6.18 ($\sigma$~=0.16)
and 6.13 ($\sigma$~=0.06) for \ion{Fe}{1} and \ion{Fe}{2}, respectively.
Again, our [Fe/H] value is somewhat less than theirs, but the
uncertainties in especially the \ion{Fe}{1} abundance are large.
We do not intend to solve the absolute scale of metallicity in this paper.
Future effort on this issue will be investigated with a wider 
range of metallicity for the RR Lyr sample.
For now, we tentatively recommend a $-0.25$~dex shift 
to the Layden abundance scale for RR Lyr stars.
This downward revision is in accord with recent investigations of 
the \ion{Fe}{2} metallicity scale for the globular clusters \citep{KI03} 
and the metal-poor horizontal branch stars of the 
Galactic field (\citealp{PrestonRHB06}; \citealp{For10}).

%%%%%%%%%%%%%%%%%%%%%%%%%%%%%%%%%%%%%%%%%%%%%%%%%%%%%%%%%%%%%%%%%%%%%%%%%%%%%%%
\subsection{Microturbulence vs Effective Temperature\label{vmicrotrend}}
%%%%%%%%%%%%%%%%%%%%%%%%%%%%%%%%%%%%%%%%%%%%%%%%%%%%%%%%%%%%%%%%%%%%%%%%%%%%%%%

We revisit the variation of \vturb\ with \teff\ along 
the horizontal branch suggested in Figure 7 of \citet{For10}.
The variation within the instability strip was uncertain in that paper, 
because the data for RR Lyr came from heterogeneous sources.  
Now with internally consistent data and analyses we can investigate 
the variation across the instability strip with more confidence.  
In the top panel of Figure~\ref{vt_teff} we show one example
by plotting the individual \vturb\ values for the stable pulsator 
RV Oct (Table~\ref{inputmodels}) in the \vturb--\teff\ diagram 
of \citet{For10}.  
Excluding one point that is much lower than the rest, the
\vturb\ values for RV~Oct all lie in a relatively narrow range:
3.0 $<$ \vturb\ $<$ 3.6~km~s$^{-1}$. 
A continuous \vturb-\teff\ relation across the horizontal branch 
is suggested, which we interpret empirically by drawing a smooth 
curve to represent the data points.

When data for all of the RR Lyr in our program are plotted in the 
bottom panel of Figure~\ref{vt_teff} we see that the microturbulence values
encompass a larger range than do those of RV Oct:
2.5 $<$ \vturb\ $<$ 4.5~km~s$^{-1}$.
However, closer inspection of the individual points reveals that the
most extreme microturbulence excursions occur in the Blazhko variables.
Five out of seven stars with \vturb~$>$ 4.0~km~s$^{-1}$ are Blazhko
stars, as are five out of six stars with \vturb~$<$ 2.6~km~s$^{-1}$.
Thus for most RRab stars in all phases $<$\vturb~$> \sim$ 3.4~km~s$^{-1}$,
with maximum excursions of $\pm$0.6~km~s$^{-1}$
The range of \vturb\ values for our RR Lyr is superficially similar 
to those reported by \citet{Clementini95} and \citet{Lambert96}. 
Evidently \vturb\ goes through a maximum in the RR Lyr instability 
strip of the halo field horizontal branch.
The range in \vturb\ for each RR Lyr is real, produced by systematic 
variation during pulsation cycles as we discuss in the next section.

%%%%%%%%%%%%%%%%%%%%%%%%%%%%%%%%%%%%%%%%%%%%%%%%%%%%%%%%%%%%%%%%%%%%%%%%%%%%%%%
\subsection{Microturbulence vs Phase\label{vmicrophase}}
%%%%%%%%%%%%%%%%%%%%%%%%%%%%%%%%%%%%%%%%%%%%%%%%%%%%%%%%%%%%%%%%%%%%%%%%%%%%%%%

Turbulent velocity variations occur during the pulsation cycles 
of RR Lyraes, as indicated by the investigation of \citet{Fokin99}, 
\citet{FG97} and \citet{Fokin99}.
The conclusions of these investigators are based on measured FWHM 
values of metallic lines profiles.  
These FWHM reach a minimum value briefly near phase 0.35 and 
then rise to a broad maximum on 0.6 $<$ $\phi$ $<$ 1.2 when two shocks 
occur above the photosphere. 
The FWHM that accompany these phenomena exceed our 
maximum \vturb\ values ($<$ 5 km~s$^{-1}$) at all phases.  
This is illustrated in Figure~\ref{xzaps_fwhm_phi}, where we plot 
the observed values of FWHM corrected for instrumental broadening of 
11~km s$^{-1}$ in quadrature term versus phase in the pulsation 
cycle of XZ Aps.  
One of these lines, \ion{Ti}{2} 4501.3~\AA, was the featured metal
line of Figures~\ref{xzapscomb1_postages} and \ref{xzapscomb2_postages},
and the variations in its line width could be seen easily by inspection.
Compare our FWHM with those in Figure 6 of \citet{CG96} and in 
Figure~4 of \cite{Kolenberg10}. 
This broadening of line profiles is a manifestation of 
macroturbulence, \ie, the bulk motions of gas volumes with dimensions 
comparable to the thickness of the metallic line-forming regions of 
the atmosphere.

In our study, we derive values of microturbulence, \vturb, by demanding 
that the abundances of individual \ion{Fe}{1} and \ion{Fe}{2} lines 
show no trend with reduced width $\log$ RW.  
Our \vturb\ are empirical descriptors of motions on length scales small 
compared to the line-forming region of the atmosphere that broaden the 
metallic line absorption coefficients and thus intensify line strengths.  
A plot of our \vturb\ values versus phase for all of our RR Lyr is 
shown in Figure~\ref{vt_phase}. 
The values of vt and FWHM, derived from independent considerations 
rise and fall together, indicating that our RR Lyr display growth 
and decay of turbulent velocities on two length scales together 
at all phases of their pulsation cycles.

%%%%%%%%%%%%%%%%%%%%%%%%%%%%%%%%%%%%%%%%%%%%%%%%%%%%%%%%%%%%%%%%%%%%%%%%%%%%%%
\section{The OPTIMAL PHASES\label{phasebest}}
%%%%%%%%%%%%%%%%%%%%%%%%%%%%%%%%%%%%%%%%%%%%%%%%%%%%%%%%%%%%%%%%%%%%%%%%%%%%%%

In this section, we discuss the optimal phases for chemical abundances analysis. 

The zero point of RR~Lyr phase is generally chosen to coincide with the 
moment of maximum light.  
Expansion of the atmosphere decelerates from this phase until the 
layers near the photosphere come to rest near phase $\phi$~$\sim$ 0.35. 
The expansion is not homologous; see the middle panels of 
Figure 2 of \citet{FG97} and the measured radial velocities of 
Balmer lines H$\alpha$, H$\beta$ and H$\gamma$ \citep{Preston11}. 
Near $\phi$~$\sim$ 0.35 the atmospheric turbulence is at a 
relative minimum. 
Spectra at this phase regime are accordingly best suited for chemical
composition analysis because atomic lines suffer minimal blending.
This most clearly evident from examination of line widths plotted in 
Figure~\ref{xzaps_fwhm_phi}. 

During the optimal $\phi$~$\sim$ 0.35 phase, the effective 
temperatures of RR Lyr are similar to those of warmer RHB stars 
(6500~K $<$ \teff\ $<$ 6000~K). 
We see many metal lines in the spectra at these temperatures, 
which make these phases ideal for abundances analyses. 
Additionally, the sharpness of the line spectra at this phase
makes it best for  performing spectrum synthesis 
calculations of complex blended features.

The line smearing and line asymmetry at other phases degrade their
value for analysis by spectrum synthesis. 
Nevertheless, we did not exclude the other phases in our study.
In fact, the descending and rising branches of RR~Lyr variations 
have their own advantages.
In the post-maximum phase interval ($\phi$~= 0.05--0.15) effective 
temperatures are similar to cooler BHB stars (7400 K $<$ \teff\ $<$ 6200 K).
Some low EP metal lines that are saturated at cooler phase 
temperatures are weaker in the hotter parts of RR Lyr cycles, 
and thus can be more useful in abundance analyses. 
Thus, we conclude that abundance analysis can be pursued 
profitably throughout most phases of the pulsation cycles of the RR Lyr.

%%%%%%%%%%%%%%%%%%%%%%%%%%%%%%%%%%%%%%%%%%%%%%%%%%%%%%%%%%%%%%%%%%%%%%%%%%%%%%%
\section{CHEMICAL ABUNDANCES\label{abunds}}
%%%%%%%%%%%%%%%%%%%%%%%%%%%%%%%%%%%%%%%%%%%%%%%%%%%%%%%%%%%%%%%%%%%%%%%%%%%%%%%

Metal-poor stars usually have chemical compositions that are 
enriched in the $\alpha$-elements (e.g., Mg, Si, S, Ca and possibly Ti),
i.e., [$\alpha$/Fe]~$>$ 0.  
The $\alpha$-rich behavior is attributed to the presumed predominance 
of short-lived massive stars that resulted in core collapse Type~II 
supernovae (SNe II) in early Galactic times.
The SN explosions contributed large amounts of light $\alpha$-elements 
(e.g., O, Ne, Mg and Si), lesser amounts of heavier $\alpha$-elements 
(e.g., Ca and Ti) and even smaller amounts of Fe-peak elements to the 
ISM \citep{WW95}. 
The detonation of neutron-rich cores also is supposed to produce 
heavy \ncap\ isotopes through rapid neutron-capture (hereafter, 
\ncap) nucleosynthesis ($r$-process) where synthesis occurs faster 
than the $\beta$-decay.
As time progressed, longer-lived, lower-mass stars begin to contribute 
their ejecta by adding more Fe-peak elements through type 
Ia supernovae (SNe Ia) which exploded, perhaps due to thermonuclear 
runaway process of accreting binary stars. 
The asymptotic giant branch (AGB) stellar winds contributed isotopes 
for slow \ncap\ nucleosynthesis ($s$-process) at later 
Galactic times. 
Eventually large amounts of iron polluted the ISM and lowered the 
$\alpha$/Fe at higher metallicity, i.e.~[Fe/H] $\simeq -1$.

Do the abundances of metal-poor RR~Lyr conform
to this general Pop~II chemical composition picture?
Using model atmospheres derived as described in \S \ref{modfin}
(listed in Table~\ref{inputmodels}), we computed chemical abundances 
for 22 species of 19 elements in $\sim$165 total phase bins for
our 11 program stars.
Abundances of most elements were derived from EW measurements, 
by adjusting abundances so that calculated EWs match observed EWs
and averaging over all lines of each species.
In the cases of \ion{Mn}{1}, \ion{Sr}{2}, \ion{Zr}{2}, \ion{Ba}{2}, 
\ion{La}{2}, and \ion{Eu}{2}, we employed spectrum syntheses 
to handle the blending, or hyperfine and/or isotopic 
substructure present in these lines. 
We computed theoretical spectra for a variety of assumed abundances 
for each line, then the assumed abundances were changed iteratively 
until the theoretical spectra match the observed ones. 
Syntheses were performed only at phase $\phi$~$\sim$ 0.35 (the 
optimal phase) of each star except for TY Gru, in which the 
spectrum at $\phi$~= 0.46 was used. 
We made this exception because it was the best available phase for 
spectrum syntheses and for the purpose of cleanly comparing our new 
abundance results with those of \citet{Preston06}. 
We caution that metal line profile distortions are slightly larger
at this part of an RR Lyr cycle than at the optimal phase, and therefore
larger uncertainties in the derived abundances can be expected. 

We show relative abundance ratios, [X/Fe], of various 
elements as a function of phase in 
Figures~\ref{rvoct_abund1_phase}--\ref{rvoct_abund4_phase} for RV Oct, a 
non-Blazhko star; and 
Figures~\ref{asvir_abund1_phase}--\ref{asvir_abund4_phase} 
for AS Vir, a Blazhko star. 
In the case of a Blazhko star, we used different colors to represent 
different series of phase bins (see discussion in \S \ref{specprep}). 
Abundances derived via spectrum synthesis are not presented as a function 
of phase because they were derived with only one phase as mentioned above. 
The error bars represent the internal error (line-to-line scatter). 
We adopted internal error of 0.2 dex for abundances derived from a
single line (for plots only). 
The mean relative abundance ratios are represented by the dashed lines. 
NLTE corrections were applied to Na, Al and Si abundances
whenever appropriate in all figures and tables.
Examining these figures, we conclude that the abundances are consistent 
throughout the pulsational cycles in both Blazhko and non-Blazhko stars.  

Tables~\ref{abund1}--\ref{abund4} give the derived [X/Fe] of each 
phase for all program stars. 
The mean [X/Fe] values of each species for each RRab variable 
star (green dots) are presented as a function of metallicity in 
Figures~\ref{xfe}--\ref{xfe2}. 
We overplot them with the results of RHB (red dots) and BHB 
(blue dots) stars presented in \citet{For10}. 
We summarize the mean [X/Fe] values of individual RR Lyr 
in Table~\ref{mabund} and mean [X/Fe] values among different HB 
groups in Table~\ref{mabundHB}. 
In the following subsections we comment on individual 
elements along with the results of RHB and BHB stars from \citet{For10}.

%%%%%%%%%%%%%%%%%%%%%%%%%%%%%%%%%%%%%%%%%%%%%%%%%%%%%%%%%%%%%%%%%%%%%%%%%%%%%%%
\subsection{Magnesium, Calcium and 
Titanium\label{ablightalp}} 
%%%%%%%%%%%%%%%%%%%%%%%%%%%%%%%%%%%%%%%%%%%%%%%%%%%%%%%%%%%%%%%%%%%%%%%%%%%%%%%

As mentioned above, metal-poor stars are generally 
overabundant in $\alpha$-elements.  
\citet{For10} showed that metal-poor non-variable HB stars 
possess standard enhancement in these elements.
The scatter of our derived light $\alpha$-elements abundances is 
small for our RRab stars over the whole metallicity range 
(see Figure~\ref{xfe}). 
We calculated $<$[\ion{Mg}{1}/Fe]$>$~$\simeq$ $+$0.48 for RRab stars, 
which is consistent with the typical $\alpha$-enhancement in 
field metal-poor stars within that metallicity range. 

An offset of [\ion{Ca}{1}/Fe] between RHB and BHB stars, $\sim0.3$ dex, 
was reported by \citet{For10}. 
Our derived [\ion{Ca}{1}/Fe] values are consistent throughout the 
cycles, both in Blazhko and non-Blazhko stars 
(see Figures~\ref{rvoct_abund2_phase} and \ref{asvir_abund2_phase}). 
The mean [Ca/Fe] ratios of our RR Lyr stars also are consistent
with those of RHB stars, as shown in Figure~\ref{xfe}.
We cannot identify the cause of the lower [Ca/Fe] values in the
BHB sample and note that we have [\ion{Ca}{1}/Fe] values of RRab 
stars covering all pulsational phases, including those that overlap 
with the coolest \teff\ range of some BHB stars ($\sim 7400$ K).  
%Unfortunately, we could not perform synthesis for the \ion{Ca}{2} 
%3933~\AA\,K-line in our RRab stars to further investigate this issue,
%because the phases with similar \teff\ as BHB stars have spectra 
%with the most severe line distortion problems. 
%In addition, this line is is extremely strong in the optimal phases of 
%RRab stars, and thus is very insensitive to the Ca abundance.
We also note that the reported trend of decreasing [Ca/Fe] with 
increasing \teff\ for BHB stars as shown in Figure~11 of \citet{For10} 
does not extend into the RR Lyr domain investigated here.

There are no \ion{Ti}{1} lines detectable in the hottest phases of 
RRab stars, \ie, during those early and late phases of a cycle when 
\teff\ overlap with the coolest \teff\ of the BHB stars (\teff$\sim$7400). 
Thus, the  $<$[\ion{Ti}{1}/Fe]$>$ values of our program stars as 
shown here (Figure~\ref{xfe}) cover a similar \teff\ range as the 
warmer RHB stars.
The overall [\ion{Ti}{2}/Fe] ratios appear to be constant with 
[Fe/H], in contrast to the increasing [X/Fe] of the other 
$\alpha$-elements as metallicity declines.
However, if we only consider abundances of \ion{Ti}{1} and 
\ion{Ti}{2} derived for RRab stars, we find that both exhibit a 
flat distribution with a relatively small scatter in this metallicity 
range (excluding the deviant [\ion{Ti}{1}/Fe] of TY Gru).
We also find no trend of [\ion{Ti}{1}/Fe] with increasing \teff\ 
(see \eg, Figure~\ref{asvir_abund2_phase} of AS Vir) in contrast to the 
previous conclusion of \citet{For10} and findings by others 
(see \citealp{Lai08} and references therein).  
Investigation of larger sample of RRab stars covering a wider metallicity 
range is needed to further explore the Ti abundance questions,
but the basic result is clear:  Ti is overabundant in RRab stars
at about the same level as it is in metal-poor stars of other
evolutionary states.

%%%%%%%%%%%%%%%%%%%%%%%%%%%%%%%%%%%%%%%%%%%%%%%%%%%%%%%%%%%%%%%%%%%%%%%%%%%%%%%
\subsection{The Alpha Element Silicon: Revisiting A Special 
Case\label{absilicon}}
%%%%%%%%%%%%%%%%%%%%%%%%%%%%%%%%%%%%%%%%%%%%%%%%%%%%%%%%%%%%%%%%%%%%%%%%%%%%%%%

Standard LTE abundance analyses find a significant dependence of 
[\ion{Si}{1}/Fe] with temperature in metal-poor field stars 
(e.g, \citealp{Cayrel04}, \citealp{Cohen04}, \citealp{PrestonRHB06}, 
\citealp{SL08}, and \citealp{Lai08}). 
The effect seems to depend solely on \teff; no trend with \logg\
has been detected so far.
To investigate this issue, \citet{Shi09} performed an analysis of 
NLTE effects in \ion{Si}{1} in warm metal-poor stars (\teff$\geq$ 6000 K). 
They concluded that the NLTE effects differ from line-to-line and are 
substantially larger in the lower-excitation blue spectral region 
transitions ($\chi$~=~1.9~eV; 3905~\AA\ and 4102~\AA) 
than in the higher-excitation red spectral region 
($\chi$~$\geq$~5~eV; \eg, 5690~\AA\ and 6155~\AA).
Departure from NLTE in warm metal-poor stars is 
also expected for the \ion{Si}{2} 6347~\AA\ and 6371~\AA\ lines. 
									
We revisit the issue of \teff\ dependence on Si lines with our 
RRab stars, because the HB samples cover a large temperature range.
The [\ion{Si}{1}/Fe] values of our program stars were derived either 
solely from the 3905~\AA\ line or lines in red spectral region 
throughout the cycle; the selection of lines depended on the \teff. 
To avoid possible blending of the 3905~\AA\ line with a weak CH transition 
\citet{Cohen04}, which is present in cool stars, we only employed the 
3905~\AA\ line during the early or late phases of a pulsational cycle 
when \teff\ is similar to the BHB stars (\teff~$\geq$ 7400~K). 
									
As shown in Figure~\ref{rvoct_abund2_phase}, the trend of 
[\ion{Si}{2}/Fe] vs phase resembles a similar ``shape'' as the 
\teff\ vs phase plot in the top panel of Figure~\ref{par_phase_rvoct}, 
which suggests a dependence on \teff.  
However, there is no such trend visible in the case of [\ion{Si}{1}/Fe] 
between phase 0--0.8 for RV Oct (see Figure~\ref{rvoct_abund1_phase}). 
Instead, we detect a significant decline of [\ion{Si}{1}/Fe] 
with increasing \teff\ for $\phi$~$\gtrsim$ 0.8 in this star.
To investigate if NLTE effects could be the cause of such trend, 
we applied the suggested NLTE corrections of +0.1 dex and $-$0.1 dex 
by \citet{Shi09} to the \ion{Si}{1} and \ion{Si}{2} abundances 
derived from 3905\AA, 6347\AA\ and 6371\AA\ lines. 
In Figures~\ref{si1_teff} and \ref{si2_teff}, we extend For $\&$ Sneden's 
Figures~14 and 15 by adding all measured [\ion{Si}{1}/Fe] and 
[\ion{Si}{2}/Fe] values that had been corrected for NLTE effects, 
whenever appropriate. 
While the scatter of [\ion{Si}{1}/Fe] is large for our program stars, 
we find a possible declining trend with increasing \teff\ if 
the two outliers (indicated with a black box in the figure) are ignored. 
In contrast, the [\ion{Si}{2}/Fe] values tend to increase with 
increasing \teff\ (as indicated by the arrow in Figure~\ref{si2_teff}).
However, we caution the reader that most [\ion{Si}{2}/Fe] values were 
derived with 1--2 lines, for which we 
anticipate errors of $\pm0.2$ dex. 

To further investigate the NLTE effects on the trends, we present the 
silicon abundances as a function of phase for RV Oct and WY Ant in 
Figure~\ref{si_phase}, where the blue and red dots represent lines 
in the blue and red spectral regions, respectively. 
To emphasize, all values of [\ion{Si}{2}/Fe] and only the blue dots 
of [\ion{Si}{1}/Fe] have been corrected for NLTE effects. 
We find that the NLTE corrections do not resolve the puzzle of \teff\ 
dependency in silicon abundances. 
In fact, even lower [\ion{Si}{1}/Fe] values (as seen in the 
obvious case of WY Ant) were obtained from the use of 3905~\AA\ line in 
warm metal-poor RRab stars.
This suggests that the NLTE computations need to be re-done. 
A discussion about the line transitions of blue and red spectral lines 
of \ion{Si}{1} is given in \citet{SL08}. 
An alternative explanation for the declining and the increasing trends of 
silicon abundances between phase 0.8--1.0 is that the neutral lines 
partially disappear during these phases due to the shock wave.
This phenomenon was first observed in the spectra of S~Arae 
by \citet{Chadid08}, which the disappearance of \ion{Ti}{1} 
and \ion{Fe}{1} lines was shown in their Figure~6.
If this is the case, we might expect to see similar effects in other 
neutral species. 
We do not see this phenomenon in our data set, and the resolution of 
this issue is unsatisfactory.

The overall silicon abundances of RRab stars exhibit a large 
star-to-star scatter, which is similar to 
the results of RHB and BHB stars (see Figure~\ref{xfe}). 
However, the mean Si abundances, $<$[\ion{Si}{1}/Fe]$>$~= +0.48 and 
$<$[\ion{Si}{2}/Fe]$>$~= +0.52 dex are consistent with the mean 
of typical $\alpha$-enhancement in metal-poor stars.

%%%%%%%%%%%%%%%%%%%%%%%%%%%%%%%%%%%%%%%%%%%%%%%%%%%%%%%%%%%%%%%%%%%%%%%%%%%%%%%
\subsection{Light Odd-Z Elements Sodium and Aluminum\label{abnaal}} 
%%%%%%%%%%%%%%%%%%%%%%%%%%%%%%%%%%%%%%%%%%%%%%%%%%%%%%%%%%%%%%%%%%%%%%%%%%%%%%%

For sodium abundances, we used the \ion{Na}{1} resonance D-lines 
(5889.9, 5895.9 \AA) and higher excitation \ion{Na}{1} lines 
(the 5682.6, 5688.2~\AA\, and the 6154.2, 6160.7~\AA\,doublets) 
whenever available. 
The resonance D-lines are generally detected and not saturated in the 
spectra of early and late phases of RRab pulsational cycles. 
The mid (cool) phases possess similar \teff\ range as the RHB stars, 
allowing the weak higher excitation \ion{Na}{1} lines to be detected 
and used in these phases.
There are only two \ion{Al}{1} lines, the resonance 3944, 3961~\AA\ 
doublet, available for this study.

It is well known that the resonance lines of \ion{Na}{1} and 
\ion{Al}{1} can be significantly influenced by NLTE effects 
(see \eg, \citealp{Baumueller98}; \citealp{BG97}). 
The NLTE corrections are particularly important for 
metal-poor stars. 
We applied the suggested NLTE corrections of $-0.5$ dex from 
Baumueller et al. and $+0.65$ dex from Baumueller \& Gehren for 
Na and Al abundances, respectively, derived from those lines.
However, we warn the reader that different NLTE corrections have 
been reported in different studies. 
For example, recent NLTE calculations by \citet{Andrievsky07} 
estimate a correction of only $\sim-0.15$ dex for Na D-lines, but 
\citet{Andrievsky08} suggest an even larger correction for the
blue \ion{Al}{1} resonance lines.

The mean [\ion{Na}{1}/Fe] and [\ion{Al}{1}/Fe] values of RRab stars 
are $-$0.18 dex and $+$0.37 dex, respectively (see Figure~\ref{xfe}). 
NLTE corrections have been applied to individual Na and Al 
abundances whenever appropriate prior to calculating the mean and 
the corrected values are presented in 
both Figure~\ref{xfe} and Table~\ref{abund1}.
Sodium abundances show a large star-to-star scatter with a dispersion 
of 0.2 dex. 
Aluminum is overabundant in RRab stars, similar to those 
derived for BHB stars. 
We warn the reader that we did not have many Na and Al 
measurements throughout the cycles of our RRab sample.
At most, they were generally derived from 1--2 lines.
We find no trend of Al abundances with \teff. 
As such, we do not have an explanation for the discrepancy of 
[\ion{Al}{1}/Fe] between RHB and BHB/RRab stars.

%%%%%%%%%%%%%%%%%%%%%%%%%%%%%%%%%%%%%%%%%%%%%%%%%%%%%%%%%%%%%%%%%%%%%%%%%%%%%%%
\subsection{The iron-peak elements: Scandium through Zinc\label{abfepeak}}. 
%%%%%%%%%%%%%%%%%%%%%%%%%%%%%%%%%%%%%%%%%%%%%%%%%%%%%%%%%%%%%%%%%%%%%%%%%%%%%%%

As noted by \citet{PM00}, scandium abundances can be affected by 
hyperfine substructure of the \ion{Sc}{2} features. 
However, tests performed in \citet{For10} suggest that the effect is 
small in lines of interest here. 
Thus, we proceeded as in that paper, using EWs to derive 
\ion{Sc}{2} abundances.
Both [\ion{Sc}{2}/Fe] and [\ion{V}{2}/Fe] values are roughly solar 
with $<$[\ion{Sc}{2}/Fe]$>$~$\simeq$ $+ $0.1 dex and 
$<$[\ion{V}{2}/Fe]$>$~$\simeq$ $+$0.2 dex for RRab 
stars (see Figure~\ref{xfe1}). 
They are also in accord with the results derived for RHB and BHB stars. 
We note that there are not many detectable \ion{V}{2} lines available 
for analysis throughout the RR Lyr cycles. 
We also find no trends of [\ion{Sc}{2}/Fe] and [\ion{V}{2}/Fe] 
with either [Fe/H] or \teff. 

The derived [\ion{Cr}{1}/Fe] and [\ion{Cr}{2}/Fe] in our RR Lyr 
sample are discrepant: abundances from the neutral lines are 
$\sim$0.2 dex lower than those from the ion lines.
This result is similar to those found for other metal-poor stars groups 
(see \citealp{Sobeck07}, and references therein). 
But even solar \ion{Cr}{1} and \ion{Cr}{2} abundances derived with 
recent reliable transition probabilities for these species cannot 
be brought into agreement; Sobeck et al. found an offset of 0.15--0.20 dex.
This suggests that the problem is not entirely due to NLTE effects.
As shown in Figure~\ref{rvoct_abund3_phase}, our chromium abundances 
are consistent throughout the cycle. 
It supports the conclusions of Sobeck et al. but is different from
the conclusion of \citet{For10}, which found a trend of increasing 
[\ion{Cr}{1}/Fe] as increasing \teff\ $<$ 7000 K.

Manganese abundances show a large star-to-star scatter with a 
dispersion of 0.17 dex for our RRab star (see Figure~\ref{xfe1}). 
In general, only 1--3 lines were employed for synthesis. 
The [\ion{Mn}{1}/Fe] values presented here are not an average value 
throughout the cycle but the abundance from the single ``optimal'' phase.
The overall manganese abundances trend of increasing [\ion{Mn}{1}/Fe] 
with higher [Fe/H] metallicities is in accord with previous studies 
(see \citealp{Sobeck06}, \citealp{Lai08}, and references therein). 

The derived [\ion{Co}{1}/Fe] values for RRab stars have smaller 
star-to-star scatter ($\sigma\simeq0.08$) compared to those 
derived for RHB stars ($\sigma\simeq0.26$); (see Figure~\ref{xfe1}). 
This is due to the fact that many [\ion{Co}{1}/Fe] values 
have been derived throughout the cycles and used to give the average 
[\ion{Co}{1}/Fe] for each star presented in Figure~\ref{xfe1}. 
Our Ni abundances were also derived in a similar manner as Co abundances. 
Formally, we derive [\ion{Ni}{1}/Fe]~= $+$0.47, but the star-to-star
scatter is large for both RRab stars and RHB stars ($\sigma$~= 0.13 and 
0.22, respectively).
There are no clean \ion{Ni}{2} lines in our spectra of RRab stars.

We caution that abundances of \ion{Co}{1} and 
\ion{Ni}{1} of each phase were determined with only 1--2 lines and 
show large phase-to-phase scatter, in particularly for [\ion{Ni}{1}/Fe] 
(see Figures~\ref{rvoct_abund4_phase} and \ref{asvir_abund3_phase}).
Determination of \ion{Ni}{2} abundances was not possible 
because the single available line at 4067~\AA\ line exhibits a
distorted profile and is only detectable in early and late phases of 
a pulsational cycle.

The dispersion of [\ion{Zn}{1}/Fe] is small, with 
$<$[\ion{Zn}{1}/Fe]$>$~$\simeq$+0.16 dex for RRab stars 
(see Figure~\ref{xfe1}). 
The enhancement of Zn abundances toward the low metallicity range 
as seen in the RHB stars is inconclusive. 
A larger sample of RRab stars in [Fe/H]$< -2.0$ regime 
might help to resolve this puzzle. 
Overall, the derived Fe-peak abundance ratios of our RRab 
stars, along with RHB and BHB stars in \citet{For10} are in agreement 
with those found in field dwarfs and giants.

%%%%%%%%%%%%%%%%%%%%%%%%%%%%%%%%%%%%%%%%%%%%%%%%%%%%%%%%%%%%%%%%%%%%%%%%%%%%%%
\subsection{The neutron capture elements: Strontium, Yttrium, Zirconium, 
            Barium, Lanthanum and Europium\label{abncap}} 
%%%%%%%%%%%%%%%%%%%%%%%%%%%%%%%%%%%%%%%%%%%%%%%%%%%%%%%%%%%%%%%%%%%%%%%%%%%%%%

We were able to derive abundances of three light \ncap\
elements (Sr, Y and Zr) and and three heavy \ncap\ 
elements (Ba, La and Eu) in most of our RRab stars.
The derived abundances of these elements show large star-to-star scatter 
with respect to Fe (see Figure~\ref{xfe2}). 

Strontium abundances were derived using the \ion{Sr}{2} resonance
lines 4077, 4215~\AA, and the higher excitation  4161~\AA\ line.
These lines are generally strong and/or blended in cool stars. 
A large dispersion of Sr abundances has been found in RHB and BHB 
stars \citep{For10}, as well as in other samples of metal-poor stars,
so we believe that the large dispersion ($\sigma$~= 0.25 dex, 
see Table~\ref{mabundHB}) derived for our RRab stars represents 
a true star-to-star intrinsic scatter. 
The overall [\ion{Sr}{2}/Fe] distribution is similar to those of RHB stars, 
which unfortunately does not aid us in explaining the presence of Sr 
abundance offset between RHB and BHB stars.

Equivalent width analysis and synthesis were performed to obtain 
Yttrium and Zirconium abundances, respectively. 
Both [\ion{Y}{2}/Fe] and [\ion{Zr}{2}/Fe] exhibit a large star-to-star 
scatter with dispersions $\simeq0.17$ dex. 
Zirconium abundances are overabundant as compared to the other 
light \ncap\ elements Sr and Y.
The \ion{Zr}{2} lines are generally very weak; 
there are not many phases per star with detected lines.
Hence, interpretation of Zr abundances should be done with caution.

Barium lines are affected by both hyperfine substructure and 
isotopic splitting (see a line list given by \citealp{McWilliam98}). 
The solar abundance ratio distribution among the 
$^{134--138}$Ba isotopes \citep{Lodders03} was adopted for synthesizing the
\ion{Ba}{2} 4554 \AA, 5853 \AA, 6141 \AA, and 6496 \AA\ lines, whenever present in the spectra.
We note that the 4554 \AA\ line is always substantially stronger than
the other lines, and Ba abundances derived from this line can also be 
larger.
Abundances derived from the  4554~\AA\ are severely affected by 
microturbulence and damping uncertainties. 
Syntheses were performed on the \ion{La}{2} 4086, 4123~\AA\ lines, 
and the \ion{Eu}{2} 4129, 4205~\AA\ lines, whenever present in the spectra. 
These lines are very weak and only 1--2 lines are available for analysis. 
The overall barium, lanthanum and europium abundances for RRab stars are 
in accord with those derived for RHB and BHB stars in the same 
metallicity range.

%%%%%%%%%%%%%%%%%%%%%%%%%%%%%%%%%%%%%%%%%%%%%%%%%%%%%%%%%%%%%%%%%%%%%%%%%%%%%%%
\section{THE RED EDGE OF THE RR LYRAE INSTABILITY STRIP 
REVISITED\label{rededge}}
%%%%%%%%%%%%%%%%%%%%%%%%%%%%%%%%%%%%%%%%%%%%%%%%%%%%%%%%%%%%%%%%%%%%%%%%%%%%%%%

A recent estimate of the effective temperature at the red edge of the 
RR Lyrae instability strip, \teff(FRE), by \cite{Hansen11} prompts us to 
investigate this quantity anew.  
Hansen et~al. adopt \teff(FRE)~= 5900 K,  the effective temperature 
derived from analysis of spectra of two metal-poor RR Lyr stars observed 
near minimum light.  
Their estimate arises from a misunderstanding of the FRE.  
This is illustrated in Figure~\ref{m68}, where we superpose ($V$, $B-V$) loops 
for two RRab stars, V14 (P~= 0.5568~d) and V35 (P~=0.7025~d), on the 
horizontal branch of the metal-poor ([Fe/H]=-2.2) globular cluster M68.  
The data are those of \cite{Walker94}.  
The schematic horizontal branch was hand drawn through the data
points in Walker's Figure~13.
Vertical blue and red lines denote boundaries of the instability strip 
defined approximately by the locations of BHB, RR Lyr, and RHB stars 
in that Figure.  
For a considerable portion of the pulsation cycles preceding minimum 
light, as can be inferred from the densities of data points at 
faintest apparent magnitudes, the colors of the RR Lyrae stars lie in 
the RHB domain, well outside of the instability strip.  
This is a general characteristic of the RRab stars.  
T(FRE) cannot be derived from observations at these phases alone. 

\cite{PrestonRHB06} obtained their higher value, T(FRE)~= 6300~K, 
by pinching the FRE between the red edge of the color (temperature) 
distribution of metal-poor RRab stars and the blue edge of the 
metal-poor RHB distribution at their disposal.  
For this purpose they used mean colors of RRab stars, employing 
the formalism of \cite{Preston61}.  
This formalism, used to locate RR Lyrae stars in CMDs for many 
decades, defines the mean color (hence mean \teff) of an RR~Lyr 
star as the color of a fictitious static star with the same \teff\ 
and absolute luminosity.  
Variants of the procedure by which this color is calculated 
are reviewed (with references) by \cite{Sandage06}.  
The variants produce small differences in the mean colors that are 
not important for the present discussion.

Here, we follow a procedure similar to that used by \cite{PrestonRHB06} 
based on effective temperatures derived from analyses of our RR Lyr spectra.  
We calculated \teff\ values at intervals of 0.05 in phase by linear interpolation 
among the data in Table~\ref{inputmodels}. 
We used these to calculate the average values of \teff\ for each star given in Table~\ref{mteff}. 
We omitted TY Gru and V1645 Sgr for which we deemed the data inadequate.  
We estimated T(FRE)~= 6310~K as the average of the two lowest values 
in Table~\ref{mteff} (for CD Vel and Z Mic).  
In similar fashion we estimated T(FRE)~= 6250~K as the average of the 
two highest \teff\ values among the RHB stars of \cite{For10}.  
We adopt \teff(FRE)~= 6280~$\pm$~30~K as the average of these two 
independent estimates.   
A histogram illustrating the distribution of these RR~Lyr and RHB 
temperatures is presented in Figure~\ref{starcount}.
Our procedure based on new data is closely equivalent to that of 
Preston et~al., and it produces virtually the same value for \teff(FRE), 
albeit for a sample of somewhat higher mean [Fe/H].  
The estimate offered here supersedes the estimate of \cite{For10}.

Two puzzles emerge from this discussion:  why is there such small 
dispersion in $<$\teff$>$ among the RR~Lyr that populate a relatively 
broad color region of the instability strip, and why do these values 
crowd the red edge?  
These are issues for future investigation.

%%%%%%%%%%%%%%%%%%%%%%%%%%%%%%%%%%%%%%%%%%%%%%%%%%%%%%%%%%%%%%%%%%%%%%%%%%%%%%%
\section{EVOLUTIONARY STATES OF THE RR~LYR SAMPLE\label{evol}}
%%%%%%%%%%%%%%%%%%%%%%%%%%%%%%%%%%%%%%%%%%%%%%%%%%%%%%%%%%%%%%%%%%%%%%%%%%%%%%%

Horizontal-branch morphology is a complex function of many parameters.
The first and most obvious parameter is metallicity, because 
metal-rich globular clusters have mostly RHB stars while
metal-poor globular clusters have mostly BHB and/or EHB stars. 
The metallicity distributions of the field RHB and BHB samples 
in \cite{For10} and RRab sample of this study have some differences.
More RHB stars agglomerate toward the lower metallicity regime
([Fe/H]$<-2.0$), more BHB stars toward the higher metallicity regime 
([Fe/H]$>-1.5$), respectively, and the RRab sample falls in between. 
These distributions, which confuse arguments
about the first parameter of HB morphology, are artificial and
arise from observational selection biases.
They cannot provide physical interpretation of HB morphology.

In \citet{For10}, the majority RHB stars were selected from 
\citet{PrestonRHB06}, which was a study specifically focused on 
metal-poor RHB stars. 
On the other hand, metal-poor BHB stars ([Fe/H]$<-2.0$) were excluded 
due to the lack of measurable \ion{Fe}{1} and \ion{Fe}{2} lines 
for spectroscopic analysis (see comment in Table~2 of \citealp{For10}). 
The RRab stars that were selected for this study partly to better 
understand the nature of a carbon-rich and $s$-process rich 
RRab star, TY Gru \citep{Preston06}. 
We refer the reader to the description of selecting RRab stars in 
this study to FPS11. 

With these cautions in mind, we compared the physical properties 
of our RRab stars with the RR Lyr samples of \citet{Lambert96} 
and \citet{Clementini95}.  
In Figure~\ref{hr}, we extend Figure~19 of \citet{For10} by adding 
the derived spectroscopic \teff\ and \logg\ values of two of our 
RRab stars, CD Vel and WY Ant, on the \teff-\logg\ plane. 
These two stars are selected due to the lower \logg\ of WY Ant throughout 
the cycle as compared to CD Vel, which provides a small vertical offset 
for easier visual inspection. 
The \teff\ and \logg\ values of field RR Lyr samples are 
based on the spectroscopic derivations of \citet{Lambert96}, and photometric 
\teff\ and Baade-Wesselink \logg\ of \citet{Clementini95} study. 
Our \logg\ values derived from spectroscopic ionization 
balance are generally lower than the Baade-Wesselink method. 
However, they follow the general physical \teff\ and \logg\ change 
with the RHB and BHB population across the \teff-\logg\ plane. 

In Figure~\ref{hr_rrlyr}, we enlarge Figure~\ref{hr} 
near the RR Lyr instability strip region. 
In this figure we have added $\alpha$-enhanced HB tracks of 
[M/H]= $-1.79$, $Z= 0.0003$ and $Y= 0.245$ with different model masses. 
These HB tracks are adopted from \cite{PCSC06}, which have been implemented 
with low $T$-opacities of \citet{Ferguson05} and an $\alpha$-enhanced 
distribution that represents typical Galactic halo and bulge stars. 
We employed Eq.~4 to convert the bolometric luminosities in the model 
to \logg\ values. 
A large star-to-star scatter for Lambert et al's data is evident,
but our RR Lyrs follow the general trend of a single mass evolutionary 
track (within \logg\ uncertainties) except near 7000--7500 K region. 
The scatter in this \teff\ range is due to the fast moving and complex 
nature of RR Lyr atmosphere during the rising and descending 
branch of the cycle. 
Accepting at face value the large spread in \logg\ implies
masses in the entire range from 0.5--0.8 \Msun, a conclusion broadly
consistent with horizontal branch theory.

%%%%%%%%%%%%%%%%%%%%%%%%%%%%%%%%%%%%%%%%%%%%%%%%%%%%%%%%%%%%%%%%%%%%%%%%%%%%%%%
\section{SUMMARY AND CONCLUSIONS\label{sumconcl}}
%%%%%%%%%%%%%%%%%%%%%%%%%%%%%%%%%%%%%%%%%%%%%%%%%%%%%%%%%%%%%%%%%%%%%%%%%%%%%%%
 
We present the first detailed chemical abundance study of field 
horizontal branch RR Lyrae variable stars throughout their pulsational cycles. 
For this work we gathered some 2300 high resolution spectra of 11 RRab
stars with the du Pont 2.5-m telescope at the Las Campanas Observatory.
The samples were selected based on the study of \citet{Preston11}. 
A new, indirect method to estimate initial \teff\ values for 
the analysis was developed.
These estimated temperatures work reasonably well for both Blazhko 
and non-Blazhko effect stars.

We derived the model stellar atmospheric parameters, \teff, 
\logg, [M/H] and \vturb\ for all our program stars throughout the 
pulsational cycles based on spectroscopic constraints. 
Variations of microturbulence as a function of \teff\ and phase were found. 
We found a variation of \vturb\ with \teff\ along the horizontal 
branch that goes through a maximum in the RR Lyr instability strip.
We also showed for the first time observationally that the  
variation of \vturb\ as a function of phase is similar to the theoretical 
\vturb\ and kinetic energy calculations of \citet{Fokin99} 
and \citet{Kolenberg10}, respectively. 

Employing the derived model stellar atmospheric parameters, we 
obtained abundance ratios, [X/Fe], of the $\alpha$-elements, 
light odd-$Z$ elements, Fe-peak elements, and \ncap\ elements.
The elemental abundance ratios show consistency throughout the pulsational 
cycles for both Blazhko and non-Blazhko effect stars. 
The mean abundance ratios vs metallicity of our program stars 
are also generally in accord with the RHB and BHB stars. 
We did not obtain satisfactory solution for the known trend of Silicon 
abundances as a function of \teff\ with our RR Lyr stars. 

Finally, we investigated the physical properties of our RR Lyr stars by 
comparing them with those presented in \citet{Lambert96} and 
\citet{Clementini95} in the \teff-\logg\ plane. 
A large star-to-star scatter on the \teff-\logg\ plane was found 
for Lambert et al's samples in contrast to our RR Lyr, which follow the 
general trend of a single mass evolutionary track. 
Clementini et al. obtained lower \logg\ values from analysis
by the BW method.

\acknowledgments

This paper covers part of BQF's dissertation, submitted 
in partial fulfillment of the requirements for the degree of Doctor of 
Philosophy at the University of Texas, Austin. 
We thank Tom Barnes, Harriet Dinerstein, Bob Rood, Craig Wheeler and 
referee for comments on this work.
The research was supported by the National Science Foundation through
grant AST-0908978.

%\appendix
\bibliographystyle{apj}
\bibliography{ref}

\begin{thebibliography}{83}
\expandafter\ifx\csname natexlab\endcsname\relax\def\natexlab#1{#1}\fi

\bibitem[{{Allende Prieto} {et~al.}(1999){Allende Prieto}, {Garc{\'{\i}}a
  L{\'o}pez}, {Lambert}, \& {Gustafsson}}]{AP99}
{Allende Prieto}, C., {Garc{\'{\i}}a L{\'o}pez}, R.~J., {Lambert}, D.~L., \&
  {Gustafsson}, B. 1999, \apj, 527, 879

\bibitem[{{Aller}(1953)}]{Aller53}
{Aller}, L.~H. 1953, {Astrophysics; the Atmospheres of the Sun and Stars} (New
  York, Ronald Press)

\bibitem[{{Alonso} {et~al.}(1996){Alonso}, {Arribas}, \&
  {Martinez-Roger}}]{Alonso96}
{Alonso}, A., {Arribas}, S., \& {Martinez-Roger}, C. 1996, \aap, 313, 873

\bibitem[{{Andrievsky} {et~al.}(2007){Andrievsky}, {Spite}, {Korotin}, {Spite},
  {Bonifacio}, {Cayrel}, {Hill}, \& {Fran{\c c}ois}}]{Andrievsky07}
{Andrievsky}, S.~M., {Spite}, M., {Korotin}, S.~A., {Spite}, F., {Bonifacio},
  P., {Cayrel}, R., {Hill}, V., \& {Fran{\c c}ois}, P. 2007, \aap, 464, 1081

\bibitem[{{Andrievsky} {et~al.}(2008){Andrievsky}, {Spite}, {Korotin}, {Spite},
  {Bonifacio}, {Cayrel}, {Hill}, \& {Fran{\c c}ois}}]{Andrievsky08}
---. 2008, \aap, 481, 481

\bibitem[{{Aoki} {et~al.}(2002){Aoki}, {Ando}, {Honda}, {Iye}, {Izumiura},
  {Kajino}, {Kambe}, {Kawanomonoto}, {Noguchi}, {Okita}, {Sadakane}, {Sato},
  {Shelton}, {Takada-Hidai}, {Takeda}, {Watanabe}, \& {Yoshida}}]{Aoki02}
{Aoki}, W., {et~al.} 2002, \pasj, 54, 427

\bibitem[{{Asplund} {et~al.}(2006){Asplund}, {Lambert}, {Nissen}, {Primas}, \&
  {Smith}}]{Asplund06}
{Asplund}, M., {Lambert}, D.~L., {Nissen}, P.~E., {Primas}, F., \& {Smith},
  V.~V. 2006, \apj, 644, 229

\bibitem[{{Baumueller} {et~al.}(1998){Baumueller}, {Butler}, \&
  {Gehren}}]{Baumueller98}
{Baumueller}, D., {Butler}, K., \& {Gehren}, T. 1998, \aap, 338, 637

\bibitem[{{Baumueller} \& {Gehren}(1997)}]{BG97}
{Baumueller}, D., \& {Gehren}, T. 1997, \aap, 325, 1088

\bibitem[{{Beers} {et~al.}(1992){Beers}, {Preston}, \& {Shectman}}]{BPS92}
{Beers}, T.~C., {Preston}, G.~W., \& {Shectman}, S.~A. 1992, \aj, 103, 1987

\bibitem[{{Bla{\v z}ko}(1907)}]{Blazhko07}
{Bla{\v z}ko}, S. 1907, Astronomische Nachrichten, 175, 325

\bibitem[{{Butler}(1975)}]{Butler75}
{Butler}, D. 1975, \apj, 200, 68

\bibitem[{{Butler} {et~al.}(1982){Butler}, {Manduca}, {Bell}, \&
  {Deming}}]{Butler82}
{Butler}, D., {Manduca}, A., {Bell}, R.~A., \& {Deming}, D. 1982, \aj, 87, 640

\bibitem[{{Carney} \& {Jones}(1983)}]{CJ83}
{Carney}, B.~W., \& {Jones}, R. 1983, \pasp, 95, 246

\bibitem[{{Castellani} {et~al.}(2005){Castellani}, {Castellani}, \&
  {Cassisi}}]{Castellani05}
{Castellani}, M., {Castellani}, V., \& {Cassisi}, S. 2005, \aap, 437, 1017

\bibitem[{{Castelli} {et~al.}(1997){Castelli}, {Gratton}, \&
  {Kurucz}}]{Castelli97}
{Castelli}, F., {Gratton}, R.~G., \& {Kurucz}, R.~L. 1997, \aap, 318, 841

\bibitem[{{Castor}(1972)}]{Castor72}
{Castor}, J.~P. 1972, in The Evolution of Population II Stars, ed.
  {A.~G.~D.~Philip}, 147--+

\bibitem[{{Cayrel} {et~al.}(2004){Cayrel}, {Depagne}, {Spite}, {Hill}, {Spite},
  {Fran{\c c}ois}, {Plez}, {Beers}, {Primas}, {Andersen}, {Barbuy},
  {Bonifacio}, {Molaro}, \& {Nordstr{\"o}m}}]{Cayrel04}
{Cayrel}, R., {et~al.} 2004, \aap, 416, 1117

\bibitem[{{Chadid} \& {Gillet}(1996)}]{CG96}
{Chadid}, M., \& {Gillet}, D. 1996, \aap, 315, 475

\bibitem[{{Chadid} {et~al.}(2008){Chadid}, {Vernin}, \& {Gillet}}]{Chadid08}
{Chadid}, M., {Vernin}, J., \& {Gillet}, D. 2008, \aap, 491, 537

\bibitem[{{Clementini}(2010)}]{Clementini10}
{Clementini}, G. 2010, in Variable Stars, the Galactic halo and Galaxy
  Formation, ed. {C.~Sterken, N.~Samus, \& L.~Szabados}, 107--+

\bibitem[{{Clementini} {et~al.}(1995){Clementini}, {Carretta}, {Gratton},
  {Merighi}, {Mould}, \& {McCarthy}}]{Clementini95}
{Clementini}, G., {Carretta}, E., {Gratton}, R., {Merighi}, R., {Mould}, J.~R.,
  \& {McCarthy}, J.~K. 1995, \aj, 110, 2319

\bibitem[{{Clementini} {et~al.}(2005){Clementini}, {Gratton}, {Bragaglia},
  {Ripepi}, {Martinez Fiorenzano}, {Held}, \& {Carretta}}]{Clementini05}
{Clementini}, G., {Gratton}, R.~G., {Bragaglia}, A., {Ripepi}, V., {Martinez
  Fiorenzano}, A.~F., {Held}, E.~V., \& {Carretta}, E. 2005, \apjl, 630, L145

\bibitem[{{Clementini} {et~al.}(1994){Clementini}, {Merighi}, {Gratton}, \&
  {Carretta}}]{Clementini94}
{Clementini}, G., {Merighi}, R., {Gratton}, R., \& {Carretta}, E. 1994, \mnras,
  267, 43

\bibitem[{{Cohen} {et~al.}(2004){Cohen}, {Christlieb}, {McWilliam}, {Shectman},
  {Thompson}, {Wasserburg}, {Ivans}, {Dehn}, {Karlsson}, \&
  {Melendez}}]{Cohen04}
{Cohen}, J.~G., {et~al.} 2004, \apj, 612, 1107

\bibitem[{{Ferguson} {et~al.}(2005){Ferguson}, {Alexander}, {Allard}, {Barman},
  {Bodnarik}, {Hauschildt}, {Heffner-Wong}, \& {Tamanai}}]{Ferguson05}
{Ferguson}, J.~W., {Alexander}, D.~R., {Allard}, F., {Barman}, T., {Bodnarik},
  J.~G., {Hauschildt}, P.~H., {Heffner-Wong}, A., \& {Tamanai}, A. 2005, \apj,
  623, 585

\bibitem[{{Fernley} {et~al.}(1998){Fernley}, {Barnes}, {Skillen}, {Hawley},
  {Hanley}, {Evans}, {Solano}, \& {Garrido}}]{Fernley98}
{Fernley}, J., {Barnes}, T.~G., {Skillen}, I., {Hawley}, S.~L., {Hanley},
  C.~J., {Evans}, D.~W., {Solano}, E., \& {Garrido}, R. 1998, \aap, 330, 515

\bibitem[{{Fitzpatrick} \& {Sneden}(1987)}]{Fitz87}
{Fitzpatrick}, M.~J., \& {Sneden}, C. 1987, in Bulletin of the American
  Astronomical Society, Vol.~19, Bulletin of the American Astronomical Society,
  1129--+

\bibitem[{{Fokin} \& {Gillet}(1997)}]{FG97}
{Fokin}, A.~B., \& {Gillet}, D. 1997, \aap, 325, 1013

\bibitem[{{Fokin} {et~al.}(1999{\natexlab{a}}){Fokin}, {Gillet}, \&
  {Chadid}}]{FGC99}
{Fokin}, A.~B., {Gillet}, D., \& {Chadid}, M. 1999{\natexlab{a}}, \aap, 344,
  930

\bibitem[{{Fokin} {et~al.}(1999{\natexlab{b}}){Fokin}, {Gillet}, \&
  {Chadid}}]{Fokin99}
---. 1999{\natexlab{b}}, \aap, 344, 930

\bibitem[{{For} {et~al.}(2011){For}, {Preston}, \& {Sneden}}]{For11}
{For}, B.-Q., {Preston}, G.~W., \& {Sneden}, C. 2011, \apjs, in press

\bibitem[{{For} \& {Sneden}(2010)}]{For10}
{For}, B.-Q., \& {Sneden}, C. 2010, \aj, 140, 1694

\bibitem[{{Gautschy}(1987)}]{Gautschy87}
{Gautschy}, A. 1987, Vistas in Astronomy, 30, 197

\bibitem[{{Gould} \& {Popowski}(1998)}]{GP98}
{Gould}, A., \& {Popowski}, P. 1998, \apj, 508, 844

\bibitem[{{Gratton}(1998)}]{Gratton98}
{Gratton}, R.~G. 1998, \mnras, 296, 739

\bibitem[{{Gratton} {et~al.}(1997){Gratton}, {Fusi Pecci}, {Carretta},
  {Clementini}, {Corsi}, \& {Lattanzi}}]{Gratton97}
{Gratton}, R.~G., {Fusi Pecci}, F., {Carretta}, E., {Clementini}, G., {Corsi},
  C.~E., \& {Lattanzi}, M. 1997, \apj, 491, 749

\bibitem[{{Greenstein}(1948)}]{Greenstein48}
{Greenstein}, J.~L. 1948, \apj, 107, 151

\bibitem[{{Greenstein}(1947)}]{Greenstein47}
{Greenstein}, J.~L. \&~Adams, W.~S. 1947, \apj, 106, 339

\bibitem[{{Hansen} {et~al.}(2011){Hansen}, {Nordstr{\"o}m}, {Bonifacio},
  {Spite}, {Andersen}, {Beers}, {Cayrel}, {Spite}, {Molaro}, {Barbuy},
  {Depagne}, {Fran{\c c}ois}, {Hill}, {Plez}, \& {Sivarani}}]{Hansen11}
{Hansen}, C.~J., {et~al.} 2011, \aap, 527, A65+

\bibitem[{{Hardorp} \& {Scholz}(1967)}]{Hardorp67}
{Hardorp}, J., \& {Scholz}, M. 1967, \zap, 67, 312

\bibitem[{{Helmi} \& {White}(1999)}]{Helmi99}
{Helmi}, A., \& {White}, S.~D.~M. 1999, \mnras, 307, 495

\bibitem[{{Hosford} {et~al.}(2009){Hosford}, {Ryan}, {Garc{\'{\i}}a P{\'e}rez},
  {Norris}, \& {Olive}}]{Hosford09}
{Hosford}, A., {Ryan}, S.~G., {Garc{\'{\i}}a P{\'e}rez}, A.~E., {Norris},
  J.~E., \& {Olive}, K.~A. 2009, \aap, 493, 601

\bibitem[{{Kolenberg} {et~al.}(2010){Kolenberg}, {Fossati}, {Shulyak},
  {Pikall}, {Barnes}, {Kochukhov}, \& {Tsymbal}}]{Kolenberg10}
{Kolenberg}, K., {Fossati}, L., {Shulyak}, D., {Pikall}, H., {Barnes}, T.~G.,
  {Kochukhov}, O., \& {Tsymbal}, V. 2010, \aap, 519, A64+

\bibitem[{{Kraft} \& {Ivans}(2003)}]{KI03}
{Kraft}, R.~P., \& {Ivans}, I.~I. 2003, \pasp, 115, 143

\bibitem[{{Kurucz}(1979)}]{Kurucz79}
{Kurucz}, R.~L. 1979, \apjs, 40, 1

\bibitem[{{Lai} {et~al.}(2008){Lai}, {Bolte}, {Johnson}, {Lucatello}, {Heger},
  \& {Woosley}}]{Lai08}
{Lai}, D.~K., {Bolte}, M., {Johnson}, J.~A., {Lucatello}, S., {Heger}, A., \&
  {Woosley}, S.~E. 2008, \apj, 681, 1524

\bibitem[{{Lambert} {et~al.}(1996){Lambert}, {Heath}, {Lemke}, \&
  {Drake}}]{Lambert96}
{Lambert}, D.~L., {Heath}, J.~E., {Lemke}, M., \& {Drake}, J. 1996, \apjs, 103,
  183

\bibitem[{{Layden}(1994)}]{Layden94}
{Layden}, A.~C. 1994, \aj, 108, 1016

\bibitem[{{Liu} \& {Janes}(1989)}]{LJ89}
{Liu}, T., \& {Janes}, K.~A. 1989, \apjs, 69, 593

\bibitem[{{Liu} \& {Janes}(1990)}]{LJ90}
---. 1990, \apj, 354, 273

\bibitem[{{Lodders}(2003)}]{Lodders03}
{Lodders}, K. 2003, \apj, 591, 1220

\bibitem[{{Manduca}(1981)}]{Manduca81}
{Manduca}, A. 1981, \apj, 245, 258

\bibitem[{{McWilliam}(1998)}]{McWilliam98}
{McWilliam}, A. 1998, \aj, 115, 1640

\bibitem[{{Pietrinferni} {et~al.}(2006){Pietrinferni}, {Cassisi}, {Salaris}, \&
  {Castelli}}]{PCSC06}
{Pietrinferni}, A., {Cassisi}, S., {Salaris}, M., \& {Castelli}, F. 2006, \apj,
  642, 797

\bibitem[{{Pojmanski}(2002)}]{Pojmanski02}
{Pojmanski}, G. 2002, ACTAA, 52, 397

\bibitem[{{Preston}(1959)}]{Preston59}
{Preston}, G.~W. 1959, \apj, 130, 507

\bibitem[{{Preston}(1961{\natexlab{a}})}]{Preston61}
---. 1961{\natexlab{a}}, \apj, 134, 633

\bibitem[{{Preston}(1961{\natexlab{b}})}]{Preston61b}
---. 1961{\natexlab{b}}, \apj, 133, 29

\bibitem[{{Preston}(2009)}]{Preston09}
---. 2009, \aap, 507, 1621

\bibitem[{{Preston}(2011)}]{Preston11}
---. 2011, \aj, 141, 6

\bibitem[{{Preston} {et~al.}(1991){Preston}, {Shectman}, \&
  {Beers}}]{Preston91}
{Preston}, G.~W., {Shectman}, S.~A., \& {Beers}, T.~C. 1991, \apjs, 76, 1001

\bibitem[{{Preston} {et~al.}(2006{\natexlab{a}}){Preston}, {Sneden},
  {Thompson}, {Shectman}, \& {Burley}}]{PrestonRHB06}
{Preston}, G.~W., {Sneden}, C., {Thompson}, I.~B., {Shectman}, S.~A., \&
  {Burley}, G.~S. 2006{\natexlab{a}}, \aj, 132, 85

\bibitem[{{Preston} {et~al.}(2006{\natexlab{b}}){Preston}, {Thompson},
  {Sneden}, {Stachowski}, \& {Shectman}}]{Preston06}
{Preston}, G.~W., {Thompson}, I.~B., {Sneden}, C., {Stachowski}, G., \&
  {Shectman}, S.~A. 2006{\natexlab{b}}, \aj, 132, 1714

\bibitem[{{Prochaska} \& {McWilliam}(2000)}]{PM00}
{Prochaska}, J.~X., \& {McWilliam}, A. 2000, \apjl, 537, L57

\bibitem[{{Ram{\'{\i}}rez} \& {Mel{\'e}ndez}(2005)}]{RM05}
{Ram{\'{\i}}rez}, I., \& {Mel{\'e}ndez}, J. 2005, \apj, 626, 465

\bibitem[{{Ryan} {et~al.}(1996){Ryan}, {Norris}, \& {Beers}}]{Ryan96}
{Ryan}, S.~G., {Norris}, J.~E., \& {Beers}, T.~C. 1996, \apj, 471, 254

\bibitem[{{Sandage}(2006)}]{Sandage06}
{Sandage}, A. 2006, \aj, 131, 1750

\bibitem[{{Sanford}(1930)}]{san30}
{Sanford}, R.~F. 1930, \apj, 72, 46

\bibitem[{{Shi} {et~al.}(2009){Shi}, {Gehren}, {Mashonkina}, \& {Zhao}}]{Shi09}
{Shi}, J.~R., {Gehren}, T., {Mashonkina}, L., \& {Zhao}, G. 2009, \aap, 503,
  533

\bibitem[{{Smith} \& {Butler}(1978)}]{SB78}
{Smith}, H.~A., \& {Butler}, D. 1978, \pasp, 90, 671

\bibitem[{{Sneden} \& {Lawler}(2008)}]{SL08}
{Sneden}, C., \& {Lawler}, J.~E. 2008, in American Institute of Physics
  Conference Series, Vol. 990, First Stars III, ed. {B.~W.~O'Shea \& A.~Heger},
  90--103

\bibitem[{{Sneden}(1973)}]{Sneden73}
{Sneden}, C.~A. 1973, PhD thesis, THE UNIVERSITY OF TEXAS AT AUSTIN.

\bibitem[{{Sobeck} {et~al.}(2006){Sobeck}, {Ivans}, {Simmerer}, {Sneden},
  {Hoeflich}, {Fulbright}, \& {Kraft}}]{Sobeck06}
{Sobeck}, J.~S., {Ivans}, I.~I., {Simmerer}, J.~A., {Sneden}, C., {Hoeflich},
  P., {Fulbright}, J.~P., \& {Kraft}, R.~P. 2006, \aj, 131, 2949

\bibitem[{{Sobeck} {et~al.}(2007){Sobeck}, {Lawler}, \& {Sneden}}]{Sobeck07}
{Sobeck}, J.~S., {Lawler}, J.~E., \& {Sneden}, C. 2007, \apj, 667, 1267

\bibitem[{{Sobeck} {et~al.}(2011){Sobeck}, {Kraft}, {Sneden}, {Preston},
  {Cowan}, {Smith}, {Thompson}, {Shectman}, \& {Burley}}]{Sobeck11}
{Sobeck}, J.~S., {et~al.} 2011, ArXiv e-prints

\bibitem[{{Stothers}(2006)}]{Stothers06}
{Stothers}, R.~B. 2006, \apj, 652, 643

\bibitem[{{Stothers}(2010)}]{Stothers10}
---. 2010, \pasp, 122, 536

\bibitem[{{Swensson}(1946)}]{Swensson46}
{Swensson}, J.~W. 1946, \apj, 103, 207

\bibitem[{{Vivas} {et~al.}(2008){Vivas}, {Jaff{\'e}}, {Zinn}, {Winnick},
  {Duffau}, \& {Mateu}}]{Vivas08}
{Vivas}, A.~K., {Jaff{\'e}}, Y.~L., {Zinn}, R., {Winnick}, R., {Duffau}, S., \&
  {Mateu}, C. 2008, \aj, 136, 1645

\bibitem[{{Walker}(1994)}]{Walker94}
{Walker}, A.~R. 1994, \aj, 108, 555

\bibitem[{{Wallerstein} \& {Huang}(2010)}]{Wallerstein10}
{Wallerstein}, G., \& {Huang}, W. 2010, \memsai, 81, 952

\bibitem[{{Woosley} \& {Weaver}(1995)}]{WW95}
{Woosley}, S.~E., \& {Weaver}, T.~A. 1995, \apjs, 101, 181

\end{thebibliography}

%FFFFFFFFFFFFFFFFFFFFFFFFFFFFFFFFFFFFFFFFFFFFFFFFFFFFFFFFFFFFFFFFFFFFFFFF
\clearpage

\begin{figure}
\plotone{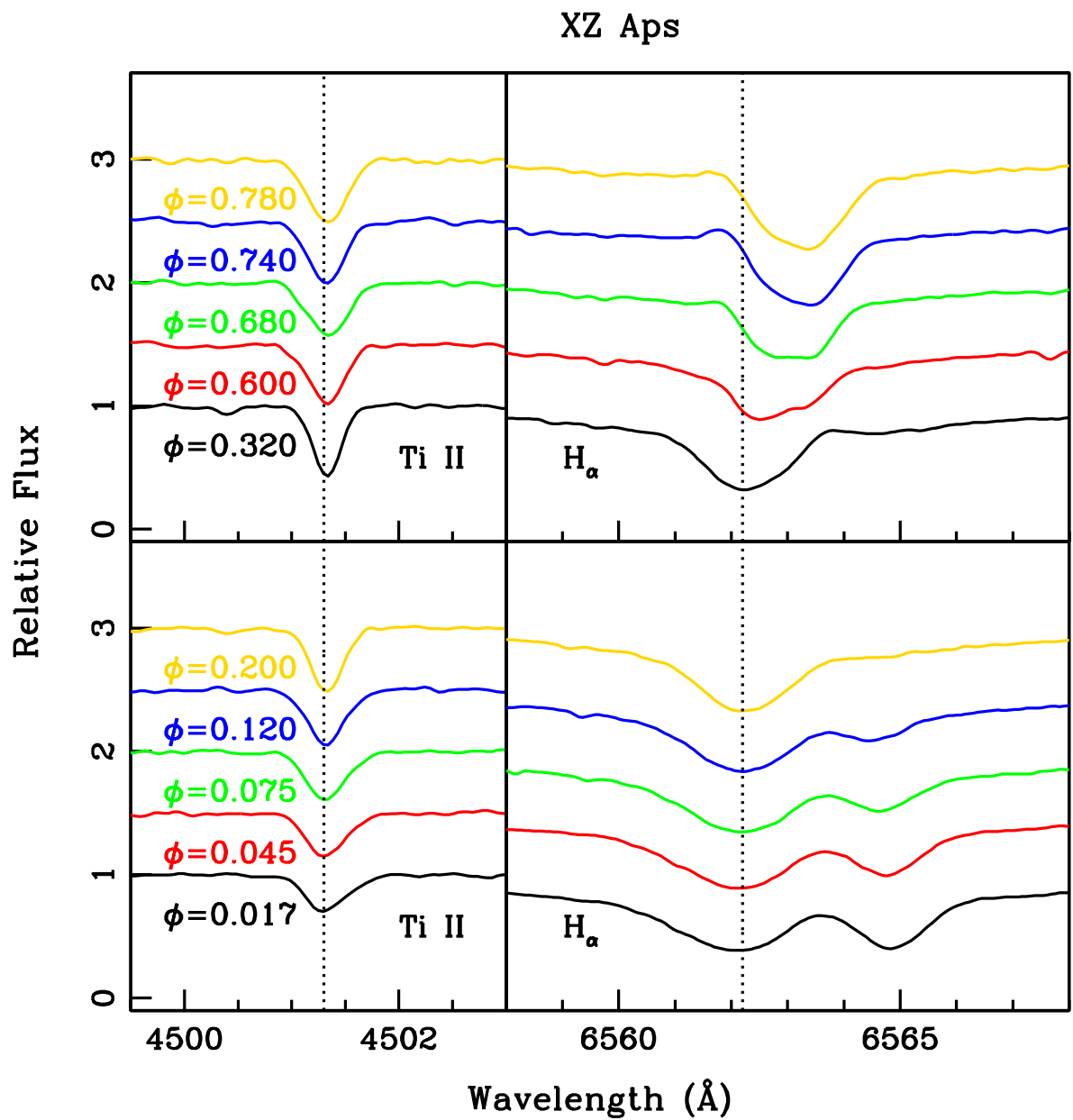}
\caption{Line profile variations of the XZ~Aps combined spectra in
         the phase range $\phi$~= 0.017--0.78 for a typical metal line, 
         \ion{Ti}{2} 4501.3~\AA\ (left-hand panels), and for
         H$\alpha$ (right-hand panels).
         The metal line appears to be sharpest near $\phi$~= 0.32. 
         However, the line profile variations are very small from 
         $\phi$~$\approx$ 0.25 to 0.55 (see 
         Figure~\ref{rvoctcomb1_postages}, and so we show only the
         $\phi$~= 0.32 spectrum here.) \label{xzapscomb1_postages}}
\end{figure}

\clearpage
\begin{figure}
\plotone{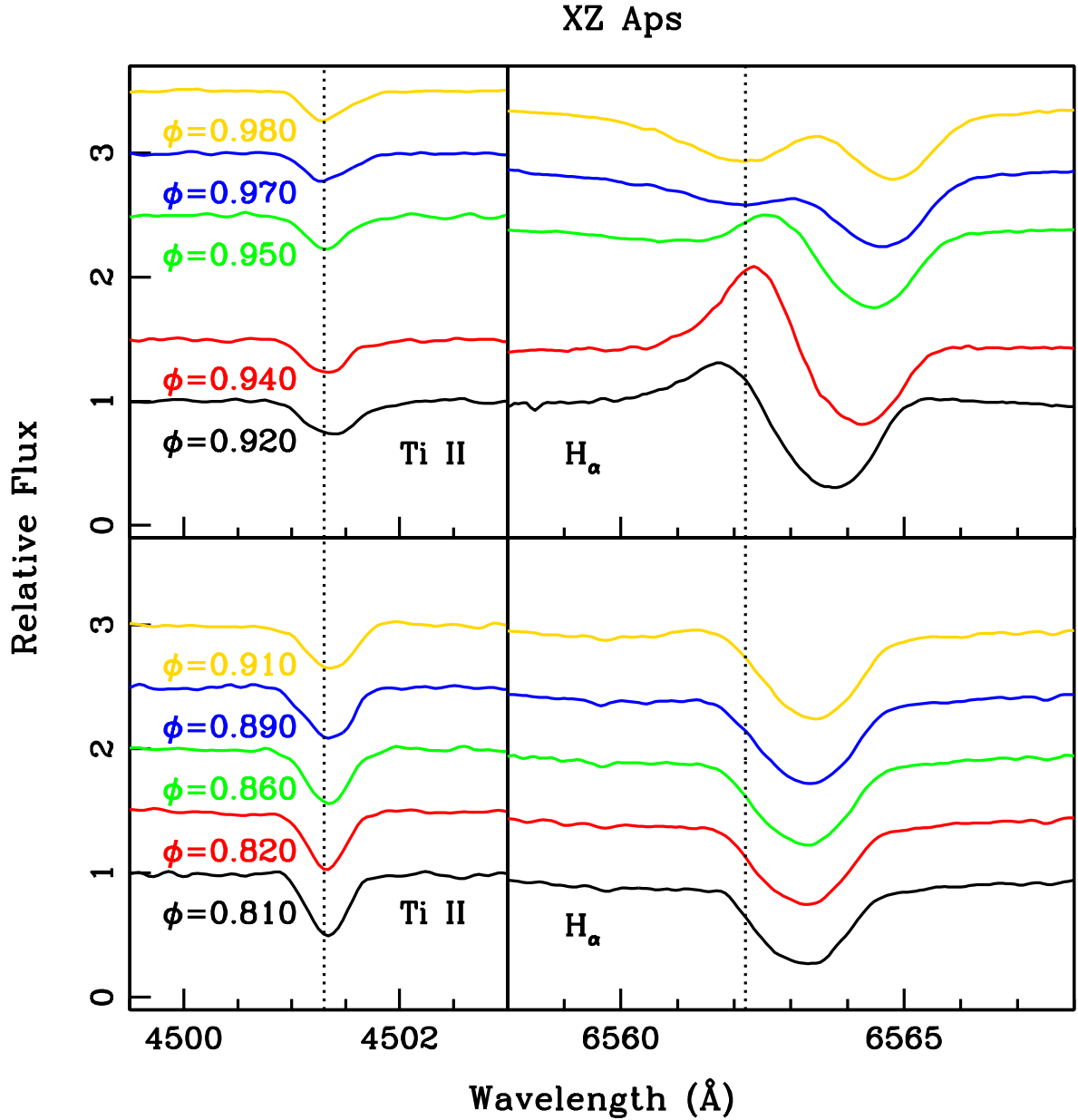}
\caption{Continuation of the Figure~\ref{xzapscomb1_postages} XZ~Aps
         line profile variations for $\phi$~= 0.81--0.98, the 
         rising-light phases of rapid variability in RR~Lyr.
         The H$\alpha$ emission occurs at its highest near 
         $\phi$~= 0.94.\label{xzapscomb2_postages}}
\end{figure}

\clearpage
\begin{figure}
\plotone{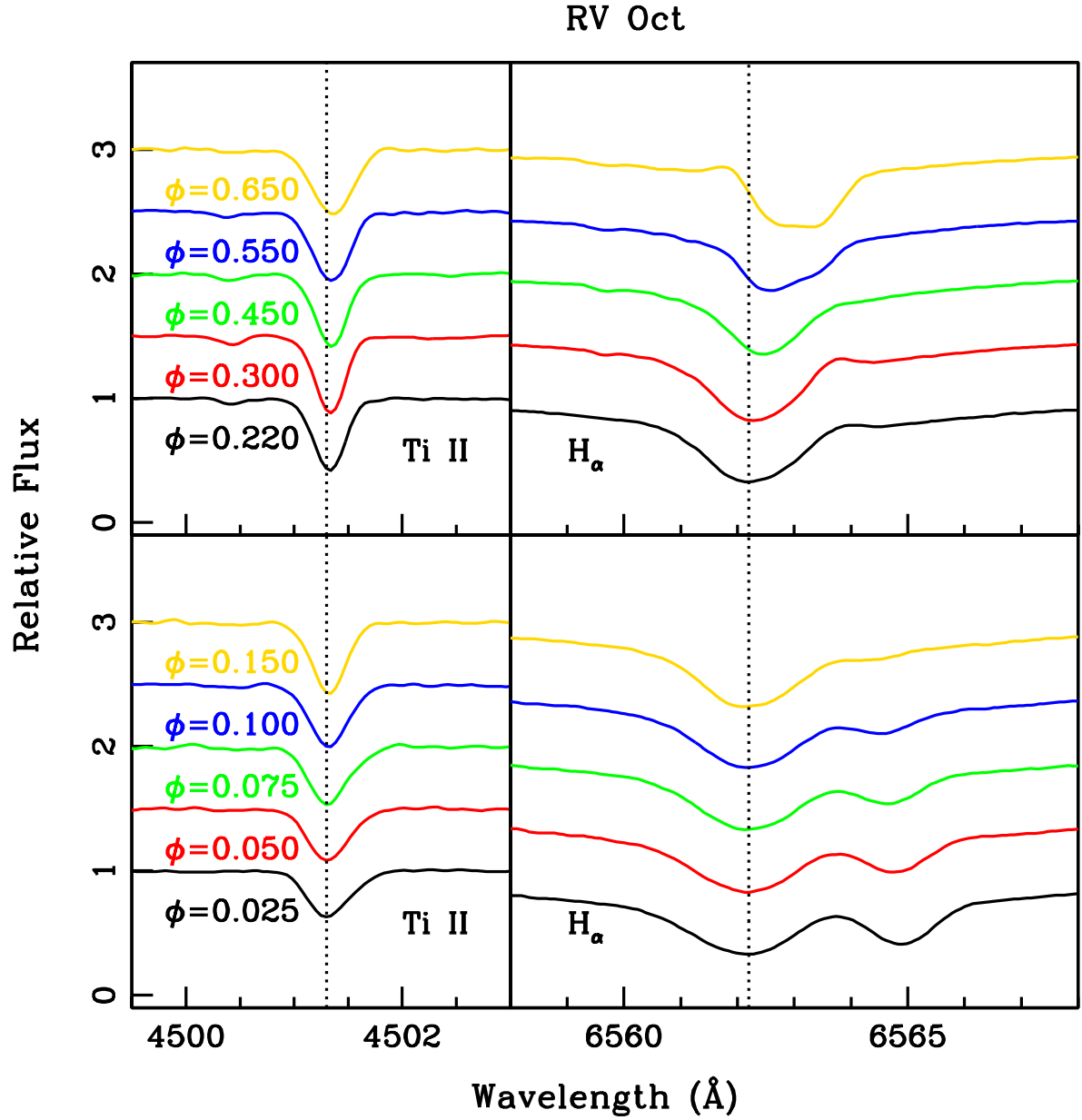}
\caption{Repeat of Figure~\ref{xzapscomb1_postages} for RV Oct, but
         showing many more combined spectra between $\phi$~= 0.2 and 0.6
         where the metal lines remain reasonably sharp with the least 
         asymmetric profile distortion.\label{rvoctcomb1_postages}}
\end{figure}

\clearpage
\begin{figure}
\plotone{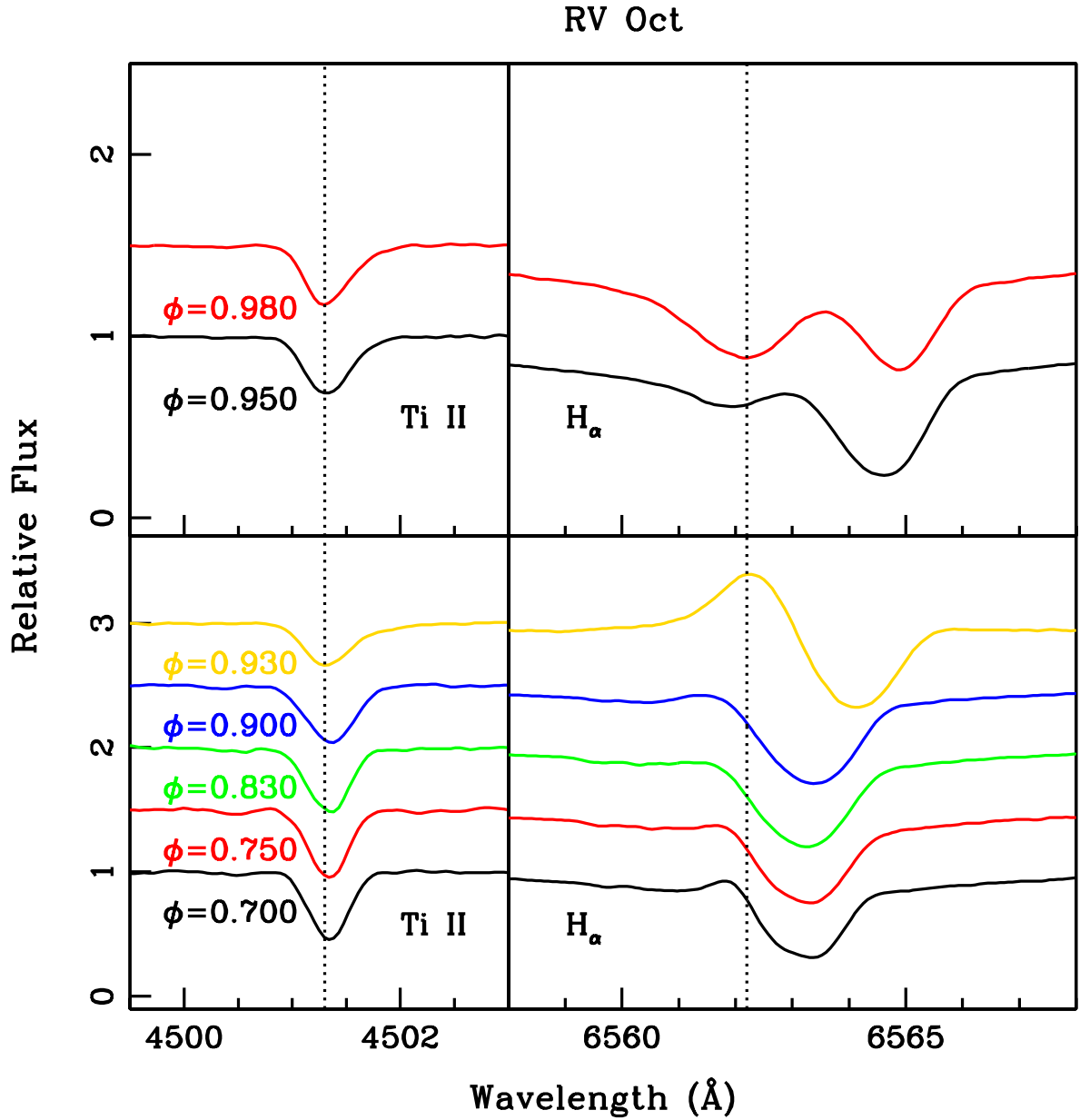}
\caption{Continuation of Figure~\ref{rvoctcomb1_postages} for 
         the RV Oct line profile variations in the rapidly changing
         phase interval from $\phi=$0.7--0.98. 
         The H$\alpha$ emission occurs at its highest near 
         $\phi$~= 0.93.\label{rvoctcomb2_postages}}
\end{figure}

\clearpage
\begin{figure}
\plotone{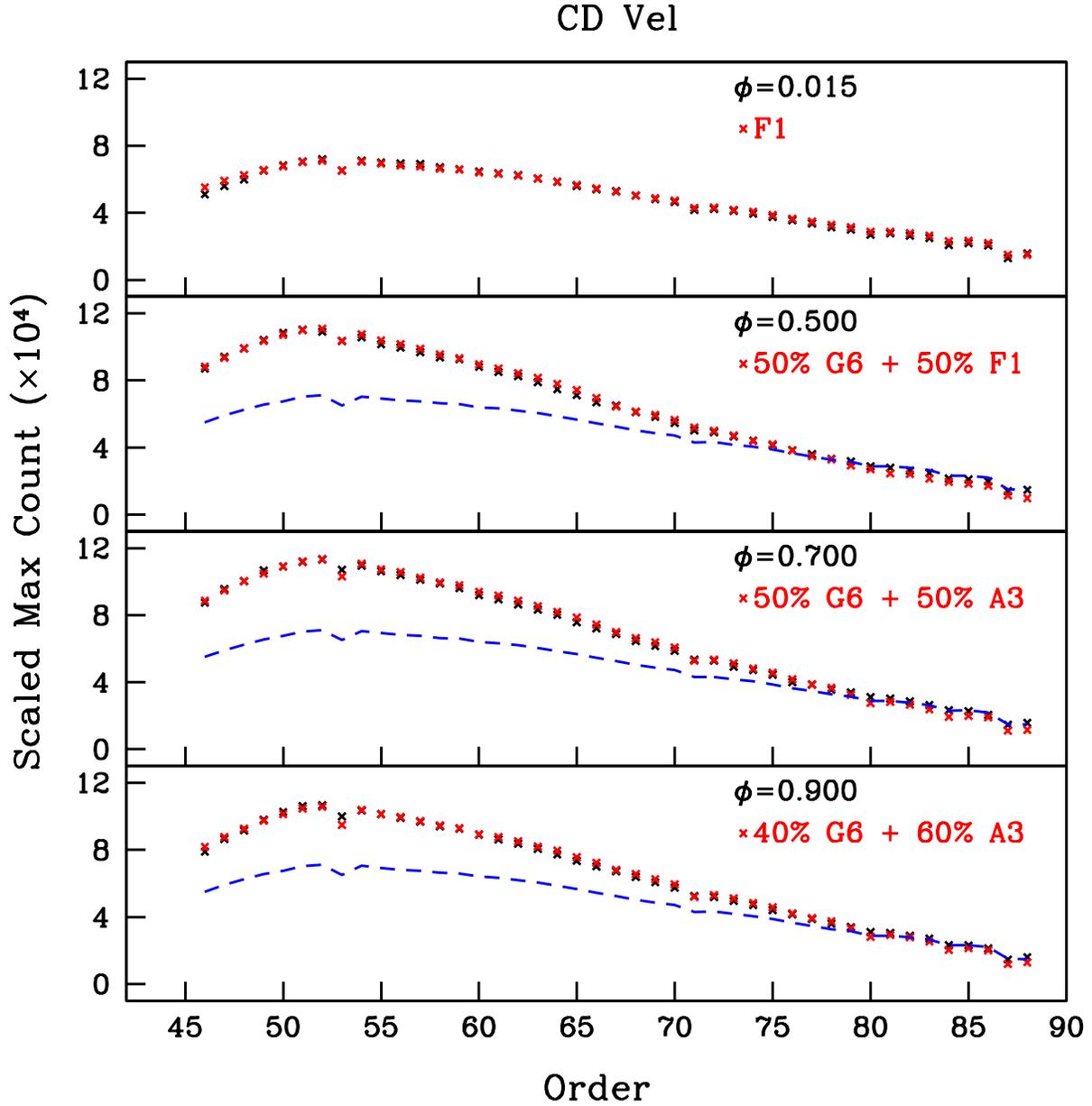}
\caption{Comparisons between the spectral energy distributions (SEDs) of 
         standard stars and/or family of their spectral combination (red 
         crosses),
         and the combined spectra of CD Vel in different phases (black 
         crosses). 
         The counts in each order were arbitrarily scaled for comparisons. 
         These comparisons were used to decide the amount of scattered 
         light correction for each order. 
         The blue dashed lines in the bottom three panels are the
         same and the points in the top panel, thus representing the 
         SED of the pure F1 spectral type for comparison with the
         ``mixed'' spectral types.\label{seds}}
\end{figure}

\clearpage
\begin{figure}
\begin{center}
\includegraphics[scale=0.65,angle=-90]{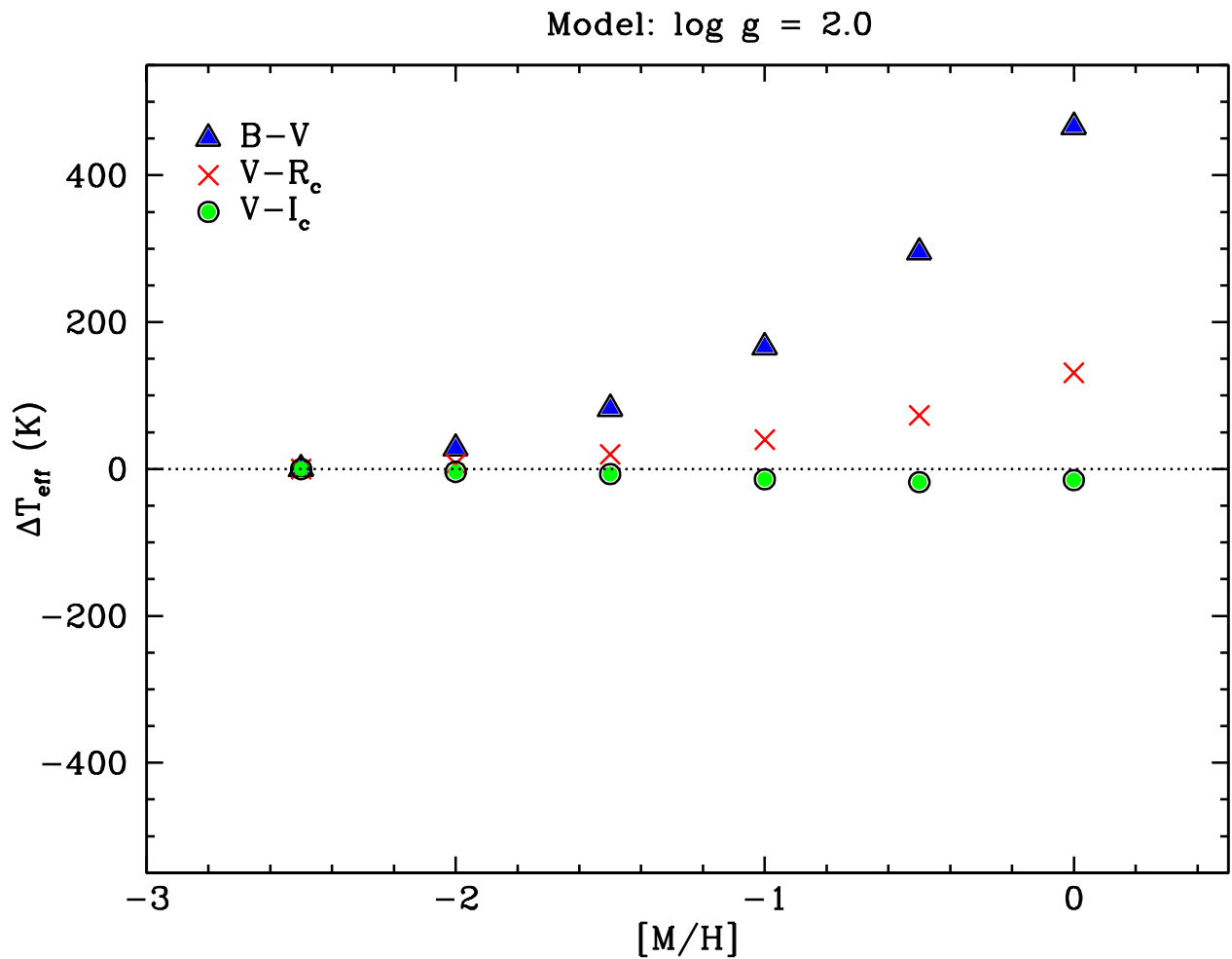}
\caption{The deviation of effective temperature calculated from 
         different synthetic color indices as a function of metallicity. 
         The color indices are computed at phase 0.3 of RR Cet 
         with a single gravity, \logg~= 2.0. 
         The temperature difference was taken between the calculated 
         \teff\ at that particular [M/H] and the \teff\ at [M/H]~=~$-$2.5. 
         Symbols represent \teff\ values derived from these color 
         indices: $B-V$ (triangles); $V-R_{\rm c}$ (crosses); and
         $V-I_{\rm c}$ (circles). \label{dev}}
\end{center}
\end{figure}

\clearpage
\begin{figure}
\begin{center}
\includegraphics[scale=0.65,angle=-90]{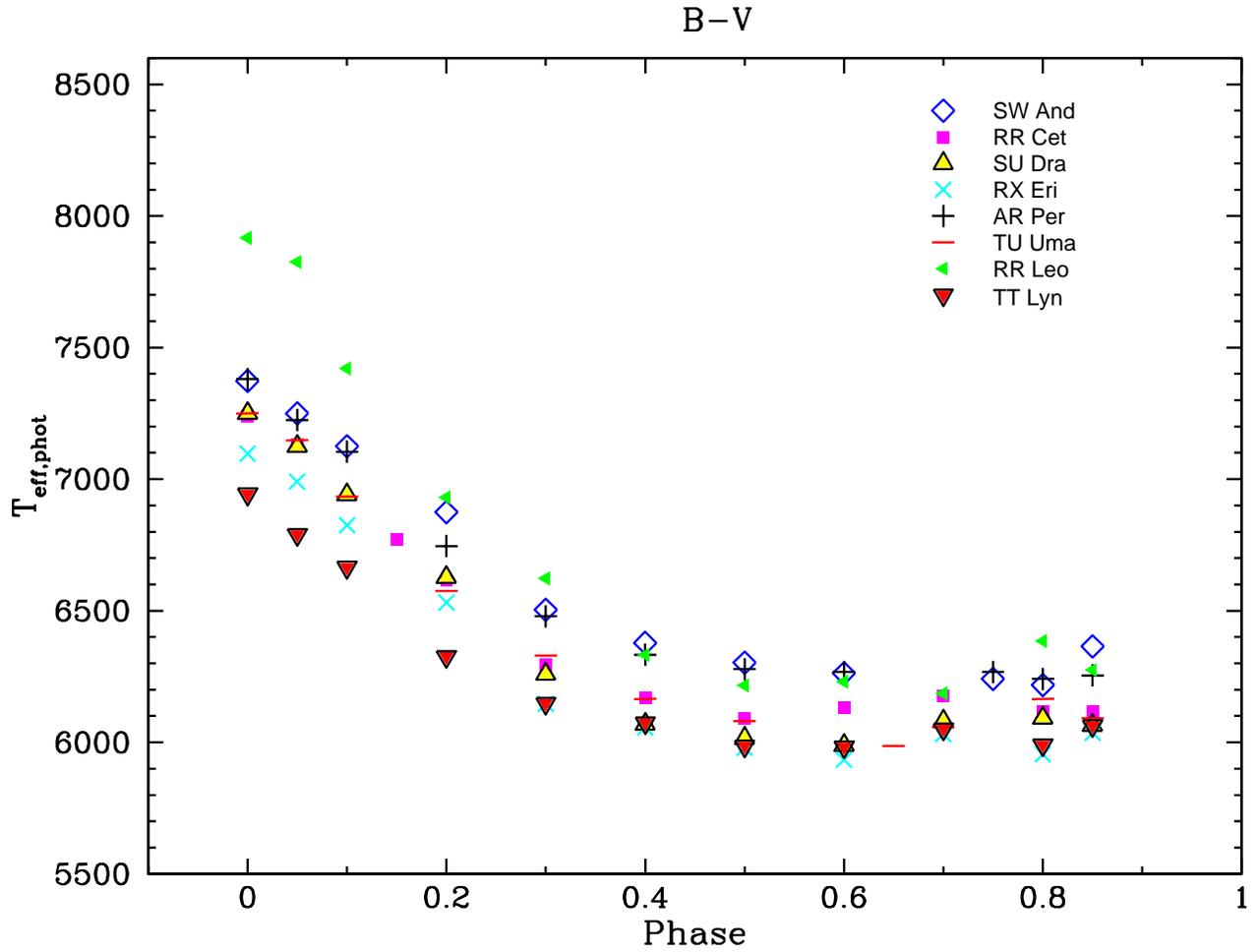}
\caption{The effective temperatures transformed from $B-V$ color 
         indices as a function of phase. 
         The different symbols represent the 8 RRab variables 
         (SW And, RR Cet, SU Dra, RX Eri, AR Per, TU Uma, 
         RR Leo and TT Lyn) selected from LJ89 and LJ90. 
         They are used as our ``calibration stars''. \label{bmv_teffphi}}
\end{center}
\end{figure}

\clearpage
\begin{figure}
\begin{center}
\includegraphics[scale=0.65,angle=-90]{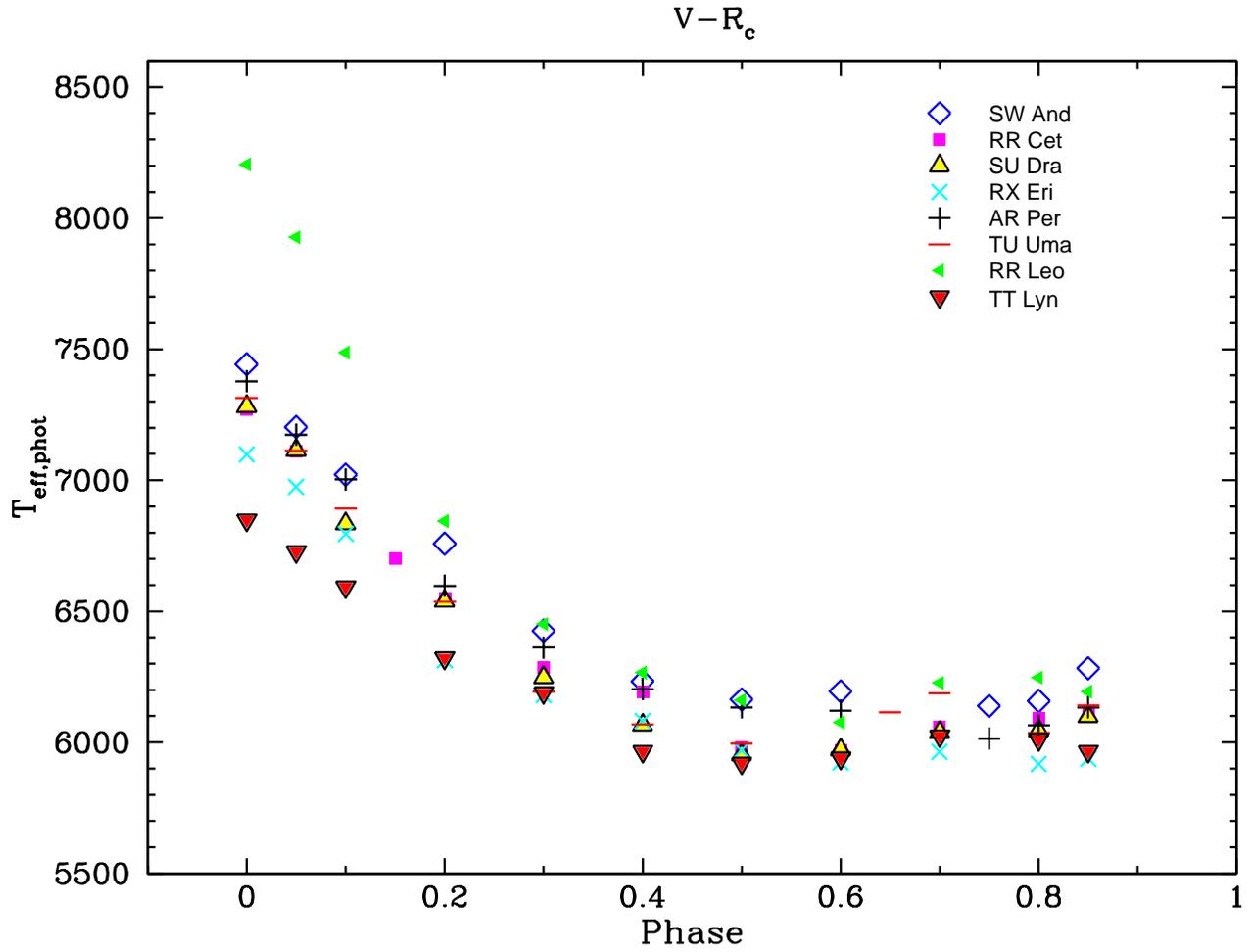}
\caption{The effective temperatures transformed from $V-R_{\rm c}$ 
         color indices as a function of phase.
         The different symbols represent the same RRab variables as 
         shown in Figure~\ref{bmv_teffphi}.\label{vmr_teffphi}}
\end{center}
\end{figure}

\clearpage
\begin{figure}
\begin{center}
\includegraphics[scale=0.65,angle=-90]{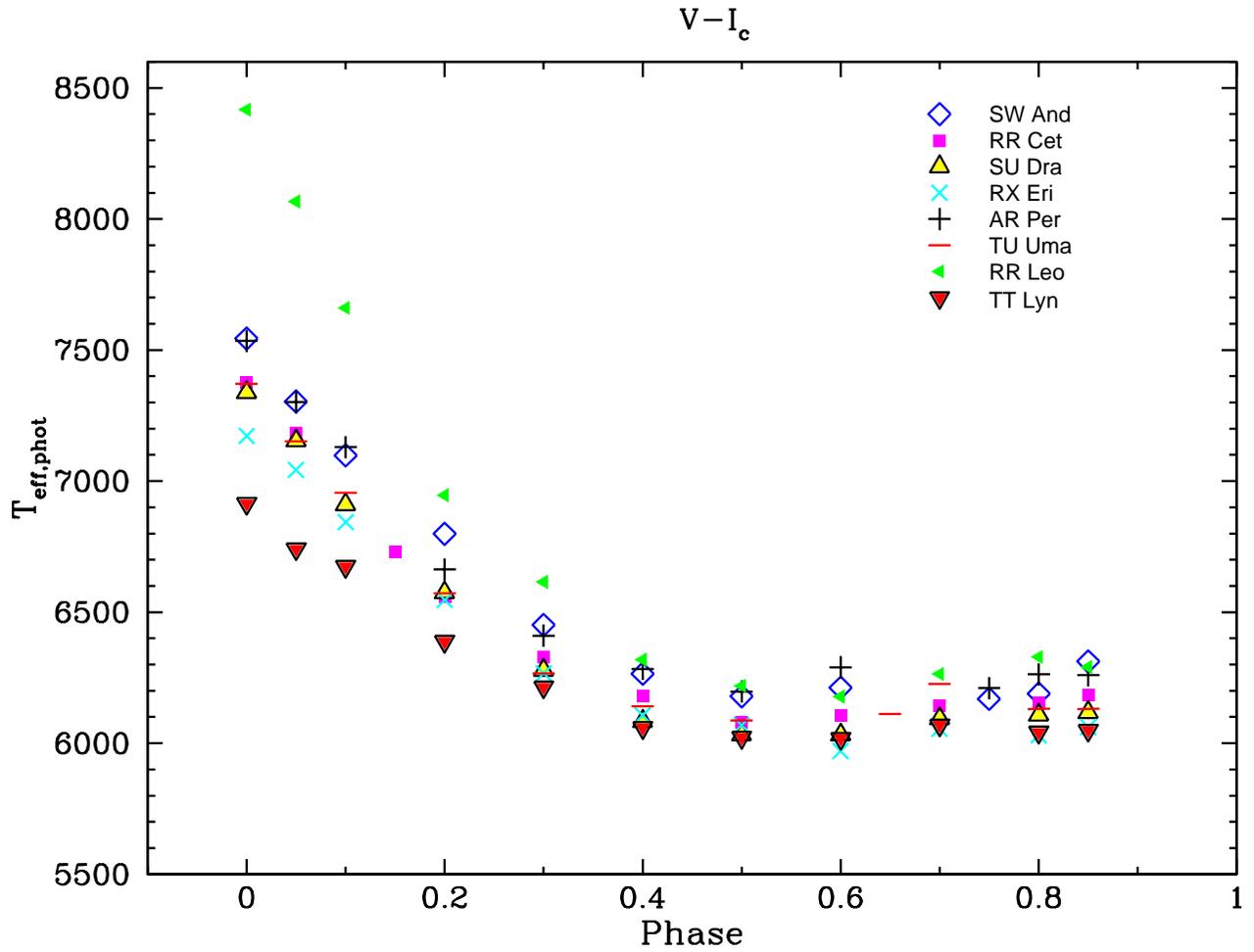}
\caption{The effective temperatures transformed from $V-R_{\rm c}$ 
         color indices as a function of phase.
         The different symbols represent the same RRab variables as 
         shown in Figure~\ref{bmv_teffphi}.  
         Individual $V-I_{\rm c}$ vs phase relations are used to fit 
         4$^{\rm th}$-order polynomial curves, which are treated as our 
         ``calibration curves''. \label{vmi_teffphi}}
\end{center}
\end{figure}

\clearpage
\begin{figure}
\plotone{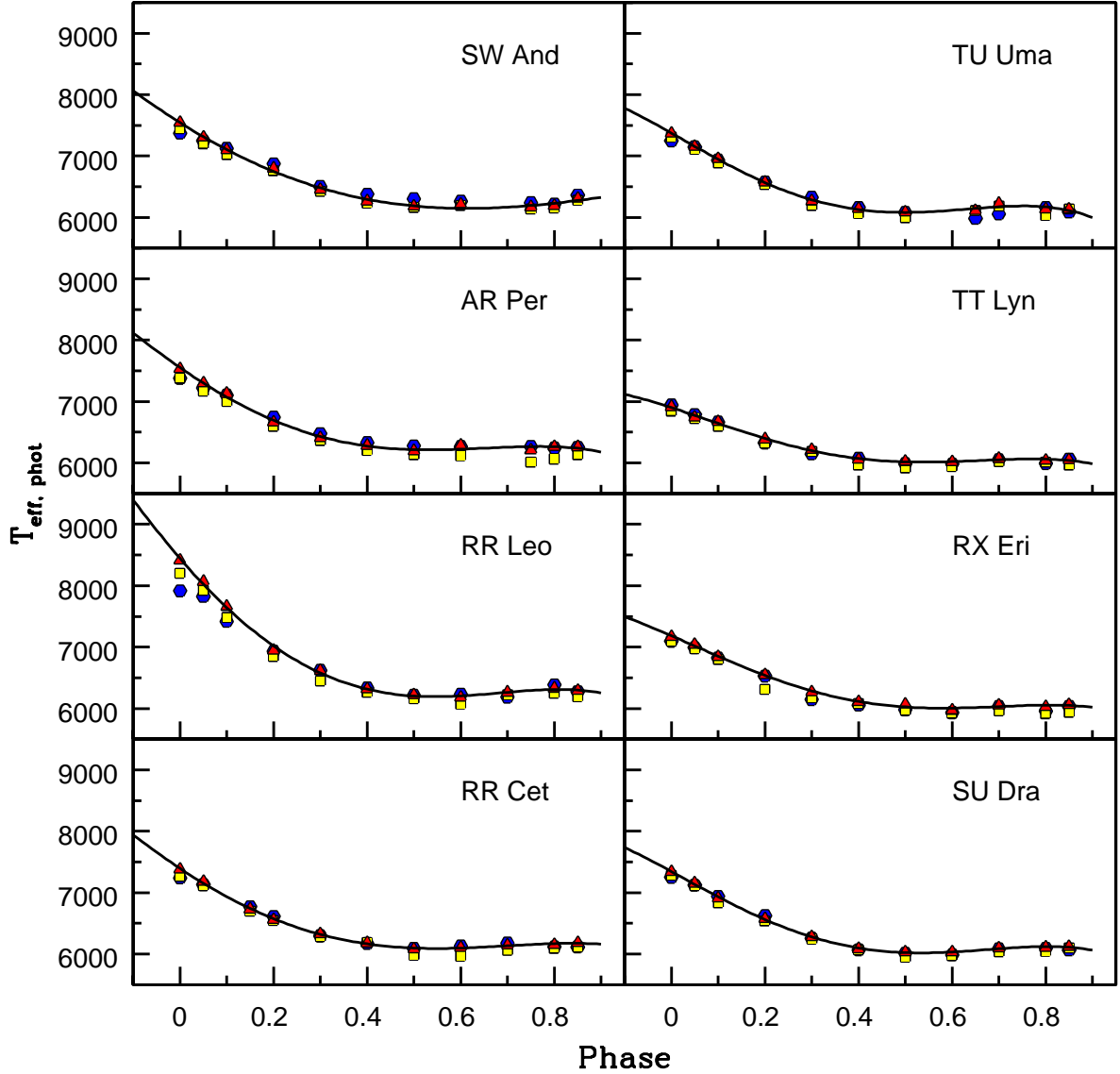}
\caption{The transformed \teff\ from different color indices as a 
         function of phase for the selected 8 RRab variables from 
         LJ89 and LJ90. 
         The solid lines are fitted 4$^{\rm th}$-order polynomials 
         to the $V-I_{\rm c}$ curves. 
         Symbols refer to \teff\ values derived from the color indices: 
         $B-V$ (blue hexagons); $V-R_{\rm c}$ (yellow squares) and 
         $V-I_{\rm c}$ (red triangles).\label{all_teffphi}}
\end{figure}

\clearpage
\begin{figure}
\begin{center}
\includegraphics[scale=0.5,angle=-90]{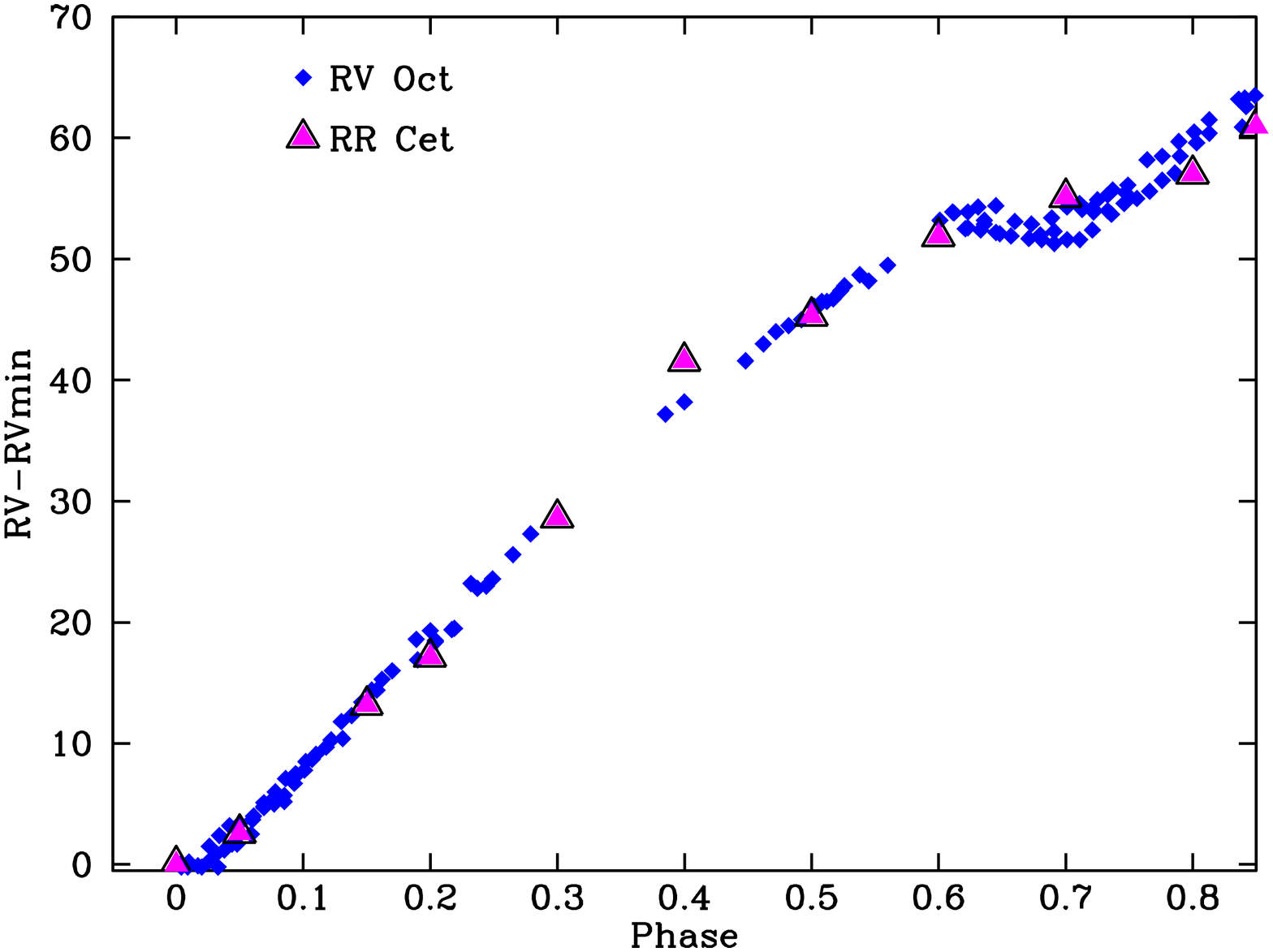}
\includegraphics[scale=0.5,angle=-90]{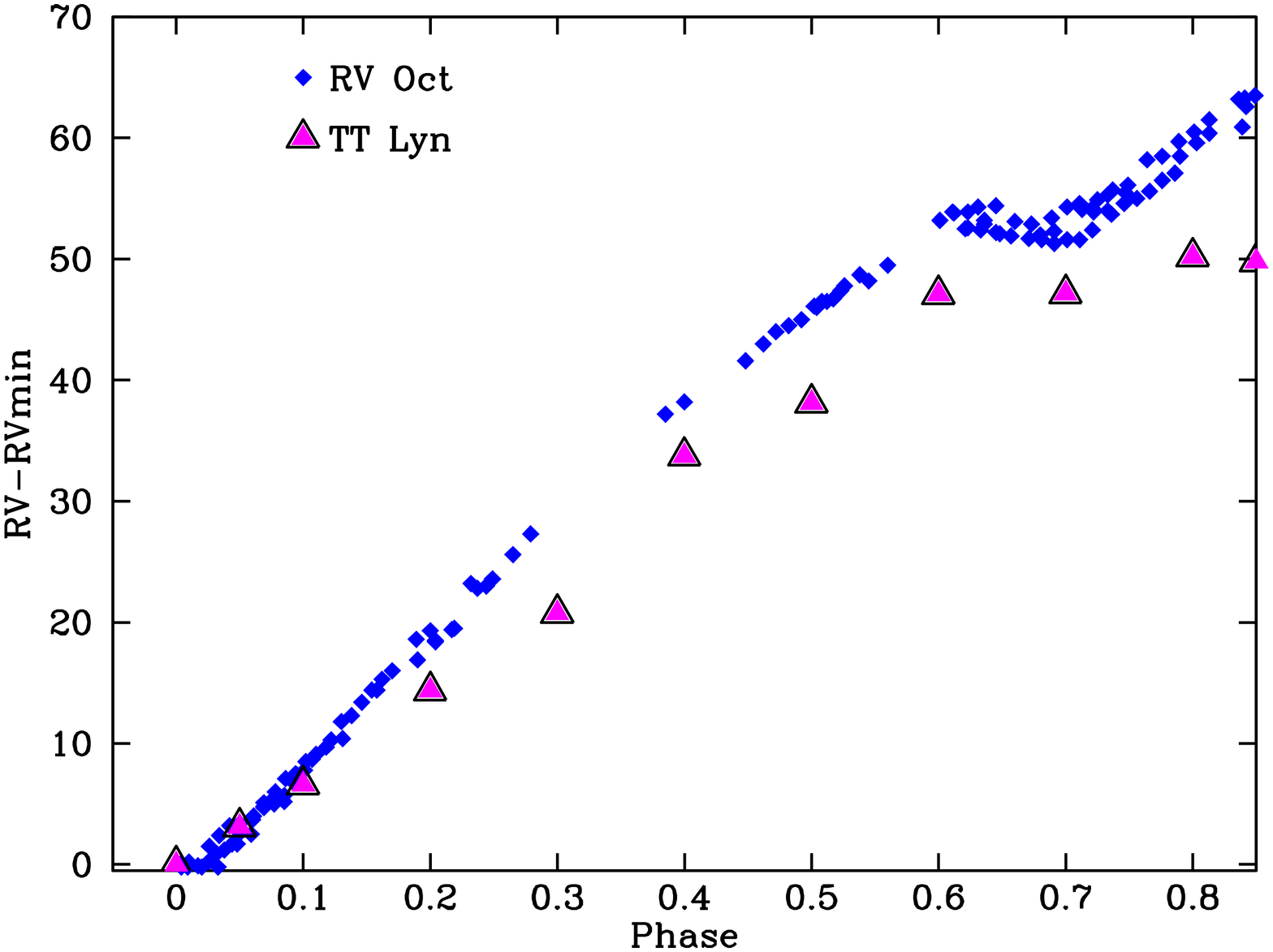}
\caption{Demonstration of selecting the best calibration curves by 
         comparing the RV$-$RVmin curve of our RV Oct to RV$-$RVmin 
         curves of RR Cet (top panel) and TT Lyn (bottom panel). 
         The top panel shows the best match pulsational behavior. 
         Symbols refer to RV Oct (blue diamonds) and RR Cet $\&$ 
         TT Lyn (magenta triangles). \label{rvmrvmin_ph}}
\end{center}
\end{figure}

\clearpage
\begin{figure}
\begin{center}
\includegraphics[scale=0.6,angle=-90]{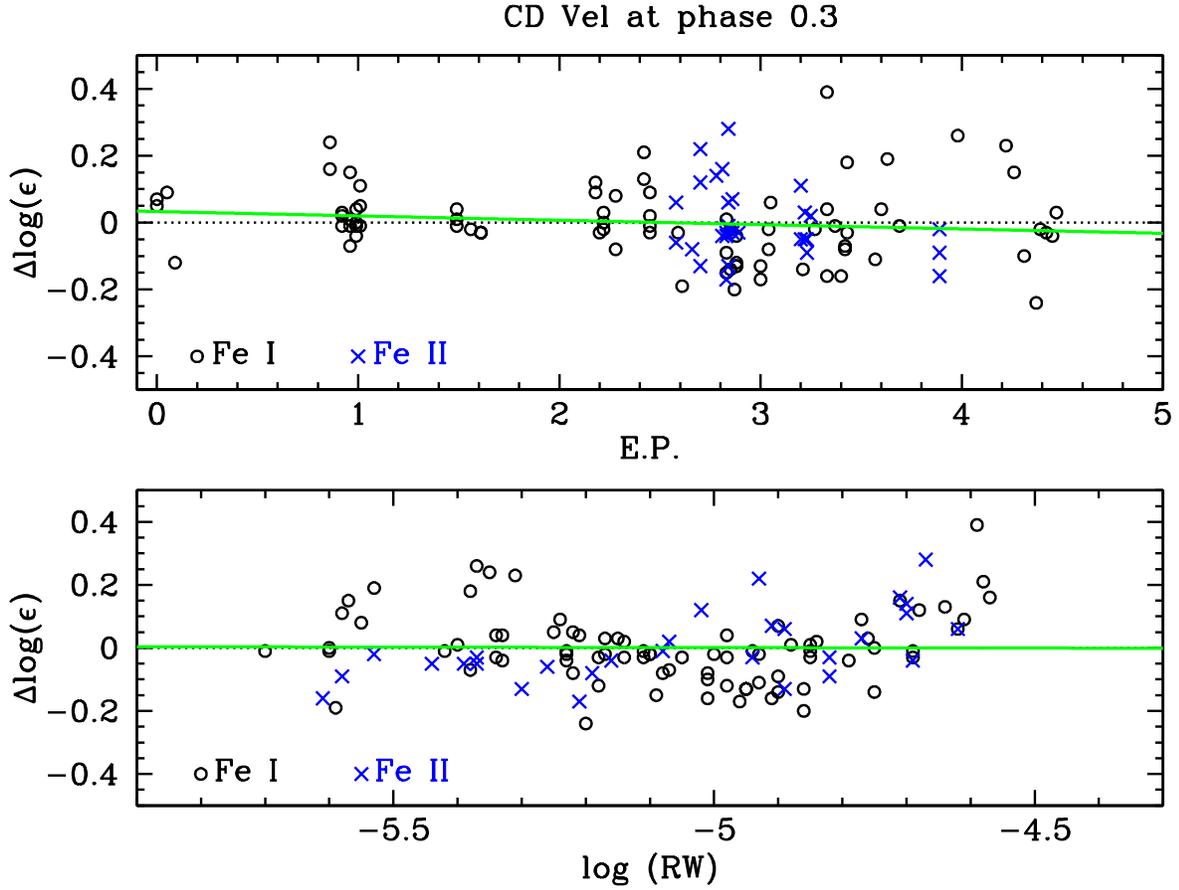}
\caption{Differences of individual \ion{Fe}{1} and \ion{Fe}{2}
         line abundances as functions of EP (top panel) and log~RW
         with the final spectroscopically-constrained model atmosphere
         parameters of CD~Vel at $\phi$~= 0.3.
         The black open circles and blue crosses represent 
         \ion{Fe}{1} and \ion{Fe}{2}, respectively, as indicated in
         the panel legends. 
         The green solid lines show the (negligible) trends of these
         abundances with EP and log~RW for \ion{Fe}{1} 
         lines.\label{cdvel0300_Feout}}
\end{center}
\end{figure}

\clearpage
\begin{figure}
\plotone{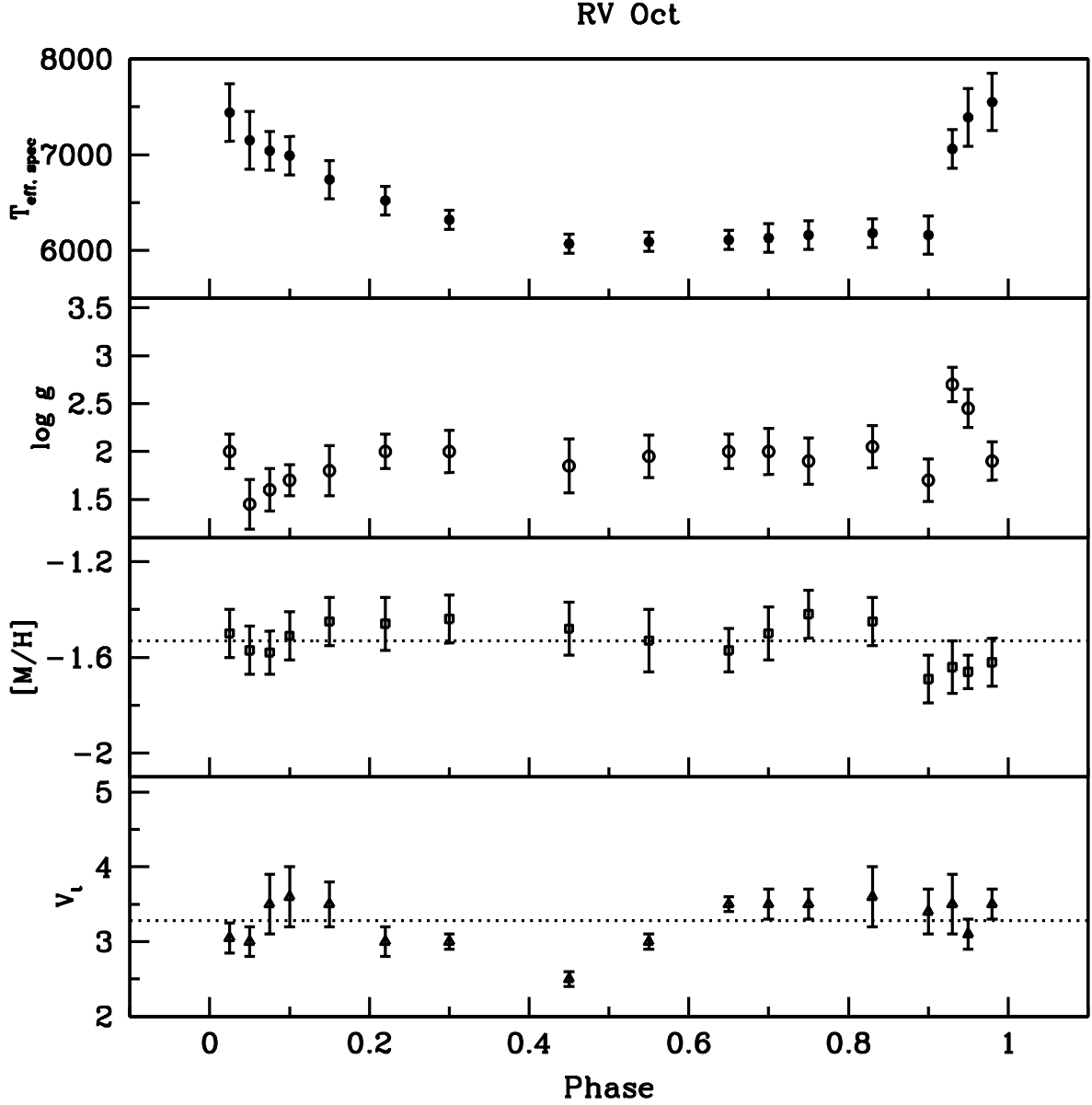}
\caption{Derived stellar parameters (\teff, \logg, [M/H], and \vturb) 
         for RV~Oct based on spectroscopic constraints as a 
         function of phase. 
         The dashed lines in the bottom two panels represent the mean 
         values of [M/H] and \vturb. \label{par_phase_rvoct}}
\end{figure}

\clearpage
\begin{figure}
\plotone{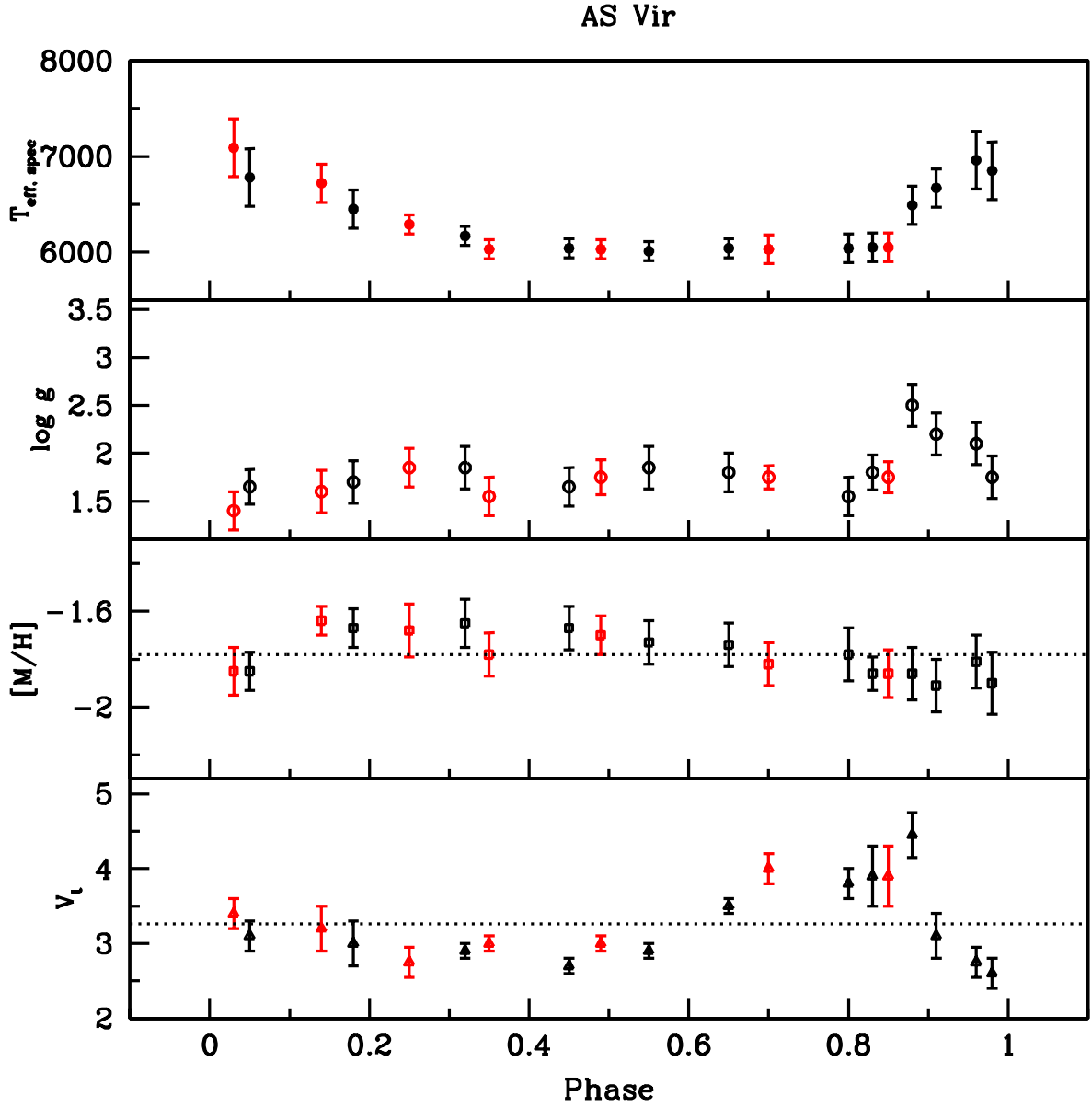}
\caption{Derived stellar parameters for AS~Vir.
         This figure is similar to Figure~\ref{par_phase_rvoct}, 
         except different color symbols represent different cycles 
         being considered for combining the spectra of this
         Blazhko variable. \label{par_phase_asvir}}
\end{figure}

\clearpage
\begin{figure}
\begin{center}
\includegraphics[scale=0.50]{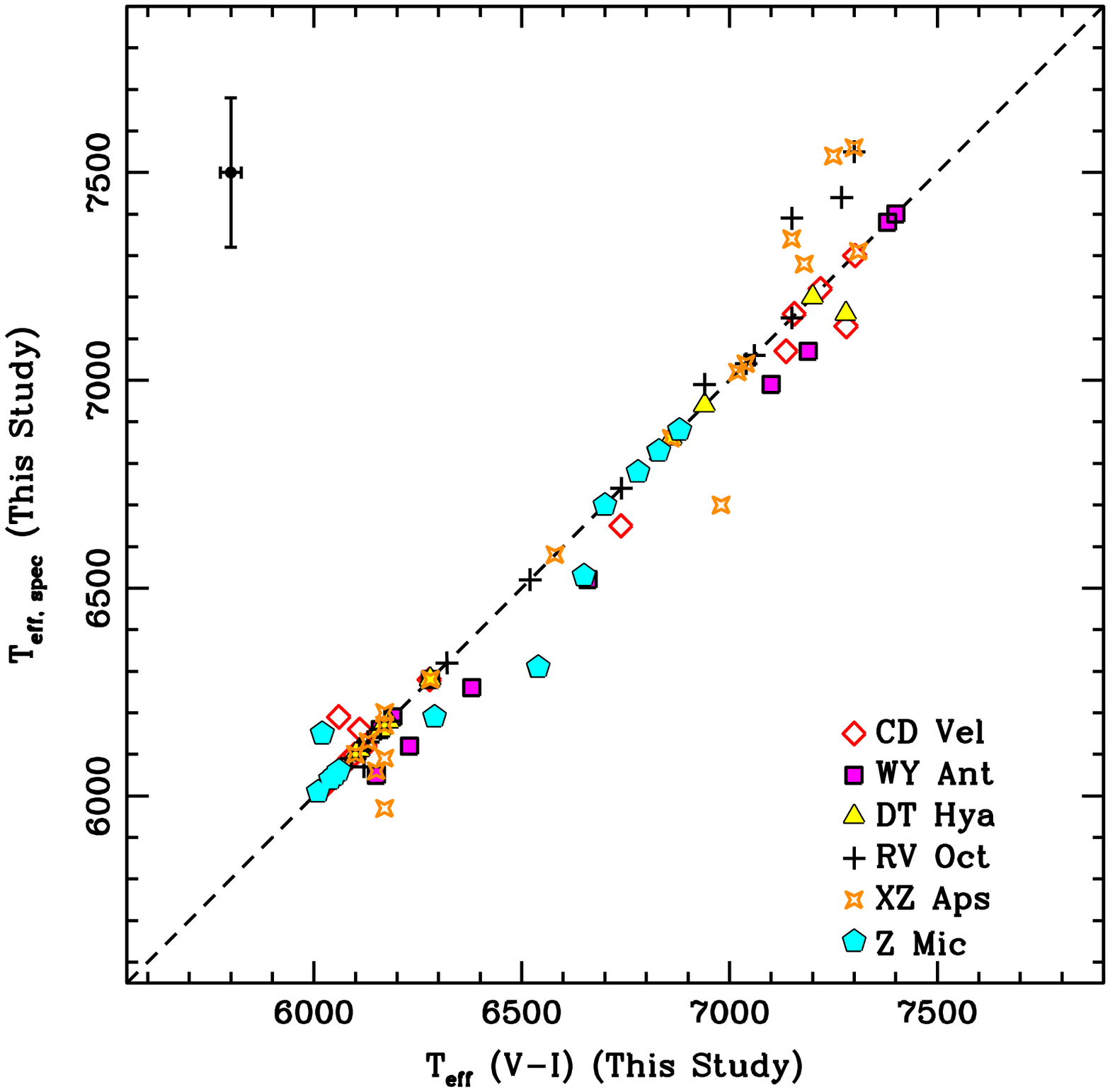}
\includegraphics[scale=0.50]{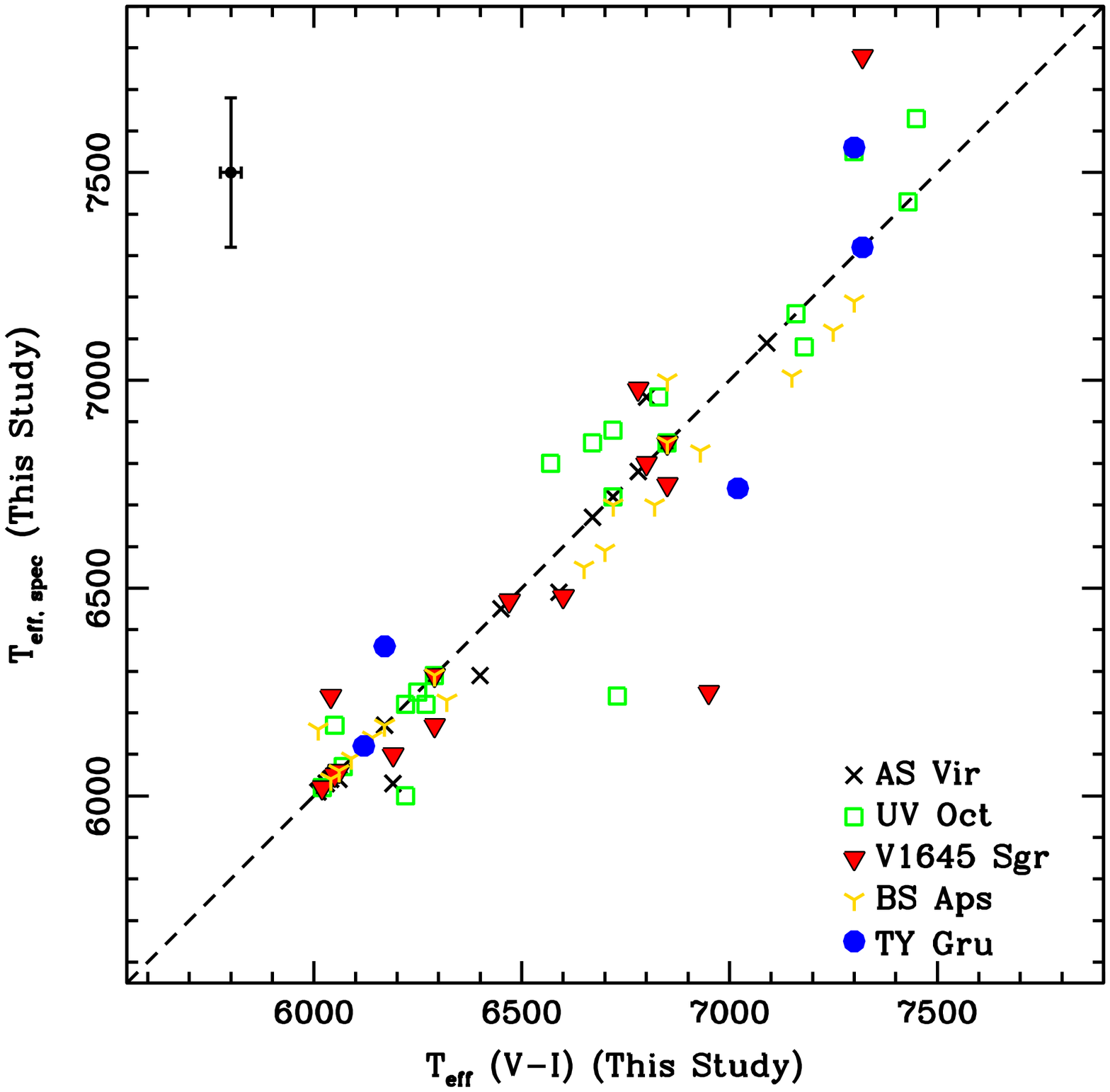}
\caption{Comparison of derived spectroscopic \teff\ with photometric \teff. 
         The top and bottom panels contain non-Blazhko and Blazhko stars, 
         respectively. 
         Symbols representing individual stars are given in the legends. 
         For the clarity in the figure, we do not plot the error bar for 
         each value, but instead indicate typical uncertainties for 
         \teff$_{,\rm spec}$ and \teff$_{,(V-I)}$.  \label{comp_teff}}
\end{center}
\end{figure}

\clearpage
\begin{figure}
\plotone{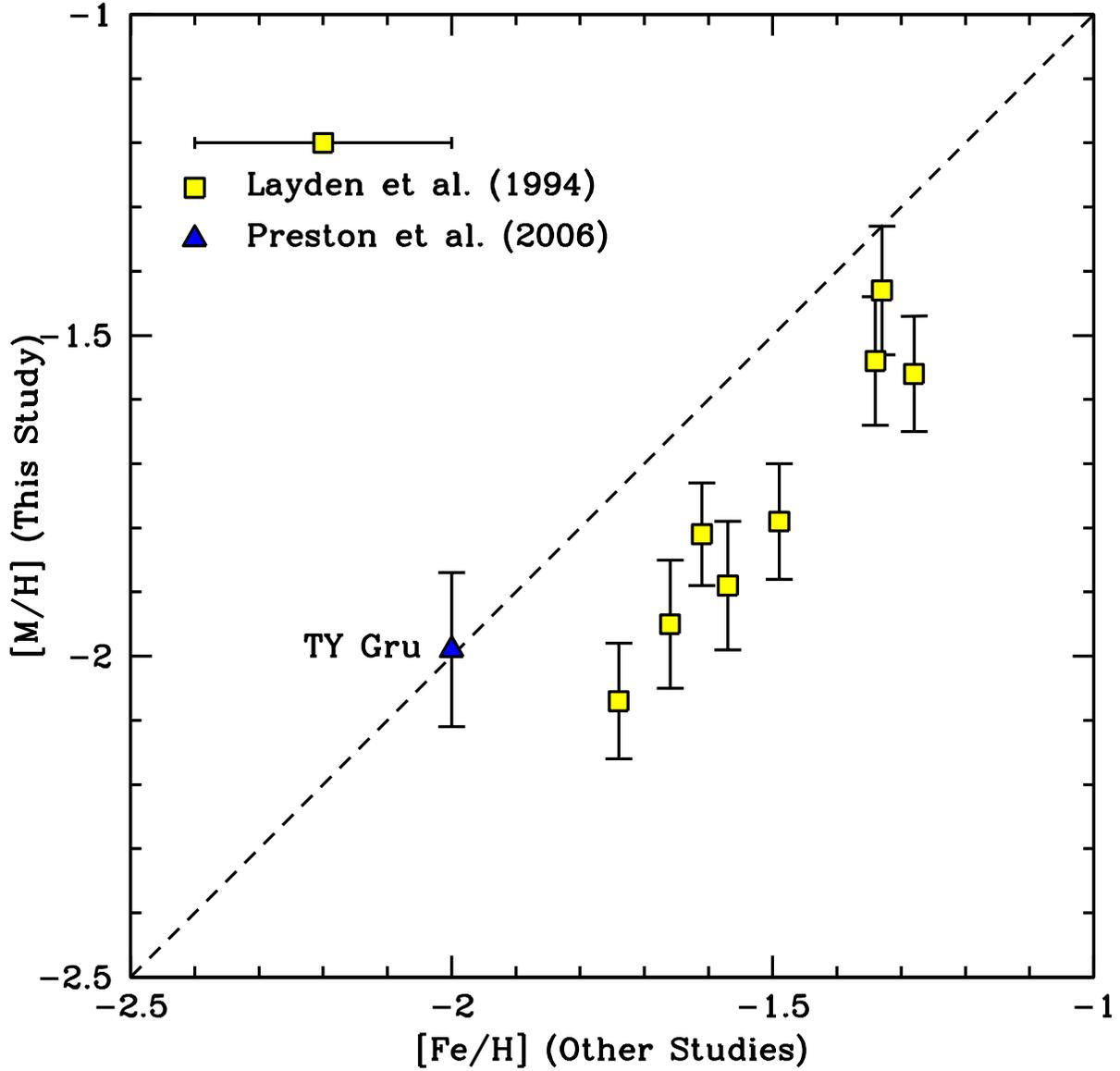}
\caption{Comparison of derived [M/H] with [Fe/H] of other studies. 
         The symbols represent the values derived from the
         from the $\Delta S$--metallicity relation by \citet{Layden94} 
         (yellow squares) and from the spectroscopic method 
         by \citet{Preston06} (single blue triangle).  
         For clarity in the figure, we do not plot error bars for 
         each star, but instead indicate typical uncertainties of 
         0.2 dex for the \citet{Layden94} paper.\label{comp_metal}}
\end{figure}

\clearpage
\begin{figure}
\begin{center}
\includegraphics[scale=0.5,angle=-90]{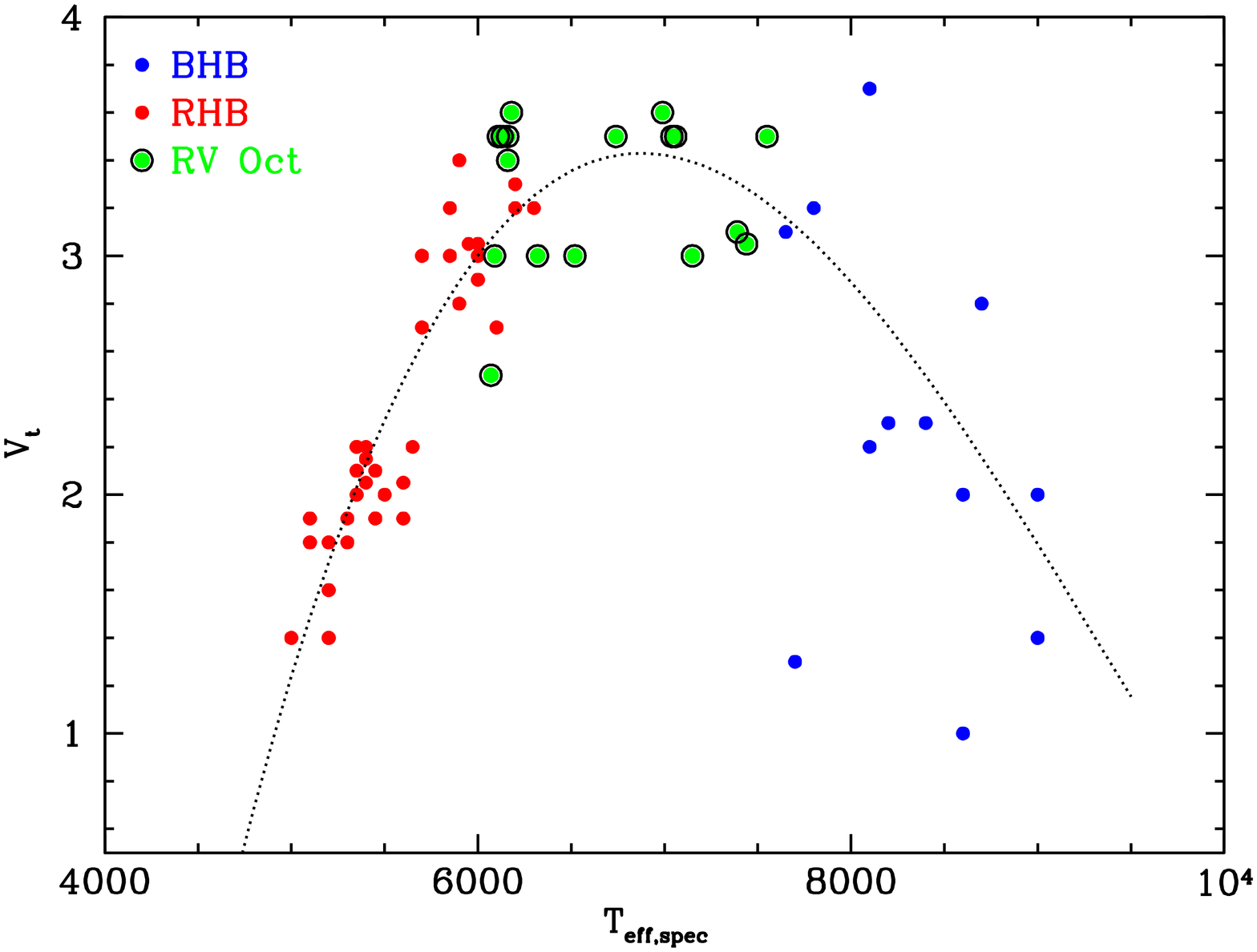}
\includegraphics[scale=0.5,angle=-90]{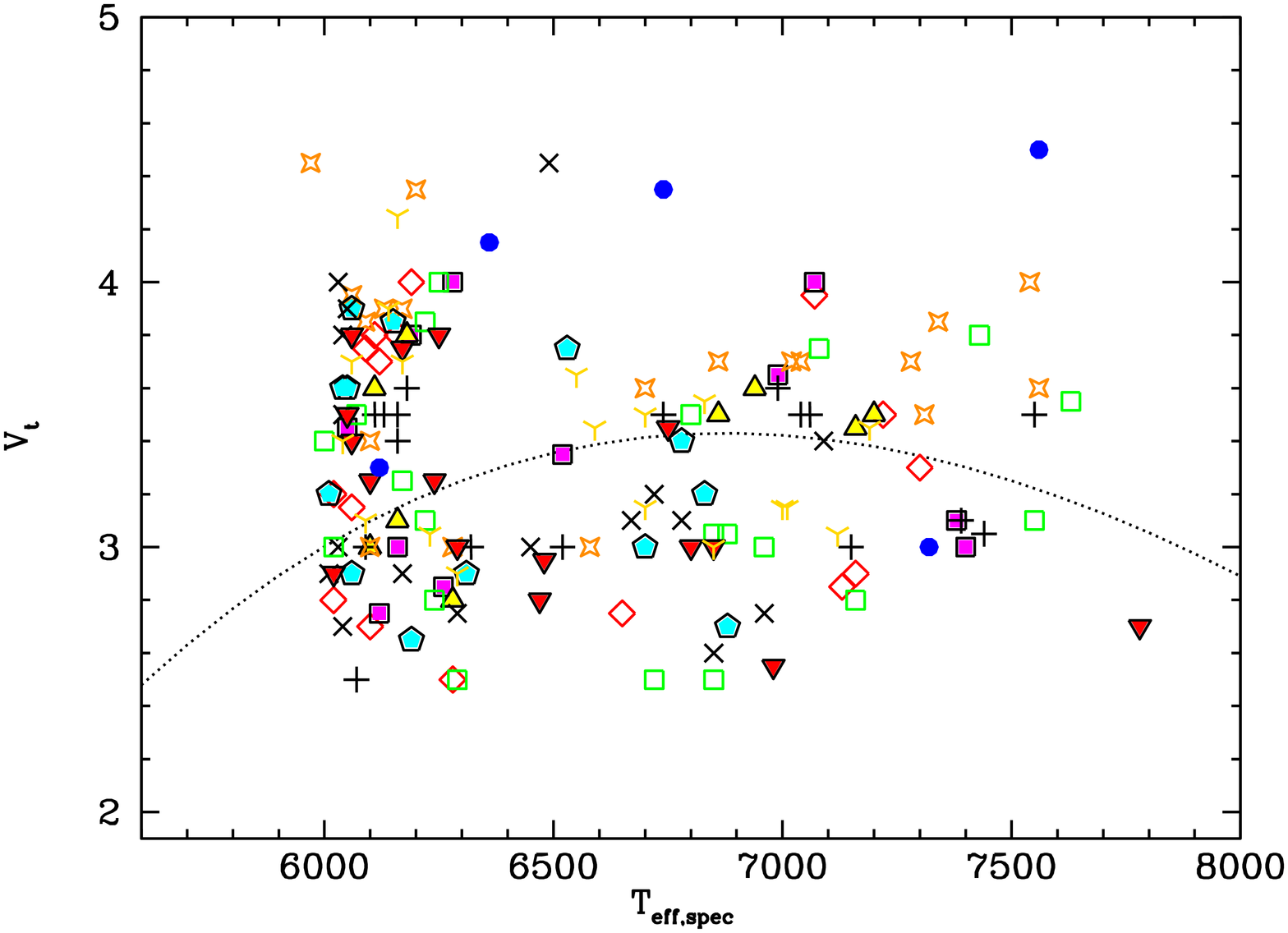}
\caption{The microturbulence as a function of \teff. 
The top panel shows \vturb\ and \teff\ of RV Oct on the \vturb-\teff\ plane, 
with additional data of RHB and BHB stars from \citet{For10}. 
The dashed curve shows a continuous \vturb--\teff\ relation across 
the HB. The bottom panel shows 
all the \vturb\ and \teff\ of all of our program stars on the 
\vturb-\teff\ plane near the instability strip region. The 
symbols represent the same stars as labeled in Figure~\ref{comp_teff}.   
\label{vt_teff}}
\end{center}
\end{figure}

\clearpage
\begin{figure}
\begin{center}
\includegraphics[scale=0.6,angle=-90]{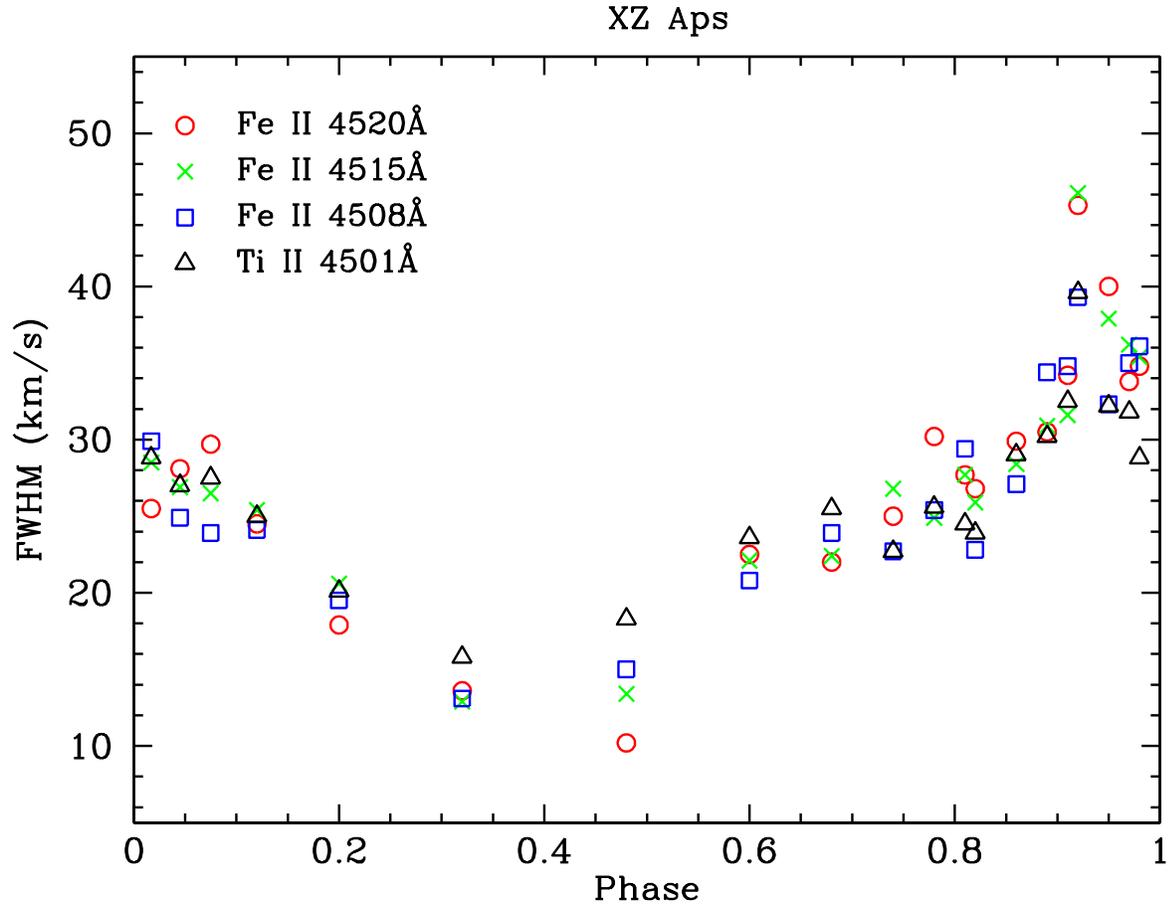}
\caption{The full width half maximum (FWHM) in km~s$^{-1}$ of four metal lines of 
         XZ~Aps throughout its pulsational cycle. The values have been corrected for 
         instrumental broadening of 11 km~s$^{-1}$. 
         The FWHM appears to be lowest near $\phi$~$\approx$ 0.3 and 
         has a peak near $\phi$~$\approx$ 0.9, probably associated 
         with the appearance of a shock wave in the photosphere.
         \label{xzaps_fwhm_phi}}
\end{center}
\end{figure}

\clearpage
\begin{figure}
\begin{center}
\includegraphics[scale=0.5,angle=-90]{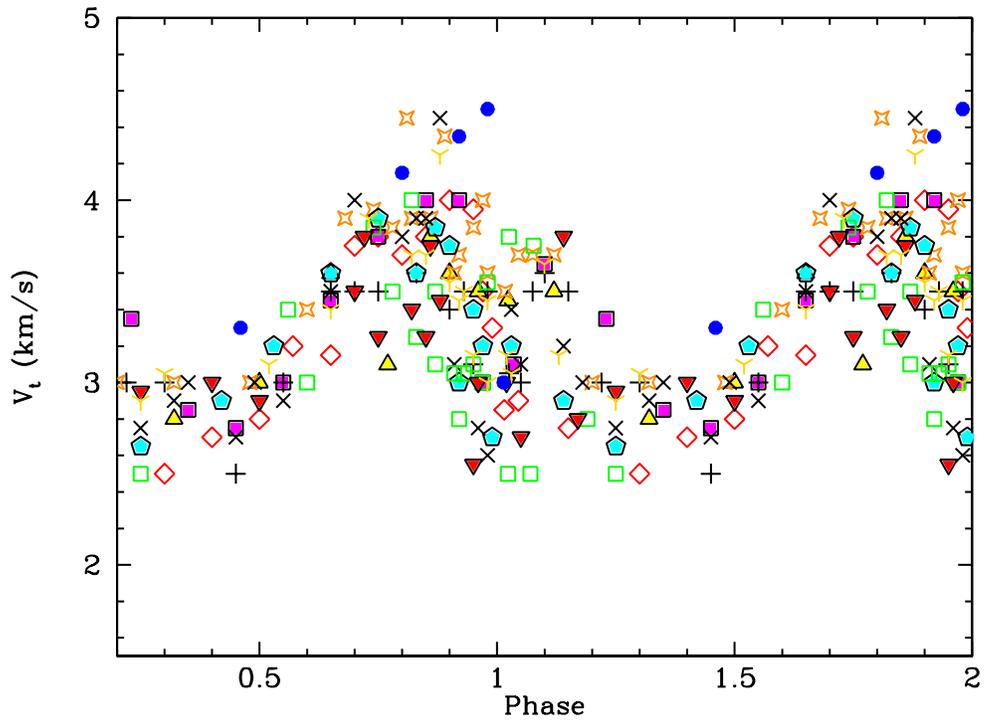}
\caption{The microturbulence as a function of phase of all of our 
         program stars. 
         The symbols represent the same stars as labeled in 
         Figure~\ref{comp_teff}. \label{vt_phase}}
\end{center}
\end{figure}

\clearpage
\begin{figure}
\plotone{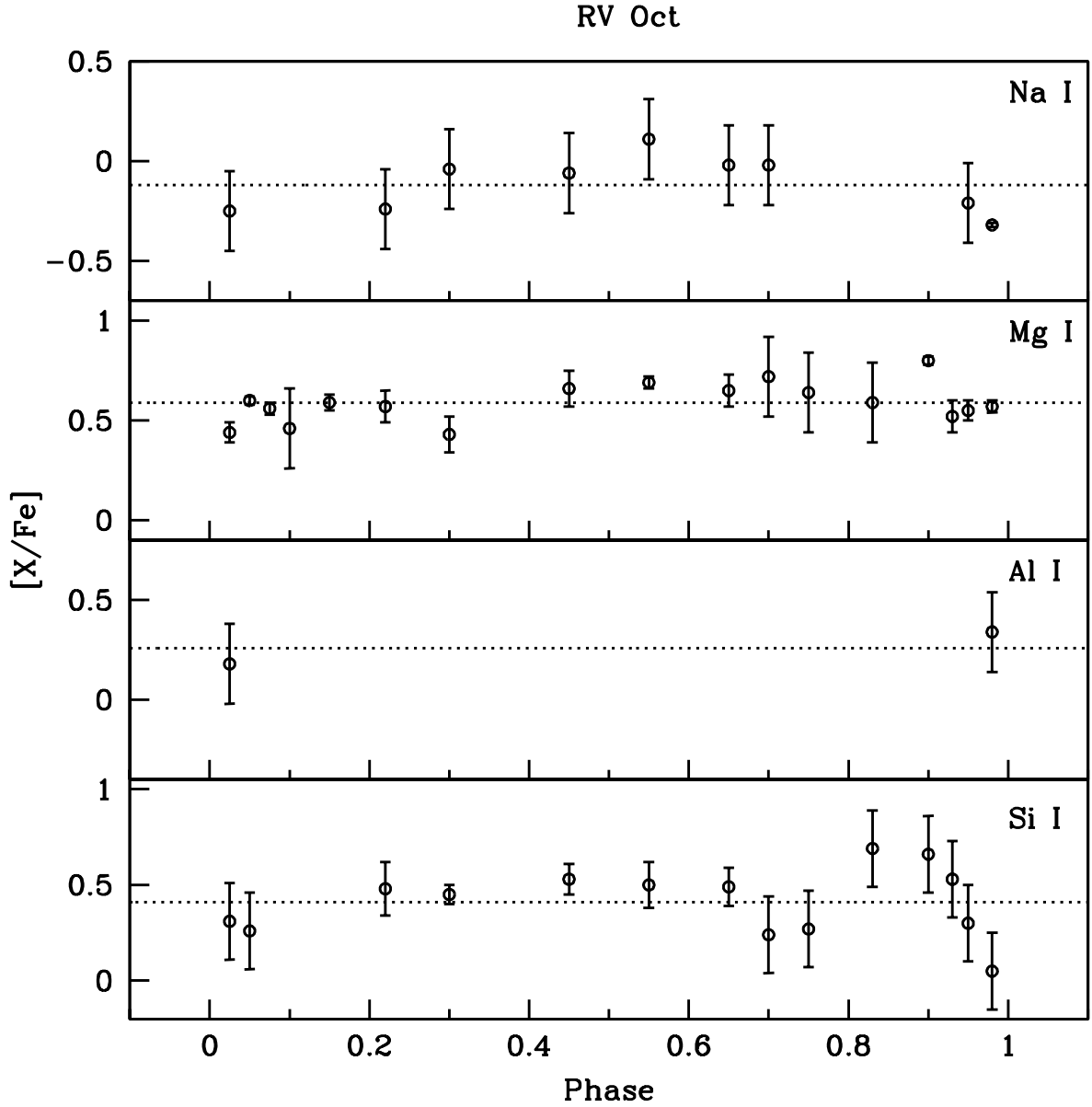}
\caption{Abundance ratios, [X/Fe], of \ion{Na}{1}, \ion{Mg}{1}, 
         \ion{Al}{1} and \ion{Si}{1} as a function of phase for the
         non-Blazhko effect star, RV~Oct.  
         The dashed lines represent the mean values. 
         The [X/Fe] values are generally consistent throughout the 
         pulsational cycle. 
         NLTE corrections were applied to Na, Al and Si abundances whenever 
         appropriate. 
         The small trend of [\ion{Si}{1}/Fe] between phase 0.8 
         and 1.0 is discussed in \S~\ref{absilicon}.
         \label{rvoct_abund1_phase}}
\end{figure}

\clearpage
\begin{figure}
\plotone{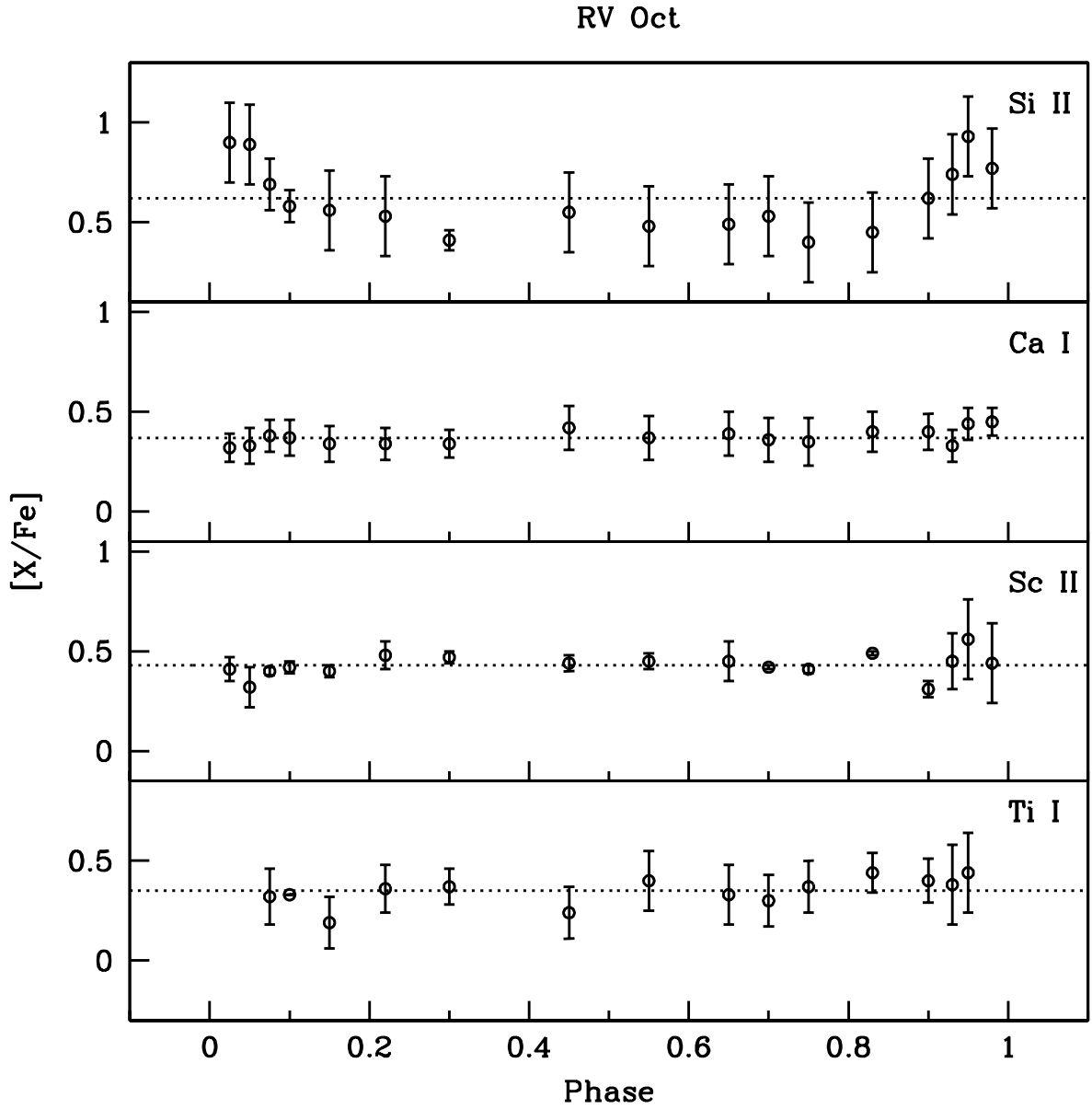}
\caption{Same as Figure~\ref{rvoct_abund1_phase}, now for 
         \ion{Si}{2}, \ion{Ca}{1}, \ion{Sc}{2} and \ion{Ti}{1}. 
         NLTE corrections applied to \ion{Si}{2} abundances whenever 
         appropriate.
         The trend of [\ion{Si}{2}/Fe] is discussed in \S~\ref{absilicon}.
         \label{rvoct_abund2_phase}}
\end{figure}

\clearpage
\begin{figure}
\plotone{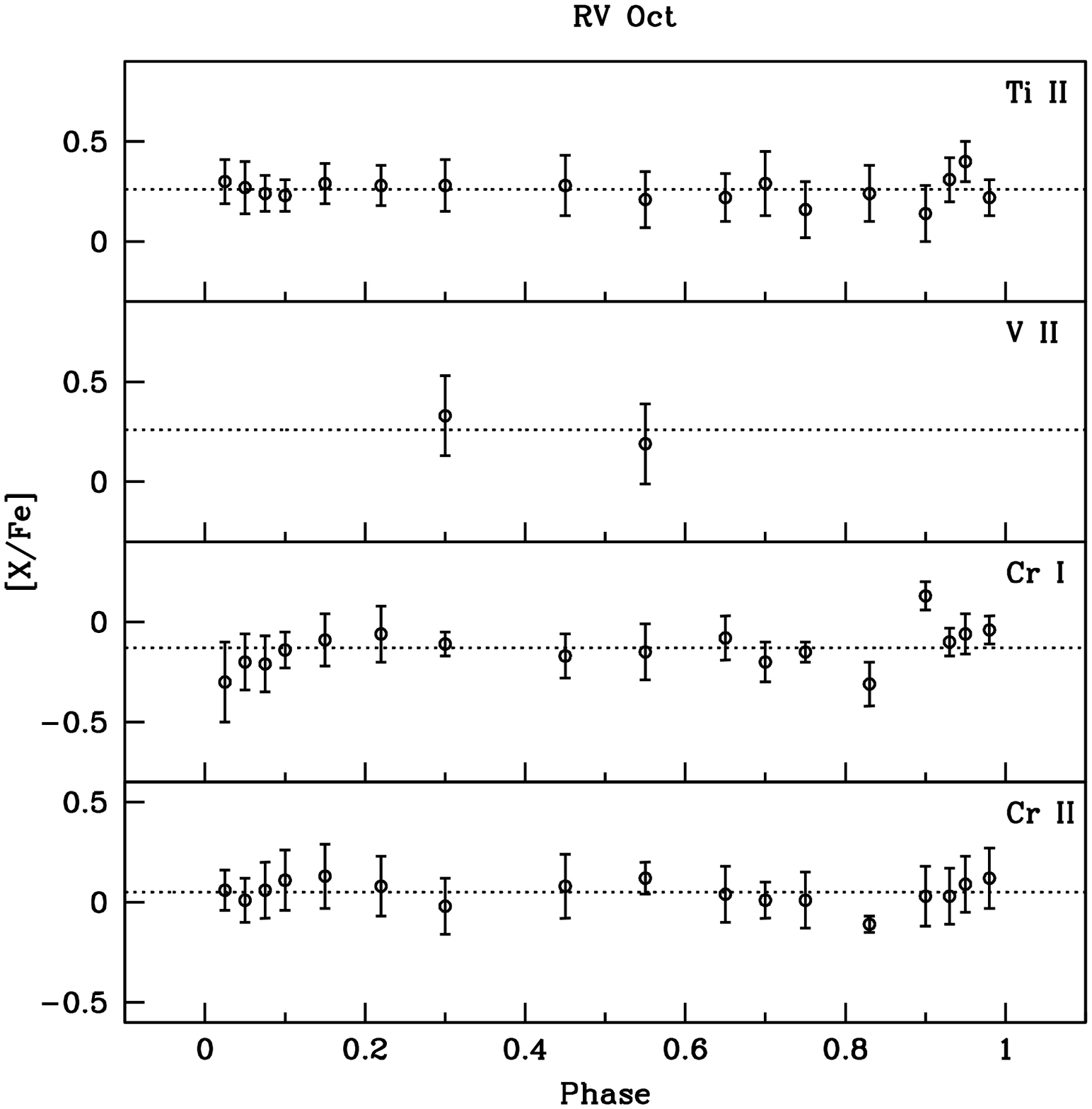}
\caption{Same as Figure~\ref{rvoct_abund1_phase}, now for \ion{Ti}{2}, 
         \ion{V}{2}, \ion{Cr}{1} and \ion{Cr}{2}. \label{rvoct_abund3_phase}}
\end{figure}

\clearpage
\begin{figure}
\plotone{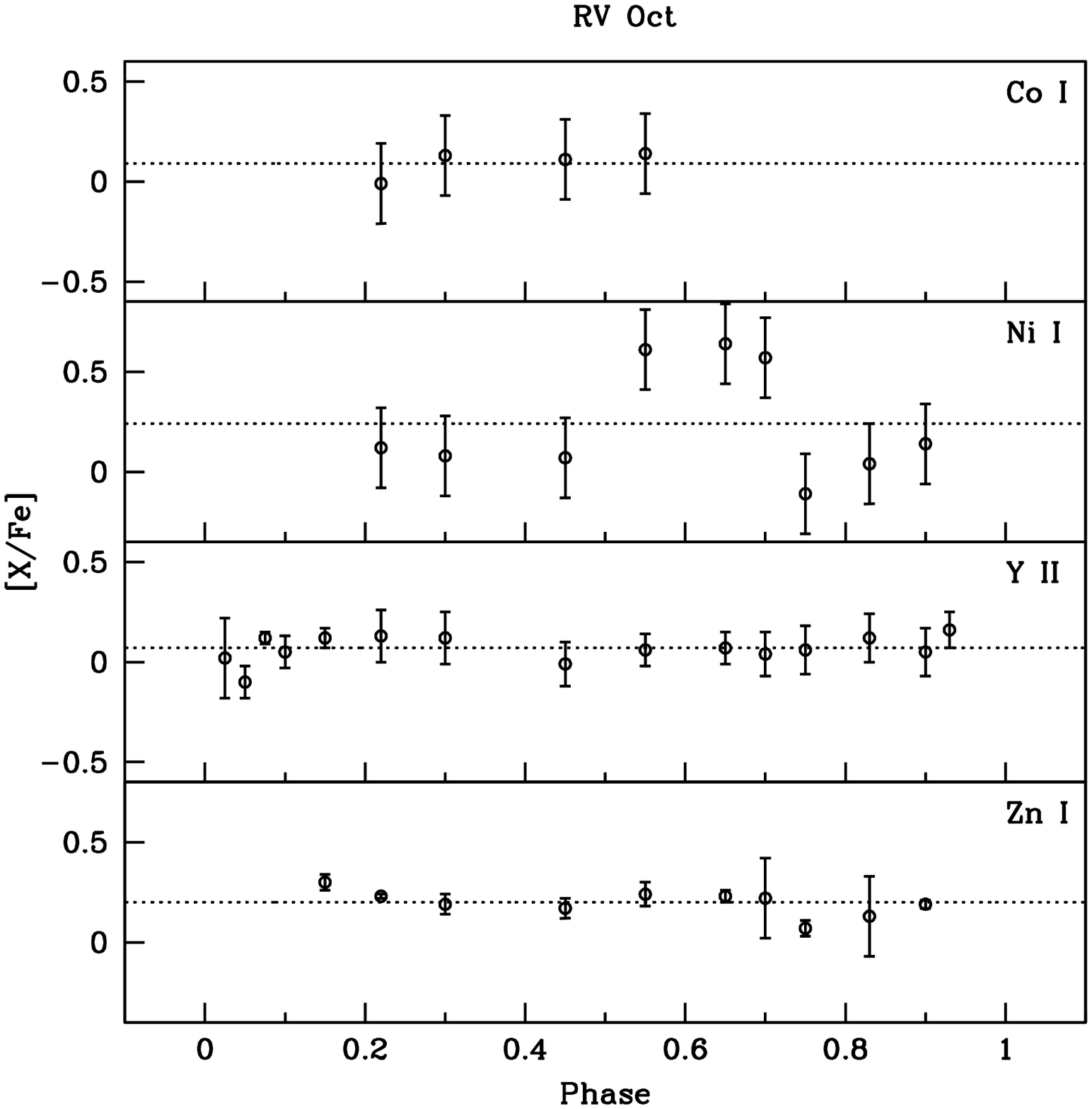}
\caption{Same as Figure~\ref{rvoct_abund1_phase}, now for \ion{Co}{1}, 
         \ion{Ni}{1}, \ion{Y}{2} and \ion{Zn}{1}. 
         The large phase-to-phase scatter of [\ion{Ni}{1}/Fe] is due to 
         the large uncertainties in the derived values. 
         \label{rvoct_abund4_phase}}
\end{figure}

\clearpage
\begin{figure}
\plotone{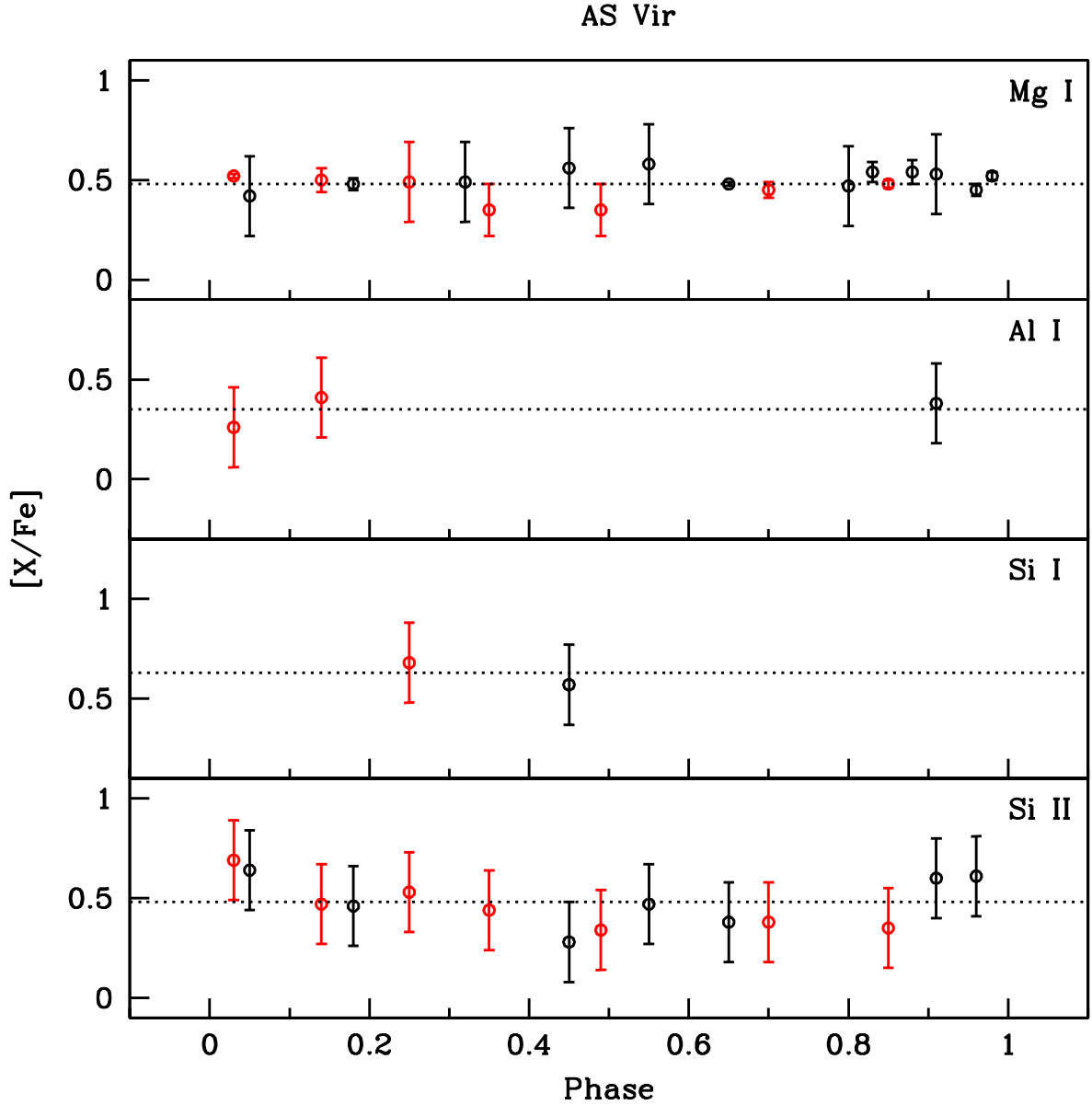}
\caption{Abundance ratios, [X/Fe], of \ion{Mg}{1}, \ion{Al}{1}, 
         \ion{Si}{1} and \ion{Si}{2} as a function of phase for a 
         Blazhko effect star, AS Vir. 
         The dashed lines and color symbols represent the mean 
         values and different cycles being considered for combining 
         the spectra, respectively. 
         The [X/Fe] values are generally consistent throughout the 
         pulsational cycle. 
         The trend in [\ion{Si}{2}/Fe] is discussed in \S~\ref{absilicon}. 
         NLTE corrections applied to Al and Si abundances whenever appropriate.
         \label{asvir_abund1_phase}}
\end{figure}

\clearpage
\begin{figure}
\plotone{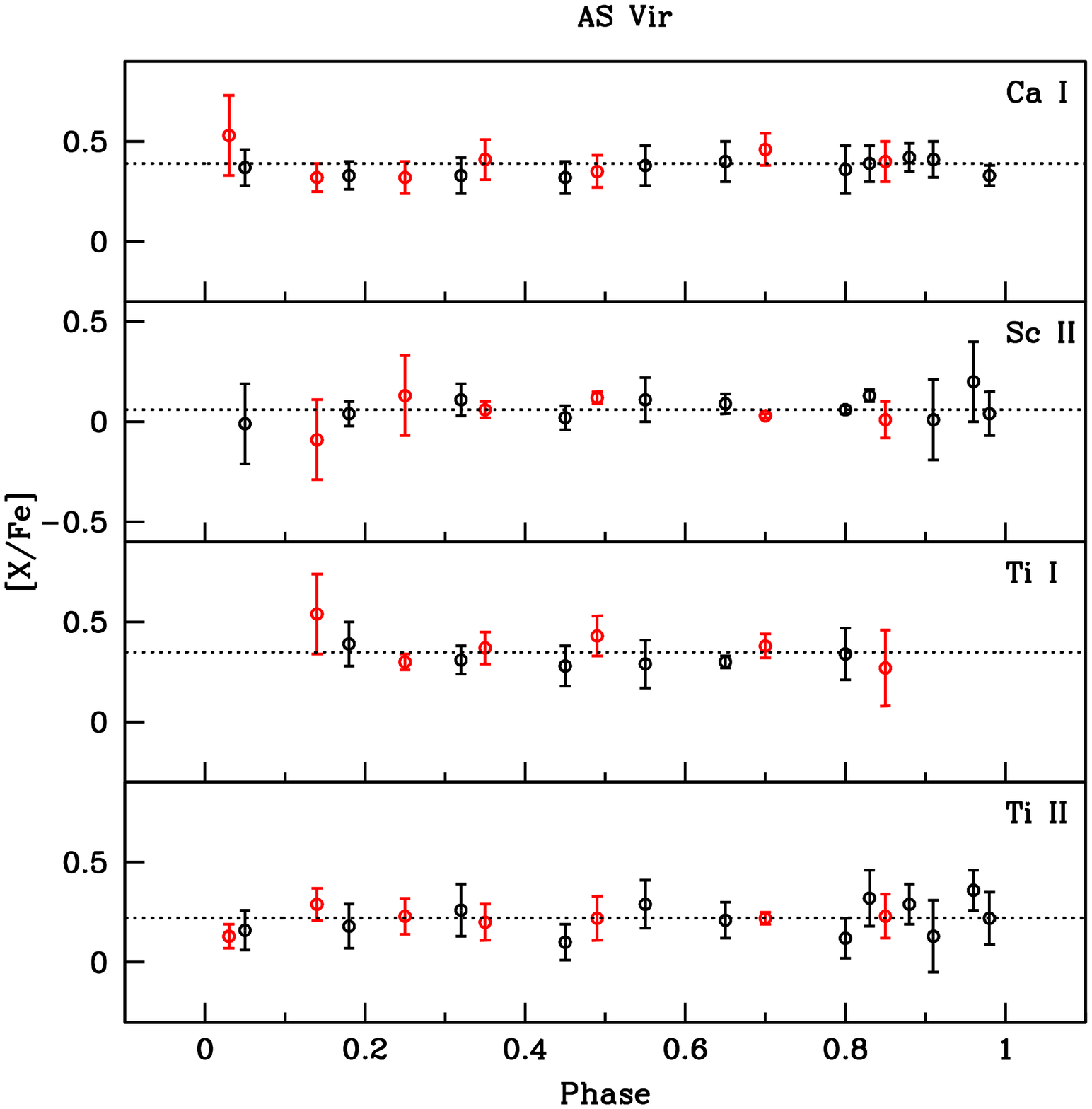}
\caption{Same as Figure~\ref{asvir_abund1_phase}, now for \ion{Ca}{1}, 
         \ion{Sc}{2}, \ion{Ti}{1} and \ion{Ti}{2}.
         \label{asvir_abund2_phase}}
\end{figure}

\clearpage
\begin{figure}
\plotone{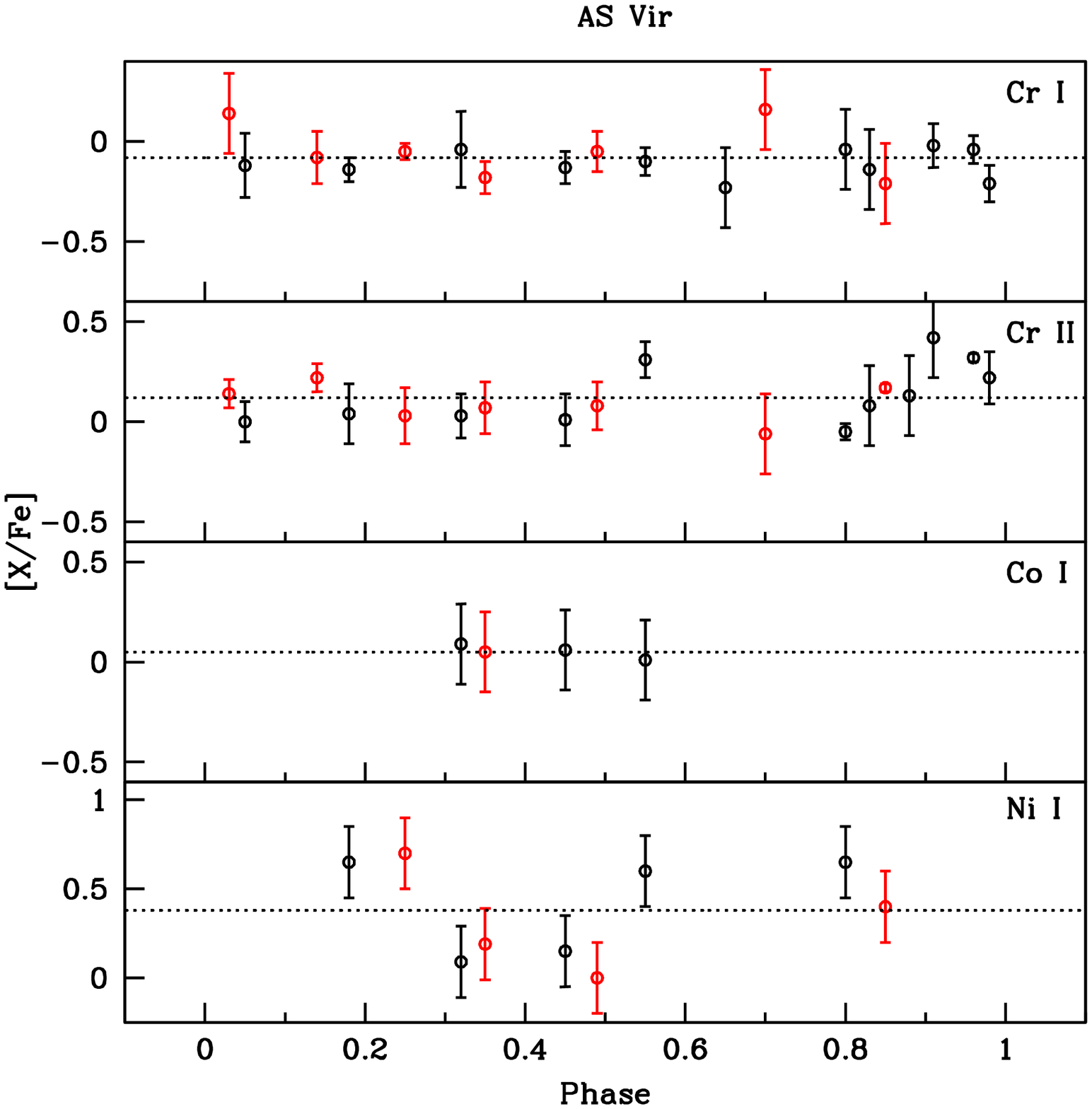}
\caption{Same as Figure~\ref{asvir_abund1_phase}, now for \ion{Cr}{1}, 
         \ion{Cr}{2}, \ion{Co}{1} and \ion{Ni}{1}. 
         The large phase-to-phase scatter of [\ion{Ni}{1}/Fe] is due to 
         the large uncertainties in the derived values.
         \label{asvir_abund3_phase}}
\end{figure}

\clearpage
\begin{figure}
\plotone{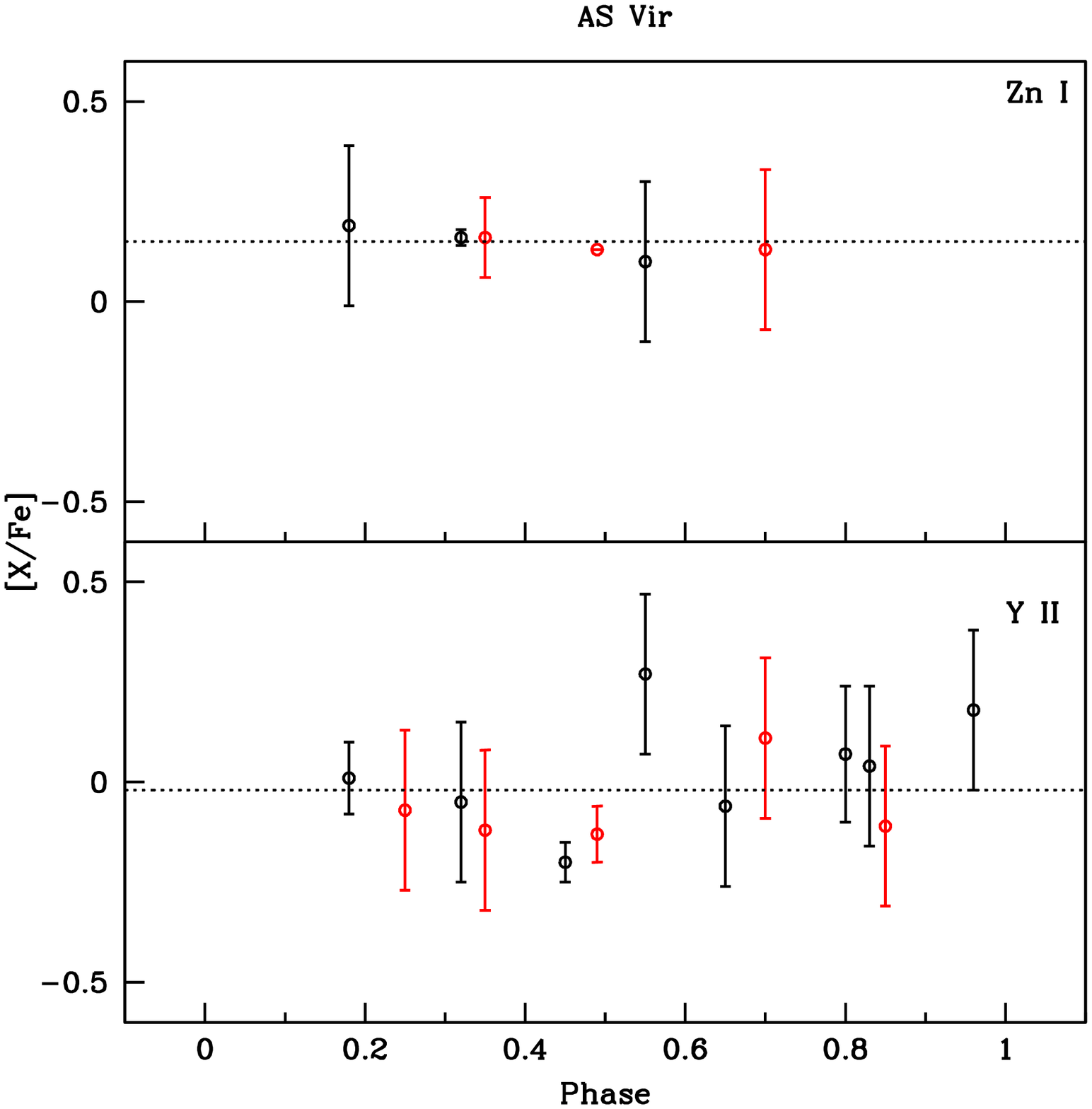}
\caption{Same as Figure~\ref{asvir_abund1_phase}, now for \ion{Zn}{1}, 
         \ion{Y}{2}. \label{asvir_abund4_phase}}
\end{figure}

\clearpage
\begin{figure}
\plotone{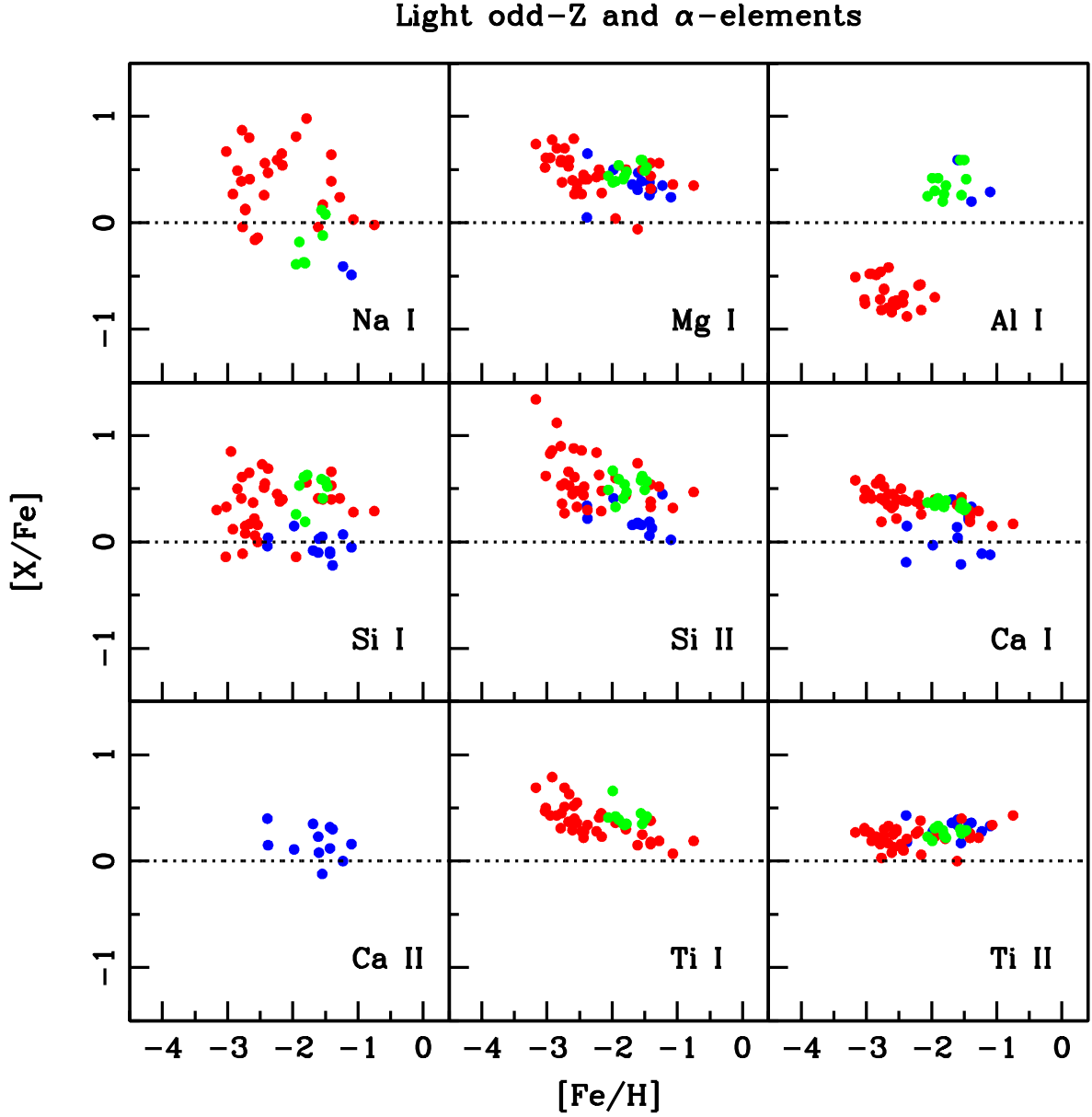}
\caption{Abundance ratios of light odd-Z and $\alpha$-elements as 
         a function of metallicity. 
         NLTE corrections applied to \ion{Na}{1}, \ion{Al}{1}, 
         \ion{Si}{1} and \ion{Si}{2} whenever appropriate. 
         The red and blue dots represent RHB and BHB stars from \citet{For10}. 
         The green dots represent the mean abundance ratios of each 
         RR Lyr in our program.
\label{xfe}}
\end{figure}

\clearpage
\begin{figure}
\plotone{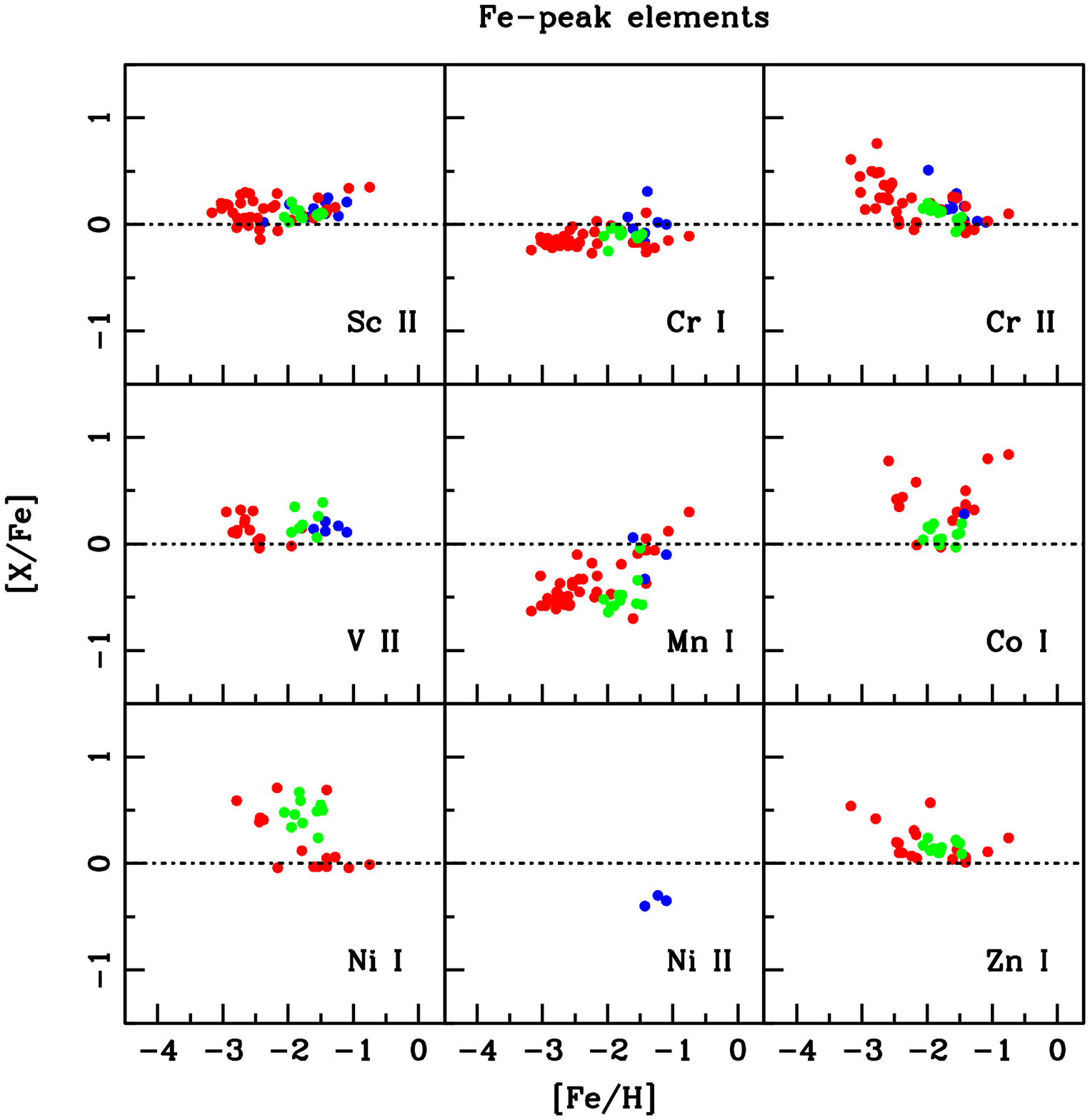}
\caption{Abundance ratios of Fe-peak elements as a function of metallicity. 
         The red and blue dots represent RHB and BHB stars from \citet{For10}. 
         The green dots represent the mean abundance ratios of each 
         RR Lyr in our program.\label{xfe1}}
\end{figure}

\clearpage
\begin{figure}
\plotone{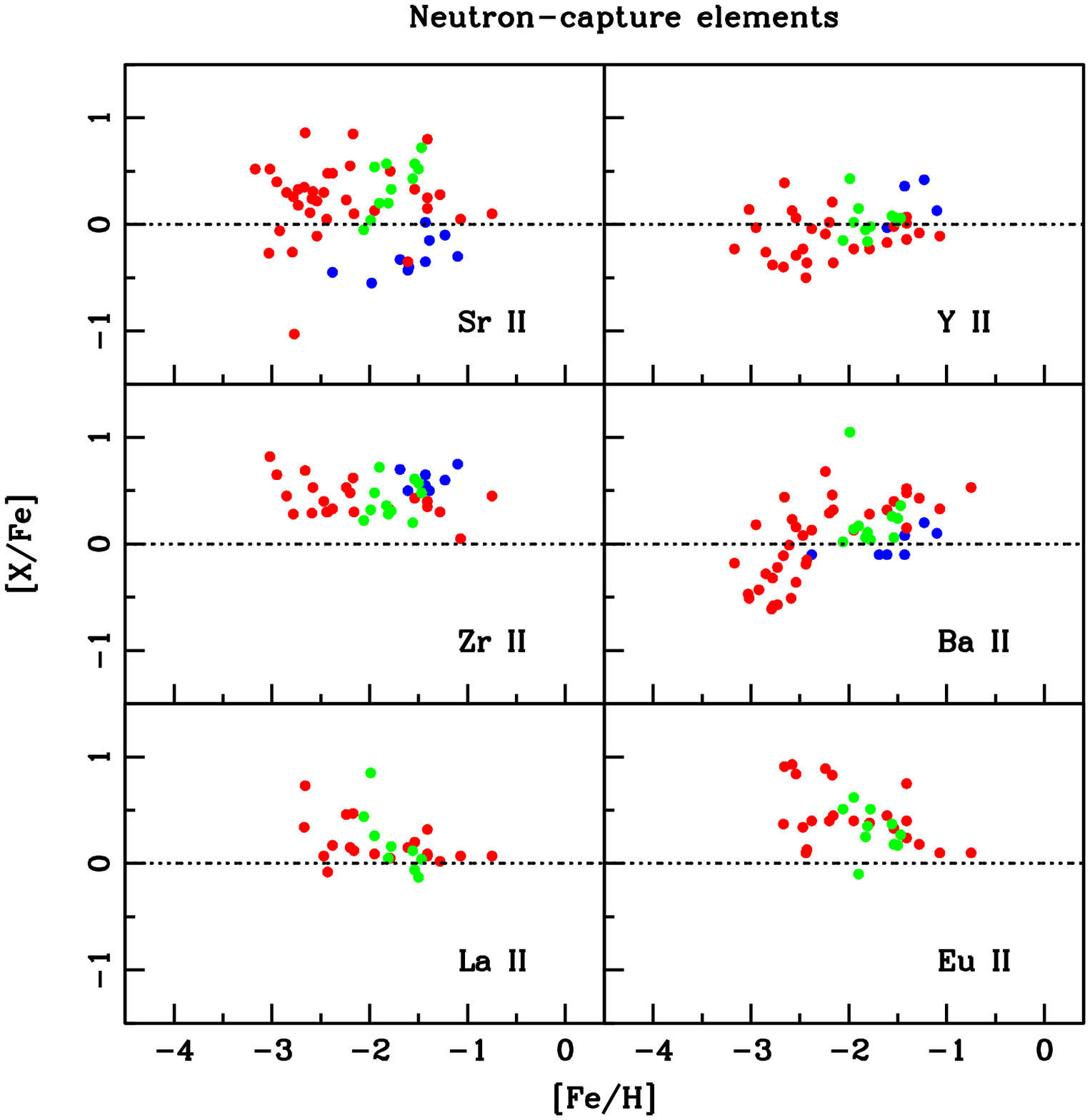}
\caption{Abundance ratios of \ncap\ elements as a 
         function of metallicity. 
         The red and blue dots represent RHB and BHB stars from \citet{For10}. 
         The green dots represent the mean abundance ratios of each 
         RR Lyr in our program. \label{xfe2}}
\end{figure}

\clearpage
\begin{figure}
\plotone{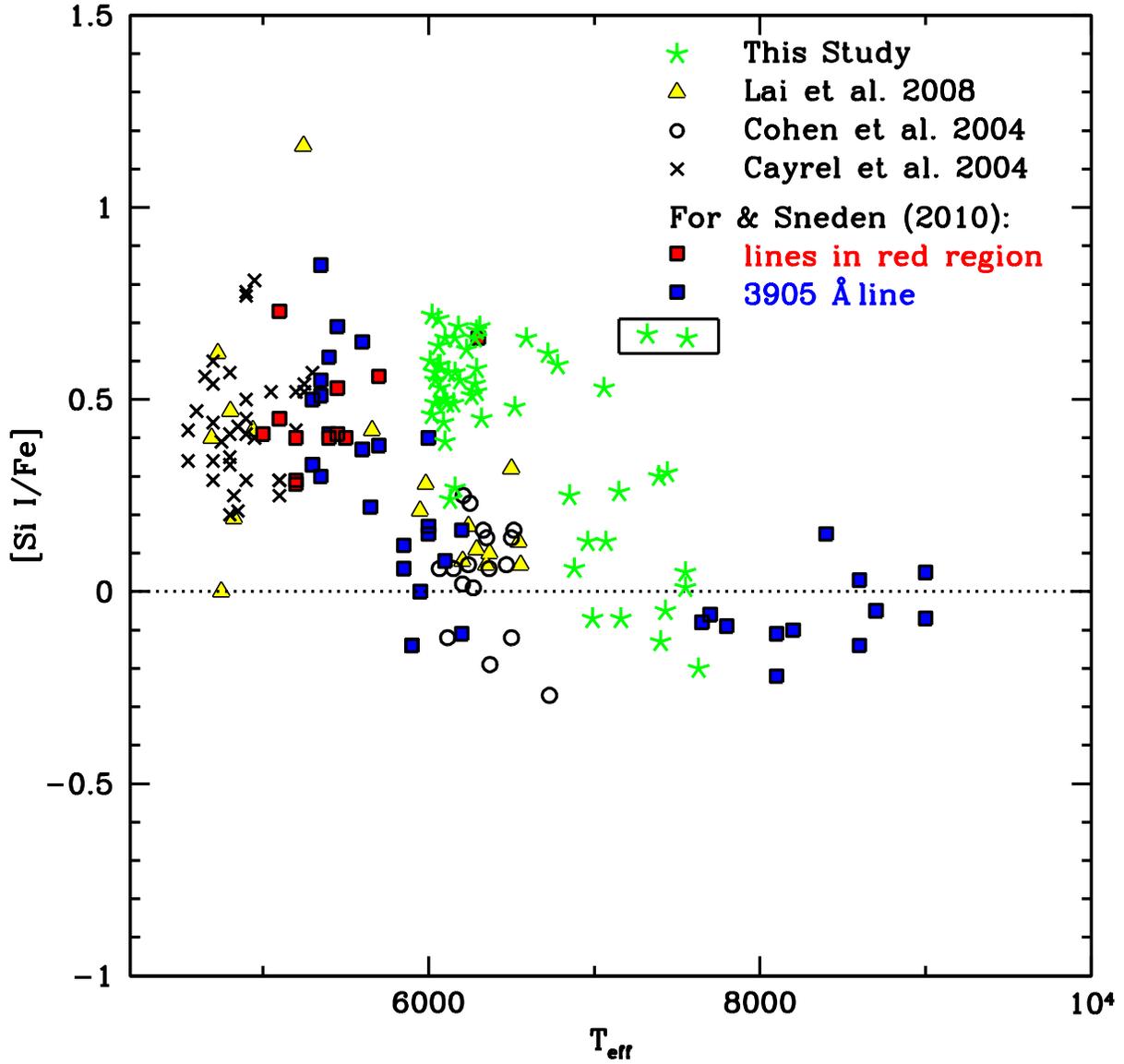}
\caption{Abundance ratios of [\ion{Si}{1}/Fe] of all our program stars 
         in all phases (green stars) vs.  spectroscopic \teff\ , with 
         additional data from \citet{Cayrel04} (crosses), \citet{Cohen04} 
         (open circles), \citet{Lai08} (yellow triangles), and \citet{For10} 
         (blue and red squares). 
         The box marks the two outliers. 
         NLTE correction applied to [\ion{Si}{1}/Fe] whenever appropriate. 
         \label{si1_teff}}
\end{figure}

\clearpage
\begin{figure}
\plotone{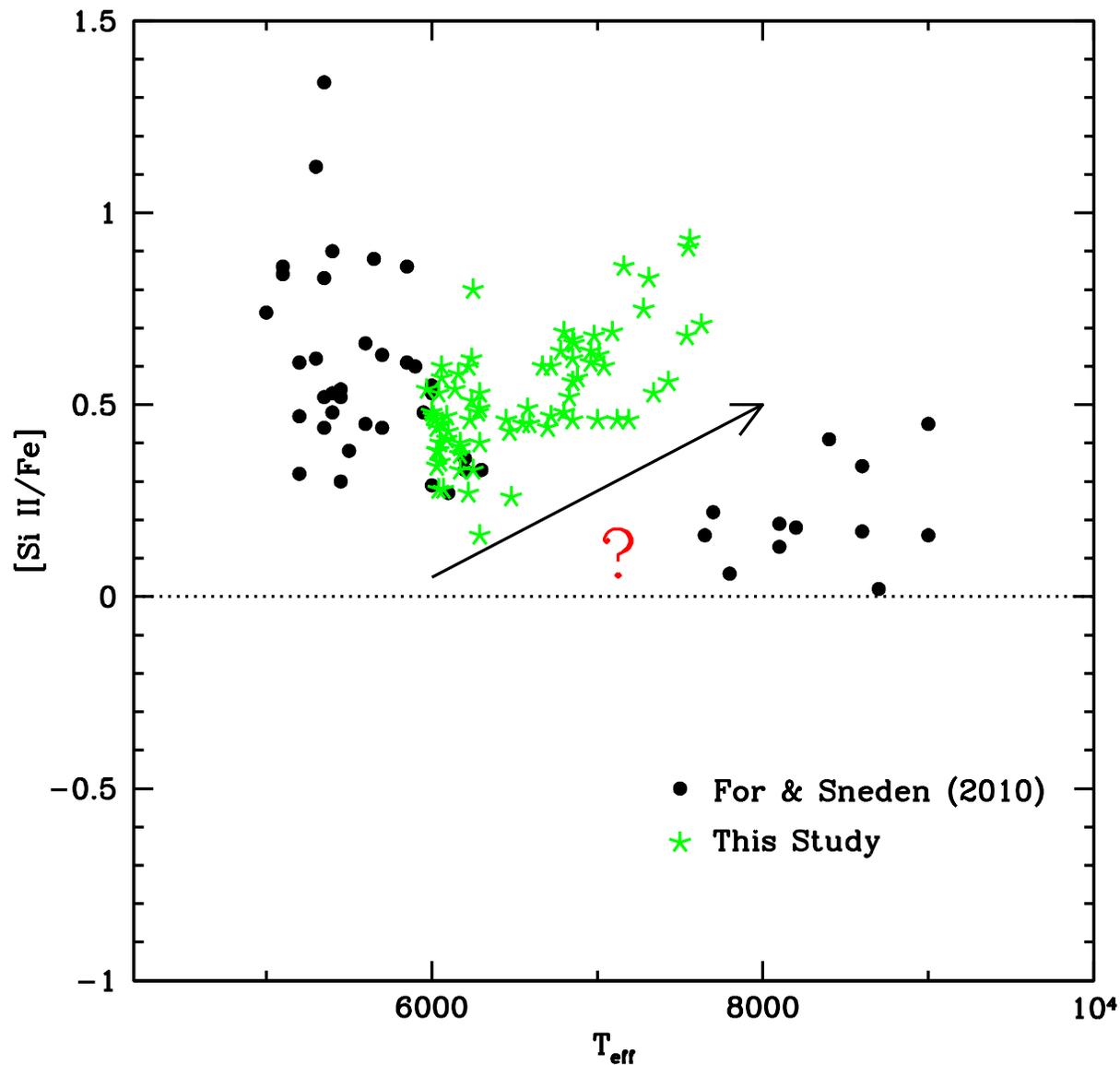}
\caption{Abundance ratios of [\ion{Si}{2}/Fe] of all our program stars 
         in all phases (green stars) vs. spectroscopic \teff. 
         NLTE correction applied to [\ion{Si}{2}/Fe] whenever appropriate. 
         The black dots denote abundances for the BHB and RHB stars 
         from \citet{For10}.
         A possible increasing [\ion{Si}{2}/Fe] trend as a function 
         of increasing \teff\ is schematically represented in this
         figure by an arrow, and is discussed in \S~\ref{absilicon}.  
         \label{si2_teff}}
\end{figure}

\clearpage
\begin{figure}
\plotone{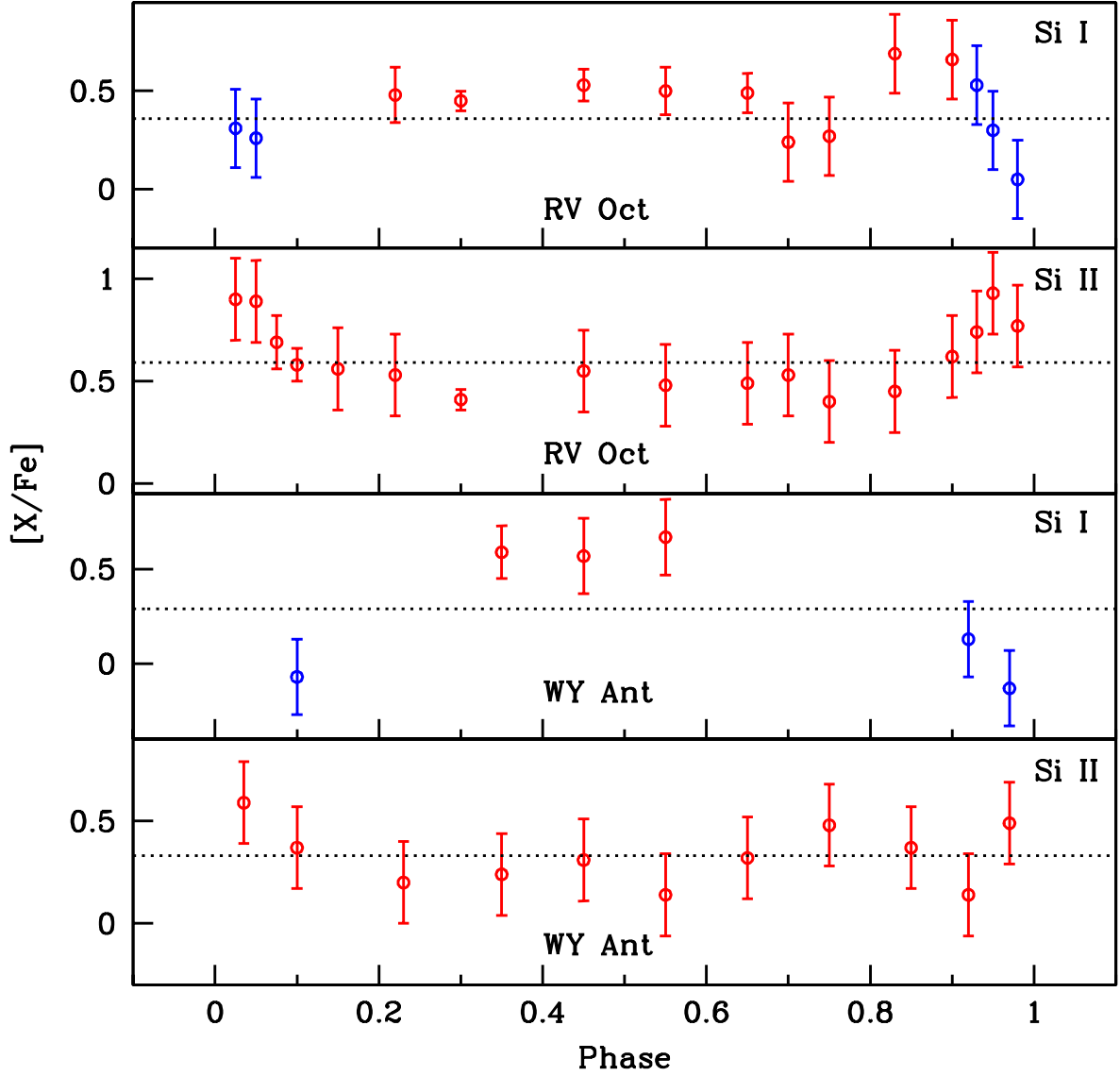}
\caption{Silicon abundance ratios as a function of phase for RV Oct 
         (first and second panels) and WY Ant (third and forth panels). 
         The blue and red open circles represent the silicon abundances 
         derived from blue and red spectral regions, respectively. 
         NLTE corrections have been applied to the values represented 
         by the blue open circles for [\ion{Si}{1}/Fe] and red open 
         circles for [\ion{Si}{2}/Fe]. \label{si_phase}}
\end{figure}

\clearpage
\begin{figure}
\plotone{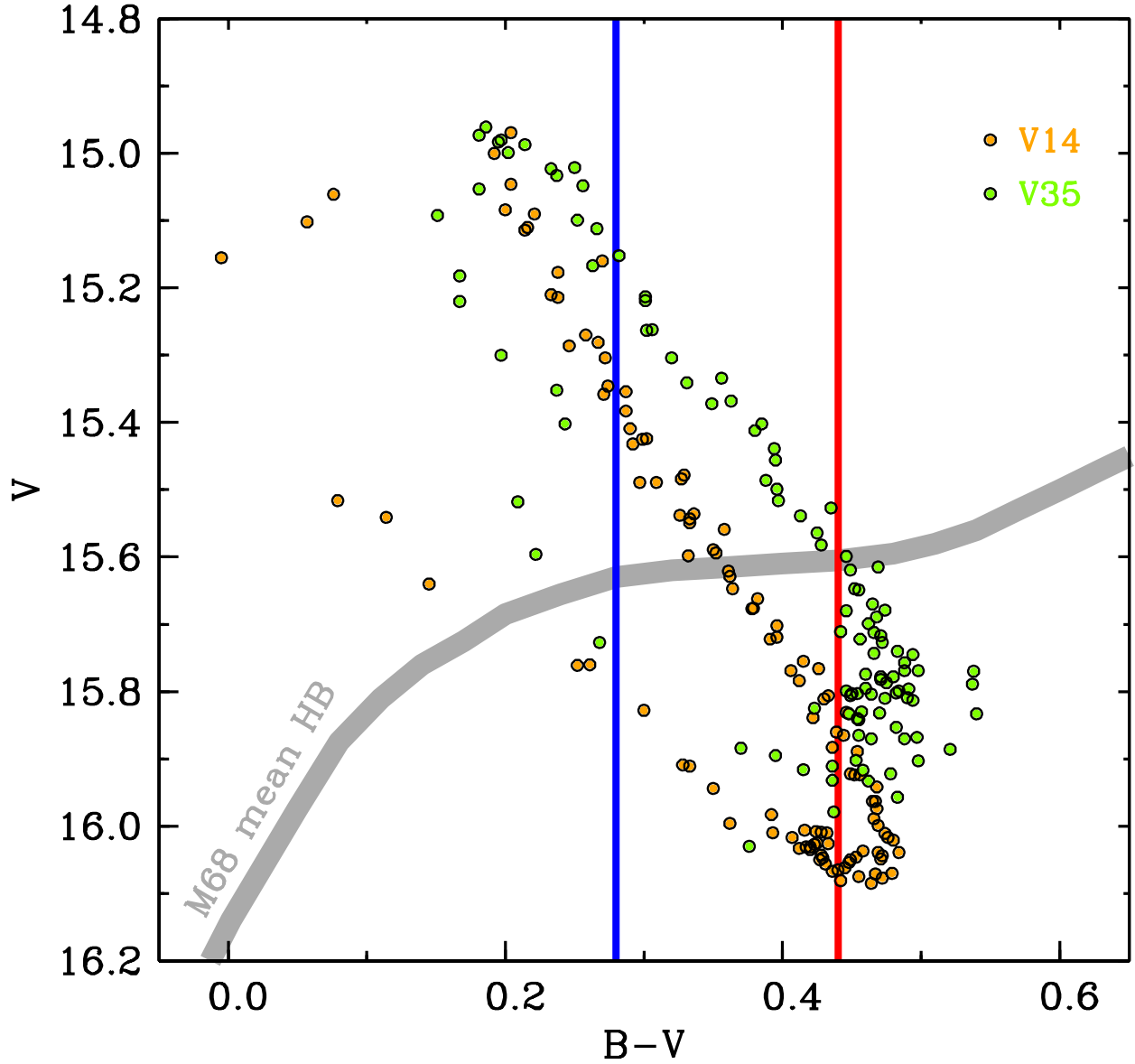}
\caption{Loops for two RRab stars in the $V$ versus $B-V$ diagram of M68.
         Orange and green symbols denote data for \cite{Walker94} 
         variables V14 (P~= 0.5568~d) and V35 (P~= 0.7025~d),
         and the grey line is a hand drawn representation of the HB population in M68.  
         Blue and red vertical lines mark the color boundaries of the RR Lyr instability 
         strip estimated from the data points in Walker's Figure 13. \label{m68}}
\end{figure}

\clearpage
\begin{figure}
\plotone{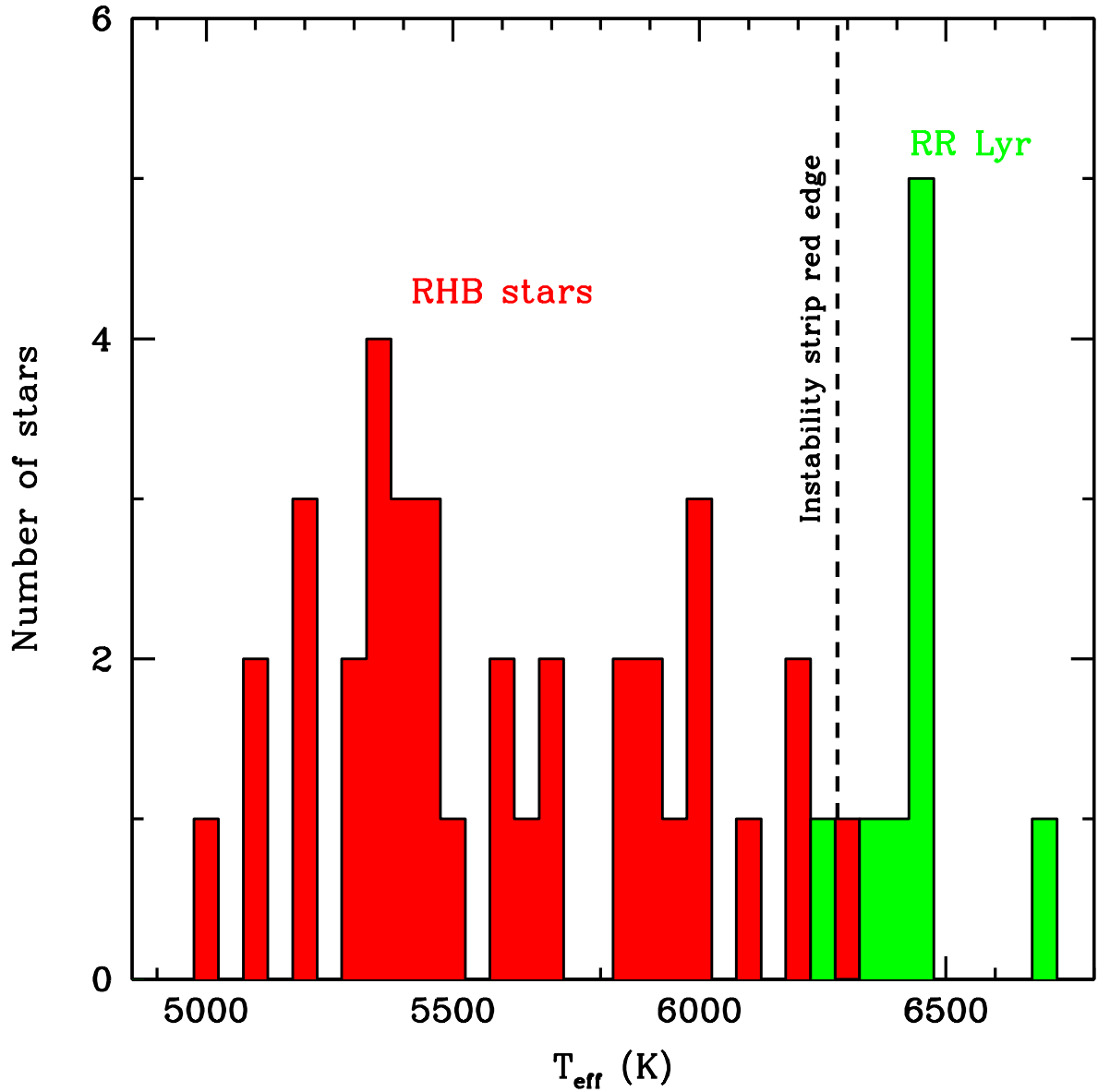}
\caption{A histogram showing the distribution of \teff\ values
         of RHB field stars (red) and mean \teff\ values for
         of RR~Lyr sample (green).
         The dashed vertical line represents our best \teff\ estimate
         (6280~K) for the FRE. \label{starcount}}
\end{figure}

\clearpage
\begin{figure}
\plotone{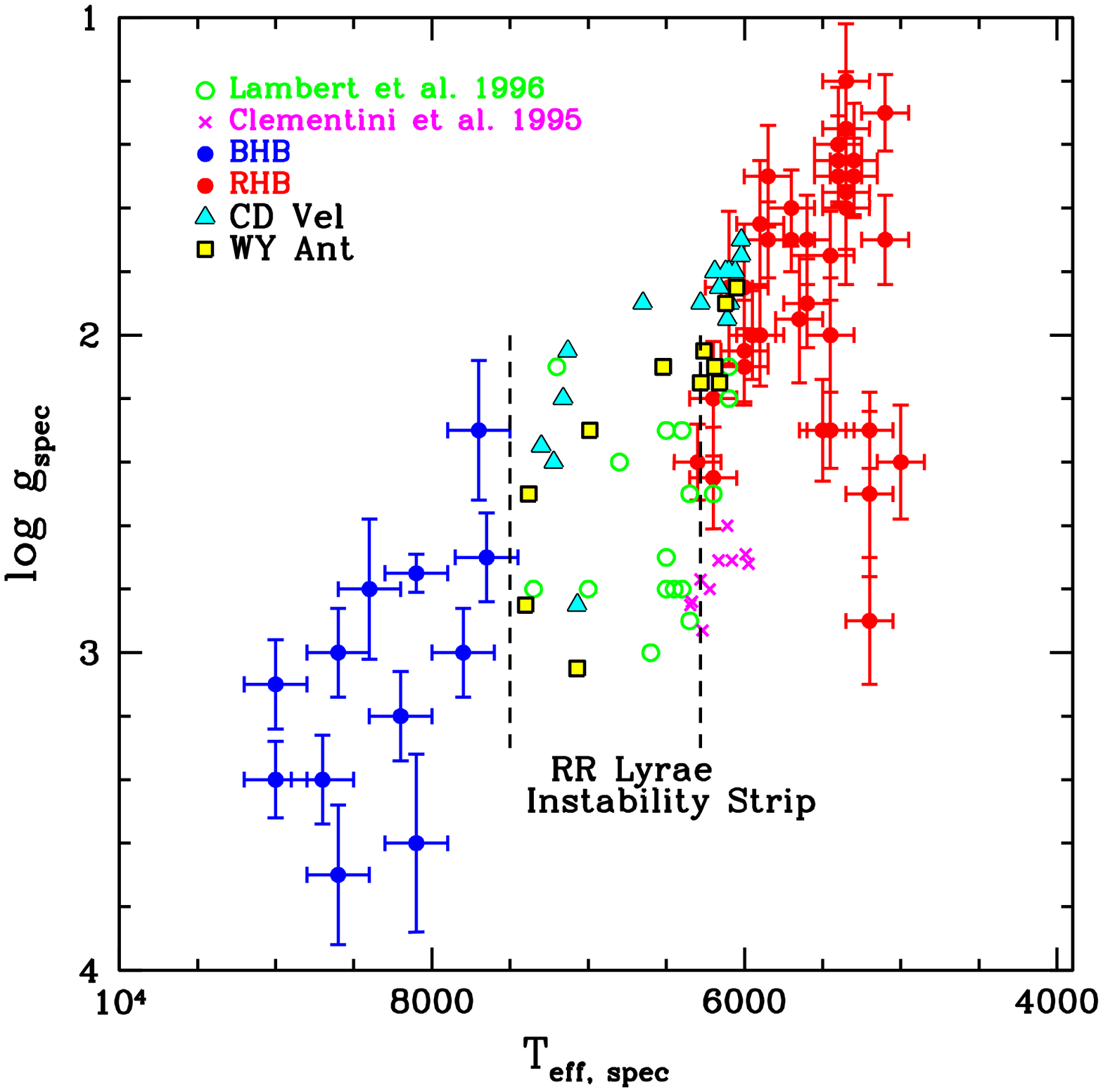}
\caption{Spectroscopic \teff\ and \logg\ of CD Vel and WY Ant, with 
         additional data from \citet{For10} (RHB: red dots; BHB: 
         blue dots), \citet{Lambert96} (green open circles) and 
         \citet{Clementini95} (magenta crosses) 
         on the \teff-\logg\ plane. \label{hr}}
\end{figure}

\clearpage
\begin{figure}
\plotone{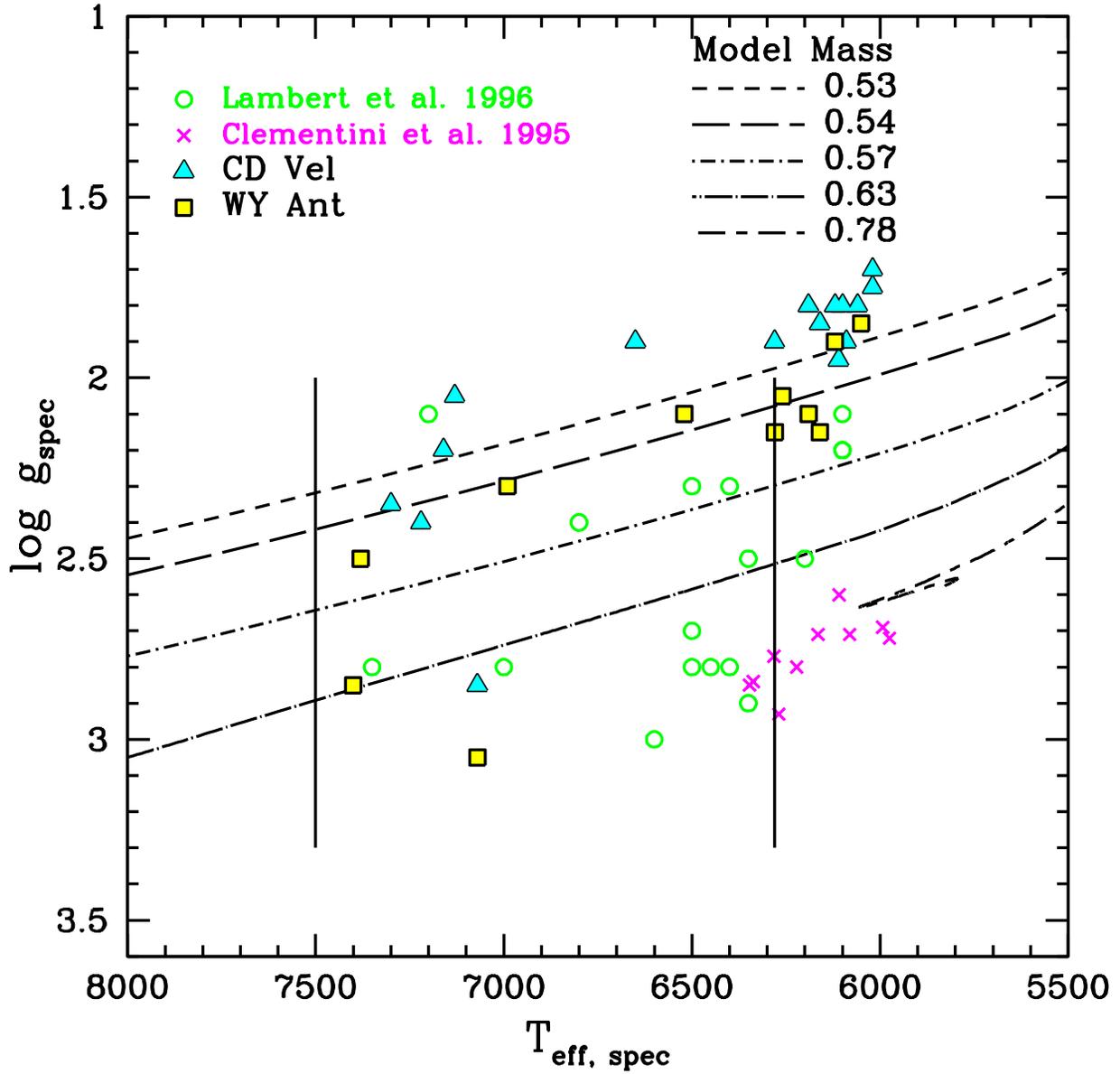}
\caption{An enlarged version of Figure~\ref{hr} near the instability strip 
         region with an overlaid of $\alpha$-enhanced HB tracks of 
         [M/H] $=-1.79$, $Z=0.0003$, $Y=0.245$. \label{hr_rrlyr}}
\end{figure}

\clearpage

\begin{deluxetable}{lcccccccc}
\rotate
\tabletypesize{\scriptsize}
\tablewidth{0pt}
\tablecolumns{9}
\tablecaption{Basic Information for the Program Stars.\label{targets}}
\tablehead{
\colhead{Star} & 
\colhead{R.A.(J2000)} & 
\colhead{Decl.(J2000)}  & 
\colhead{Data Used\tablenotemark{a}}  & 
\colhead{$T_{0}$}  &
\colhead{err}  & 
\colhead{Period}  & 
\colhead{err}  & 
\colhead{$N$\tablenotemark{b}}  \\
\colhead{} & 
\colhead{(hr m s)} & 
\colhead{($\degr$ $\arcmin$ $\arcsec$)}  & 
\colhead{(HJD 2450000+)}  & 
\colhead{(HJD 2450000+)}  &
\colhead{(HJD 2450000+)}  & 
\colhead{(day)}  & 
\colhead{(day)}  & 
\colhead{}  
}
\startdata
CD Vel	&	09 44 38.24	&	$-$45 52 37.2	&	all	&	3837.632	&	0.0003	&	0.573510	&	0.000003	&	208	\\
WY Ant	&	10 16 04.95	&	$-$29 43 42.4	&	all	&	4191.685	&	0.0097	&	0.574344	&	0.000002	&	136	\\
DT Hya	&	11 54 00.18	&	$-$31 15 40.0	&	all	&	4583.637	&	0.0089	&	0.567978	&	0.000001	&	102	\\
AS Vir	&	12 52 45.86	&	$-$10 15 36.4	&	all	&	4907.709	&	0.0098	&	0.553412	&	0.000002	&	262	\\
RV Oct	&	13 46 31.75	&	$-$84 24 06.4	&	all	&	3841.602	&	0.0016	&	0.571170	&	0.000002	&	222	\\
XZ Aps	&	14 52 05.43	&	$-$79 40 46.6	&	all	&	3842.735	&	0.0052	&	0.587264	&	0.000002	&	289	\\
BS Aps	&	16 20 51.51	&	$-$71 40 15.8	&	all	&	4583.785	&	0.0045	&	0.582561	&	0.000007	&	252	\\
UV Oct	&	16 32 25.53	&	$-$83 54 10.5	&	3836.84--3842.91, 4306.46--5021.84	&	3837.875	&	0.0072	&	0.542578	&	0.000003	&	323	\\
\enddata
\tablenotetext{a}{Data with the corresponding HJDs were used to derive the $T_{0}$.}
\tablenotetext{b}{Total number of observed spectra.}
\end{deluxetable}

\begin{deluxetable}{lrrrrrrrrrrrrr}
\tabletypesize{\scriptsize}
\tablewidth{0pt}
\tablecolumns{14}
\tablecaption{Synthetic Colors for Models with $\log g=2.0$.\label{grids}}
\tablehead{
\colhead{} & 
\multicolumn{13}{c}{Effective Temperature (K)}  \\
\colhead{Color Indices} & 
\colhead{5500}  & 
\colhead{5750}  & 
\colhead{6000}  &
\colhead{6250}  & 
\colhead{6500}  & 
\colhead{6750}  & 
\colhead{7000}  & 
\colhead{7250}  &
\colhead{7500}  & 
\colhead{7750}  & 
\colhead{8000}  &
\colhead{8250}  & 
\colhead{8500}  
}
\startdata
\cutinhead{[M/H]$=-0.10$ (SW And)}
$B-V$	        & 0.755	& 0.650	& 0.555	& 0.470	& 0.394	& 0.323	& 0.249	& 0.177	& 0.124	& 0.081	& 0.044	& 0.014	& $-$0.007  \\
$V-R_{c}$	& 0.396	& 0.352	& 0.310	& 0.270	& 0.230	& 0.193	& 0.156	& 0.120	& 0.090	& 0.065	& 0.044	& 0.028	& 0.018   \\
$V-I_{c}$	& 0.779	& 0.698	& 0.620	& 0.544	& 0.470	& 0.397	& 0.328	& 0.263	& 0.206	& 0.158	& 0.118	& 0.087	& 0.065	  \\
\cutinhead{[M/H]$=-0.30$ (AR Per)}
$B-V$	        & 0.722	& 0.619	& 0.528	& 0.447	& 0.375	& 0.307	& 0.237	& 0.167	& 0.115	& 0.074	& 0.039	& 0.011	& $-$0.008  \\
$V-R_{c}$	& 0.390	& 0.347	& 0.306	& 0.266	& 0.227	& 0.190	& 0.153	& 0.118	& 0.088	& 0.064	& 0.043	& 0.027	& 0.017	  \\
$V-I_{c}$	& 0.776	& 0.697	& 0.619	& 0.543	& 0.470	& 0.398	& 0.329	& 0.264	& 0.207	& 0.159	& 0.119	& 0.088	& 0.066	  \\
\cutinhead{[M/H]$=-1.15$ (RR Leo)}
$B-V$	        & 0.612	& 0.522	& 0.446	& 0.380	& 0.319	& 0.262	& 0.204	& 0.140	& 0.092	& 0.056	& 0.026	& 0.003	& $-$0.011  \\
$V-R_{c}$	& 0.375	& 0.334	& 0.294	& 0.256	& 0.218	& 0.182	& 0.146	& 0.112	& 0.083	& 0.059	& 0.040	& 0.026	& 0.016	  \\
\enddata
\end{deluxetable}

\clearpage

\begin{deluxetable}{crccccccccc}
\tabletypesize{\scriptsize}
\tablewidth{0pt}
\tablecolumns{11}
\tablecaption{Basic Data for Deriving the \teff--Phase Relations. \label{colorteff}}
\tablehead{
\colhead{Phase} &
\colhead{RV$-$RV$_{\rm min}$} &
\colhead{$B-V$} &
\colhead{$c(B-V)$} &
\colhead{\teff} &
\colhead{$V-R_{c}$} &
\colhead{$c(V-R_{c})$} &
\colhead{\teff} &
\colhead{$V-I_{c}$} &
\colhead{$c(V-I_{c})$} &
\colhead{\teff} 
}
\startdata
\cutinhead{SW And}
0.00	&	0	&	0.211	&	0.151	&	7373	&	0.131	&	0.097	&	7443	&	0.272	&	0.198	&	7544	\\
0.05	&	2.72	&	0.237	&	0.177	&	7250	&	0.161	&	0.127	&	7203	&	0.325	&	0.251	&	7304	\\
0.10	&	7.28	&	0.273	&	0.213	&	7125	&	0.187	&	0.153	&	7022	&	0.377	&	0.303	&	7098	\\
0.20	&	16.83	&	0.346	&	0.286	&	6875	&	0.226	&	0.192	&	6758	&	0.458	&	0.384	&	6799	\\
0.30	&	26.46	&	0.453	&	0.393	&	6504	&	0.276	&	0.242	&	6426	&	0.559	&	0.485	&	6451	\\
0.40	&	34.47	&	0.491	&	0.431	&	6378	&	0.307	&	0.273	&	6233	&	0.614	&	0.540	&	6265	\\
0.50	&	44.89	&	0.514	&	0.454	&	6303	&	0.318	&	0.284	&	6164	&	0.640	&	0.566	&	6179	\\
0.60	&	48.73	&	0.526	&	0.466	&	6263	&	0.313	&	0.279	&	6195	&	0.630	&	0.556	&	6212	\\
\enddata
\end{deluxetable}

\clearpage
\begin{deluxetable}{clrrrrr}
%\tabletypesize{\scriptsize}
\tablewidth{0pt}
\tablecolumns{7}
\tablecaption{Coefficients for the Temperature$-$Phase 
Relationships\tablenotemark{a}.\label{caleqs}}
\tablehead{
\colhead{Eq} &
\colhead{Star} & 
\colhead{$a_{4}$}  & 
\colhead{$a_{3}$}  & 
\colhead{$a_{2}$}  &
\colhead{$a_{1}$}  & 
\colhead{$a_{0}$}  
}
\startdata
1 &	SW And	&	$-$1049.6	&	600.08	&	4153.8	&	$-$4808.7	&	7542.7	\\
2 &	AR Per	&	$-$5174.6	&	4653.8	&	4167.1	&	$-$5275.8	&	7554.7	\\
3 &	RR Leo	&	$-$6583.7	&	5248.0	&	7718.5	&	$-$8830.7	&	8444.6	\\
4 &	RR Cet	&	$-$3483.1	&	2780.7	&	4418.3	&	$-$5061.1	&	7394.3	\\
5 &	TU Uma	&	$-$10916        &	14340	&	$-$960.06	&	$-$4324.7	&	7373.8	\\
6 &	TT Lyn	&	$-$7213.9	&	10633	&	$-$2121.4	&	$-$2464.3	&	6902.5	\\
7 &	RX Eri	&	$-$6602.1	&	9883.2	&	$-$1303.9	&	$-$3315.4	&	7186.2	\\
8 &	SU Dra	&	$-$8545.3	&	12001	&	$-$860.84	&	$-$4142.2	&	7343.4	
\enddata

\tablenotetext{a}{\teff~$= a_{4}\phi^{4} + a_{3}\phi^{3} + a_{2}\phi^{2}
                  + a_{1}\phi+a_{0}$, where $\phi$ is the phase}

\end{deluxetable}

\clearpage

\begin{deluxetable}{ccccccccccrccr}
\tabletypesize{\scriptsize}
\tablewidth{0pt}
\tablecolumns{14}
\tablecaption{Stellar Atmosphere Parameters and Fe Abundances
throughout the Pulsational Cycles.\label{inputmodels}}
\tablehead{
\colhead{Mid Phase} &
\colhead{\teff} &
\colhead{err} &
\colhead{\logg} &
\colhead{err} &
\colhead{[M/H]} &
\colhead{\vturb} &
\colhead{err} &
\colhead{[Fe I/H]} &
\colhead{err} &
\colhead{$N$} &
\colhead{[Fe II/H]} &
\colhead{err} &
\colhead{$N$} \\
\colhead{} &
\colhead{(K)} &
\colhead{} &
\colhead{(dex)} &
\colhead{} &
\colhead{(dex)} &
\colhead{(km~s$^{-1}$)} &
\colhead{(dex)} &
\colhead{} &
\colhead{} &
\colhead{} &
\colhead{(dex)} &
\colhead{} &
\colhead{} 
}
\startdata
\cutinhead{CD Vel}
0.015	&	7130	&	300	&	2.05	&	0.20	&	$-1.80	$	&	2.85	&	0.20	&	$-1.80$	&	0.09	&	30	&	$-1.81$	&	0.10	&	25	\\
0.045	&	7160	&	300	&	2.20	&	0.24	&	$-1.63	$	&	2.90	&	0.20	&	$-1.63$	&	0.11	&	26	&	$-1.62$	&	0.12	&	16	\\
0.150	&	6650	&	200	&	1.90	&	0.16	&	$-1.74	$	&	2.75	&	0.30	&	$-1.74$	&	0.10	&	68	&	$-1.73$	&	0.08	&	23	\\
0.300	&	6280	&	100	&	1.90	&	0.22	&	$-1.73	$	&	2.50	&	0.10	&	$-1.73$	&	0.11	&	82	&	$-1.73$	&	0.11	&	29	\\
0.400	&	6100	&	100	&	1.80	&	0.20	&	$-1.80	$	&	2.70	&	0.10	&	$-1.80$	&	0.09	&	80	&	$-1.80$	&	0.10	&	23	\\
0.500	&	6020	&	100	&	1.75	&	0.20	&	$-1.83	$	&	2.80	&	0.10	&	$-1.83$	&	0.11	&	76	&	$-1.83$	&	0.10	&	29	\\
0.570	&	6020	&	100	&	1.70	&	0.20	&	$-1.89	$	&	3.20	&	0.10	&	$-1.89$	&	0.09	&	63	&	$-1.89$	&	0.10	&	20	\\
0.650	&	6060	&	100	&	1.80	&	0.14	&	$-1.88	$	&	3.15	&	0.10	&	$-1.88$	&	0.10	&	55	&	$-1.89$	&	0.07	&	25	\\
\enddata
\end{deluxetable}

\clearpage

\begin{deluxetable}{llccc}
%\tabletypesize{\scriptsize}
\tablewidth{0pt}
\tablecolumns{5}
\tablecaption{HD~140283 \teff\ and \logg\ Values from Various Studies. \label{methodlogg}}
\tablehead{
\colhead{Reference} & 
\colhead{Method} & 
\colhead{\teff}  & 
\colhead{\logg}  & 
\colhead{[Fe/H]} \\
\colhead{}  &
\colhead{}  & 
\colhead{(K)}  & 
\colhead{(dex)} &
\colhead{(dex)}
}
\startdata
\citet{Aoki02}    &  Spectroscopic (Ti $\&$ Fe) & 5750\tablenotemark{a} & 3.3  & $-2.58$\\
\citet{Hosford09} &  Spectroscopic (Fe)         & 5573$\pm$75   & 3.1$\pm$0.15 & $-2.54$\\
\citet{Asplund06} &  Balmer line wing fitting & 5753$\pm$30   & 3.7$\pm$0.04 & $-2.40$\\
\citet{Ryan96}    &  Photometry               & 5750          & 3.4          & $-2.54$\\
\citet{Alonso96}  &  Infrared flux            & 5691$\pm$69   & 4.0$\pm$0.50 & \nodata \\
This study\tablenotemark{b} &  Spectroscopic (Fe) & 5400$\pm$150  & 2.6$\pm$0.16 & $-2.80$\\
This study\tablenotemark{c} &  Spectroscopic (Fe) & 5400$\pm$150  & 2.6$\pm$0.16 & $-2.80$ \\
This study                  &  Trigonometric\tablenotemark{d} & 5400\tablenotemark{e}  & 3.7 & $-2.94$ \\          
\enddata
\tablenotetext{a}{Adopted from \citet{Ryan96}}
\tablenotetext{b}{Without scattered light correction.}
\tablenotetext{c}{With scattered light correction.}
\tablenotetext{d}{Assuming $M=0.8$~\Msun, $\pi=17.44$ mas and $E(B-V)=0$.}
\tablenotetext{e}{Adopted spectroscopic \teff.}
\end{deluxetable}

\clearpage

\begin{deluxetable}{crcrrcrrcrrcrrcrrcr}
\rotate
\tabletypesize{\tiny}
\tablewidth{0pt}
\tablecolumns{19}
\tablecaption{Abundance ratios of Na, Mg, Al, Si and Ca throughout the 
Pulsational Cycles.\label{abund1}}
\tablehead{
\colhead{Phase}   &
\colhead{[Na I/Fe]} &
\colhead{$\sigma$} &
\colhead{$N$}     &
\colhead{[Mg I/Fe]} &
\colhead{$\sigma$} &
\colhead{$N$}      &
\colhead{[Al I/Fe]}  &
\colhead{$\sigma$} &
\colhead{$N$}      &
\colhead{[Si I/Fe]}  &
\colhead{$\sigma$} &
\colhead{$N$}      &
\colhead{[Si II/Fe]} &
\colhead{$\sigma$}  &
\colhead{$N$}       &
\colhead{[Ca I/Fe]}   &
\colhead{$\sigma$}  &
\colhead{$N$}       
}
\startdata
\cutinhead{CD Vel}																																					
0.015	&	$-$0.37\tablenotemark{a}	&	0.03	&	2	&	0.40	&	0.02	&	3	&	0.03\tablenotemark{a}	&	\nodata	&	1	&	\nodata	&	\nodata	&	\nodata	&	0.58\tablenotemark{a}	&	\nodata	&	1	&	0.19	&	0.09	&	8	\\
0.045	&	$-$0.23\tablenotemark{a}	&	0.07	&	2	&	0.31	&	0.04	&	2	&	0.26\tablenotemark{a}	&	\nodata	&	1	&	\nodata	&	\nodata	&	\nodata	&	0.31\tablenotemark{a}	&	\nodata	&	1	&	0.22	&	0.10	&	7	\\
0.150	&	\nodata	&	\nodata	&	\nodata	&	0.44	&	0.03	&	2	&	0.18\tablenotemark{a}	&	0.08	&	2	&	\nodata	&	\nodata	&	\nodata	&	0.47\tablenotemark{a}	&	\nodata	&	1	&	0.26	&	0.06	&	11	\\
0.300	&	\nodata	&	\nodata	&	\nodata	&	0.41	&	0.06	&	2	&	0.48\tablenotemark{a}	&	0.22	&	2	&	0.66	&	\nodata	&	1	&	\nodata	&	\nodata	&	\nodata	&	0.33	&	0.10	&	13	\\
.    	&	.	&	.	&	.	&	.	&	.	&	.	&	.	&	.	&	.	&	.	&	.	&	.	&	.	&	.	&	.	&	.       &	.       &	.       \\
.    	&	.	&	.	&	.	&	.	&	.	&	.	&	.       &	.       &	.       &	.	&	.	&	.	&	.	&	.	&	.	&	.       &	.       &	.       \\
\enddata
\tablenotetext{a}{NLTE corrections.}
\tablecomments{Table~\ref{abund1} is published in its entirety in the
electronic edition of the {\it Astrophysical Journal Supplement Series}.  
A portion is shown here for guidance regarding its form and content.}
\end{deluxetable}

\clearpage
\begin{deluxetable}{crcrrcrrcrrcrrcrrcr}
\rotate
\tabletypesize{\tiny}
\tablewidth{0pt}
\tablecolumns{19}
\tablecaption{Abundance Ratios of Sc, Ti, V and Cr throughout the 
Pulsational Cycles.\label{abund2}}
\tablehead{
\colhead{Phase}   &
\colhead{[Sc II/Fe]} &
\colhead{$\sigma$} &
\colhead{$N$}     &
\colhead{[Ti I/Fe]} &
\colhead{$\sigma$} &
\colhead{$N$}      &
\colhead{[Ti II/Fe]}  &
\colhead{$\sigma$} &
\colhead{$N$}      &
\colhead{[V II/Fe]}  &
\colhead{$\sigma$} &
\colhead{$N$}      &
\colhead{[Cr I/Fe]} &
\colhead{$\sigma$}  &
\colhead{$N$}       &
\colhead{[Cr II/Fe]}   &
\colhead{$\sigma$}  &
\colhead{$N$}       
}
\startdata
\cutinhead{CD Vel}																																					
0.015	&	0.24	&	0.10	&	3	&	\nodata	&	\nodata	&	\nodata	&	0.41	&	0.11	&	10	&	\nodata	&	\nodata	&	\nodata	&	$-$0.09	&	0.12	&	3	&	0.21	&	0.11	&	3	\\
0.045	&	0.15	&	0.01	&	2	&	\nodata	&	\nodata	&	\nodata	&	0.31	&	0.12	&	13	&	\nodata	&	\nodata	&	\nodata	&	$-$0.12	&	0.13	&	4	&	0.16	&	0.13	&	6	\\
0.150	&	0.06	&	0.05	&	3	&	0.30	&	0.05	&	3	&	0.27	&	0.14	&	17	&	0.08	&	\nodata	&	1	&	$-$0.03	&	0.11	&	3	&	0.05	&	0.15	&	7	\\
0.300	&	0.07	&	0.04	&	3	&	0.29	&	0.13	&	4	&	0.35	&	0.18	&	17	&	0.18	&	\nodata	&	1	&	0.07	&	0.14	&	5	&	0.03	&	0.13	&	7	\\
.    	&	.	&	.	&	.	&	.	&	.	&	.	&	.	&	.	&	.	&	.	&	.	&	.	&	.	&	.	&	.	&	.       &	.       &	.       \\
.    	&	.	&	.	&	.	&	.	&	.	&	.	&	.       &	.       &	.       &	.	&	.	&	.	&	.	&	.	&	.	&	.       &	.       &	.       \\
.    	&	.	&	.	&	.	&	.	&	.	&	.	&	.	&	.	&	.	&	.	&	.	&	.	&	.       &	.       &	.       &	.	&	.	&	.
\enddata
\tablecomments{Table~\ref{abund1} is published in its entirety in the
electronic edition of the {\it Astrophysical Journal Supplement Series}.  
A portion is shown here for guidance regarding its form and content.}
\end{deluxetable}

\clearpage
\begin{deluxetable}{crcrrcrrcrrcr}
\rotate
\tabletypesize{\tiny}
\tablewidth{0pt}
\tablecolumns{13}
\tablecaption{Abundance ratios of Co, Ni, Zn and Y throughout the 
Pulsational Cycles.\label{abund3}}
\tablehead{
\colhead{Phase}   &
\colhead{[Co I/Fe]} &
\colhead{$\sigma$} &
\colhead{$N$}     &
\colhead{[Ni I/Fe]} &
\colhead{$\sigma$} &
\colhead{$N$}      &
\colhead{[Zn I/Fe]}  &
\colhead{$\sigma$} &
\colhead{$N$}      &
\colhead{[Y II/Fe]}  &
\colhead{$\sigma$} &
\colhead{$N$}      
}
\startdata
\cutinhead{CD Vel}																									
0.015	&	\nodata	&	\nodata	&	\nodata	&	\nodata	&	\nodata	&	\nodata	&	\nodata	&	\nodata	&	\nodata	&	\nodata	&	\nodata	&	\nodata	\\
0.045	&	\nodata	&	\nodata	&	\nodata	&	\nodata	&	\nodata	&	\nodata	&	\nodata	&	\nodata	&	\nodata	&	$-$0.06	&	\nodata	&	1	\\
0.150	&	0.05	&	\nodata	&	1	&	\nodata	&	\nodata	&	\nodata	&	\nodata	&	\nodata	&	\nodata	&	0.11	&	0.14	&	2	\\
0.300	&	$-$0.01	&	\nodata	&	1	&	0.70	&	\nodata	&	1	&	0.10	&	0.01	&	2	&	$-$0.18	&	0.15	&	2	\\
.    	&	.	&	.	&	.	&	.	&	.	&	.	&	.	&	.	&	.	&	.	&	.	&	.	\\
.    	&	.	&	.	&	.	&	.	&	.	&	.	&	.       &	.       &	.       &	.	&	.	&	.	\\
.    	&	.	&	.	&	.	&	.	&	.	&	.	&	.	&	.	&	.	&	.	&	.	&	.	
\enddata
\tablecomments{Table~\ref{abund1} is published in its entirety in the
electronic edition of the {\it Astrophysical Journal Supplement Series}.  
A portion is shown here for guidance regarding its form and content.}
\end{deluxetable}

\clearpage

\begin{deluxetable}{lccrrcrrcrrcrrcrrcrc}
\rotate
\tabletypesize{\tiny}
\tablewidth{0pt}
\tablecolumns{20}
\tablecaption{Abundance Ratios of Mn, Sr, Zr, Ba, La and Eu Derived from
Synthetic Spectra.\label{abund4}}
\tablehead{
\colhead{Star} &
\colhead{Phase}   &
\colhead{[Mn I/Fe]} &
\colhead{$\sigma$} &
\colhead{$N$}     &
\colhead{[Sr II/Fe]} &
\colhead{$\sigma$} &
\colhead{$N$}      &
\colhead{[Zr II/Fe]}  &
\colhead{$\sigma$} &
\colhead{$N$}      &
\colhead{[Ba II/Fe]}  &
\colhead{$\sigma$} &
\colhead{$N$}      &
\colhead{[La II/Fe]} &
\colhead{$\sigma$}  &
\colhead{$N$}       &
\colhead{[Eu II/Fe]}   &
\colhead{$\sigma$}  &
\colhead{$N$}       
}
\startdata
CD Vel	&	0.30	&	$-$0.48	&	0.15	&	3	&	0.57	&	0.21	&	2	&	0.36	&	0.14	&	2 &	0.06	&	0.17	&	3	&	\nodata	&	\nodata	&	\nodata	&	0.25	&	\nodata	&	2	\\
WY Ant	&	0.35	&	$-$0.59	&	\nodata	&	2	&	0.54	&	0.07	&	2	&	0.48	&	0.14	&	2 &	0.14	&	0.24	&	3	&	0.26	&	0.14	&	2	&	0.62	&	\nodata	&	2	\\
DT Hya	&	0.32	&	$-$0.57	&	0.14	&	2	&	0.72	&	0.14	&	2	&	0.48	&	\nodata	&	1 &	0.36	&	0.13	&	4	&	0.04	&	0.13	&	2	&	0.27	&	\nodata	&	1	\\
AS Vir	&	0.32,0.35	&$-$0.48&	0.23	&	6	&	0.33	&	0.33	&	4	&	0.31	&	0.22	&	5 &	0.04	&	0.12	&	8	&	0.16	&	0.20	&	4	&	0.51	&	0.16	&	3	\\
RV Oct	&	0.30	&	$-$0.34	&	0.15	&	4	&	0.57	&	0.21	&	2	&	0.61	&	0.12	&	3 &	0.06	&	0.15	&	4	&	$-$0.06	&	\nodata	&	1	&	0.18	&	\nodata	&	1	\\
XZ Aps	&	0.32	&	$-$0.58	&	0.15	&	3	&	0.20	&	\nodata	&	2	&	0.72	&	\nodata	&	1 &	0.17	&	0.15	&	3	&	\nodata	&	\nodata	&	\nodata	&	$-$0.10	&	\nodata	&	1	\\
BS Aps	&	0.30	&	$-$0.04	&	0.15	&	4	&	0.52	&	0.21	&	2	&	0.57	&	0.21	&	2 &	0.24	&	0.21	&	2	&	$-$0.13	&	\nodata	&	1	&	0.17	&	0.07	&	2	\\
UV Oct	&	0.25	&	$-$0.53	&	0.14	&	2	&	0.20	&	\nodata	&	2	&	0.28	&	0.14	&	2 & 	0.11	&	0.21	&	2	&	0.05	&	\nodata	&	2	&	0.35	&	\nodata	&	2	\\
\enddata
\end{deluxetable}

\clearpage

\begin{deluxetable}{lrrrrrrrrrrr}
\tabletypesize{\scriptsize}
\tablewidth{0pt}
\tablecolumns{12}
\tablecaption{Mean Abundance Ratios for Each Star.\label{mabund}}
\tablehead{
\colhead{} & 
\colhead{CD Vel}  & 
\colhead{WY Ant}  & 
\colhead{DT Hya}  &
\colhead{AS Vir}  &
\colhead{RV Oct}  & 
\colhead{XZ Aps}  & 
\colhead{BS Aps}  & 
\colhead{UV Oct}  &
\colhead{V1645 Sgr}  &
\colhead{Z Mic}   &
\colhead{TY Gru}
}
\startdata
Na I       	&		$-$0.37		&		$-$0.39		&	\nodata		        &	\nodata		        &		$-$0.12	&		$-$0.18		&	0.08		&		$-$0.38		&	\nodata		&		0.12		&       \nodata		\\
$N$\tablenotemark{a}	&	3		&		3		&	\nodata		        & \nodata	                &		9	       &		8		&		2		&		9		&	\nodata	&		2		&	\nodata	\\
Mg I                &	0.41		&		0.39		&		0.52		&		0.48		&		0.59		&		0.54		&		0.49		&		0.44		&		0.44		&		0.59		&		0.38		\\
$N$\tablenotemark{a}	&	16		&		11		&		9		&		19		&		17		&		20		&		18		&		20		&		16		&		14		&		4		\\
Al I                &	0.20		&		0.30		&		0.41		&		0.35		&		0.26		&		0.42		&		0.59		&		0.27		&		0.25		&		0.59		&		0.42		\\
$N$\tablenotemark{a}	&	10		&		9		&		2		&		3		&		2		&		4		&		3		&		12		&		6		&		3		&		1		\\
Si I                &	0.61		&		0.26		&		0.52		&		0.63		&		0.41		&		0.53		&		0.59		&		0.19		&	\nodata	&		0.59		&	\nodata	\\
$N$\tablenotemark{a}	&	3		&		6		&		2		&		2		&		14		&		2		&		6		&		10		&  \nodata	&		8		&       \nodata	\\
\enddata
\tablenotetext{a}{Total number of phases for averaging.}
\tablenotetext{b}{Total number of lines for averaging.}
\tablenotetext{c}{Averaged with two phases.}
\tablenotetext{d}{Total number of lines in two phases.}
\end{deluxetable}

\clearpage

\begin{deluxetable}{lrrrrrrr}
%\tabletypesize{\scriptsize}
\tablewidth{0pt}
\tablecolumns{8}
\tablecaption{Mean Abundance Ratios among Different HB groups.\label{mabundHB}}
\tablehead{
\colhead{Element} & 
\colhead{RR Lyr}  & 
\colhead{$\sigma$}  & 
\colhead{$N$}  &
\colhead{RHB\tablenotemark{a}}  &
\colhead{$N$}  & 
\colhead{BHB\tablenotemark{a}}  & 
\colhead{$N$}  
}
\startdata
Na I        &	$-$0.18	&	0.22	&	7	&	0.37	&	27	&	$-$0.45	&	2	\\
Mg I        &	0.48	&	0.07	&	11	&	0.47	&	36	&	0.36	&	12	\\
Al I        &	0.37	&	0.13	&	11	&	$–$0.67	&	25	&	0.36	&	3	\\
Si I        &	0.48	&	0.16	&	9	&	0.35	&	36	&	$–$0.03	&	12	\\
Si II       &	0.52	&	0.10	&	11	&	0.59	&	35	&	0.21	&	12	\\
Ca I        &	0.35	&	0.03	&	11	&	0.37	&	36	&	0.07	&	12	\\
Sc II       &	0.10	&	0.05	&	11	&	0.13	&	35	&	0.14	&	10	\\
Ti I        &	0.33	&	0.29	&	11	&	0.37	&	35	&	\nodata	&	\nodata	\\
Ti II       &	0.27	&	0.04	&	11	&	0.23	&	36	&	0.31	&	12	\\
V II       &	0.21	&	0.12	&	7	&	0.14	&	14	&	0.15	&	5	\\
Cr I        &	$-$0.10	&	0.06	&	11	&	$–$0.14	&	36	&	0.02	&	7	\\
Cr II       &	0.09	&	0.08	&	11	&	0.23	&	35	&	0.15	&	10	\\
Mn I        &	$-$0.48	&	0.17	&	11	&	$–$0.37	&	36	&	$–$0.13	&	3	\\
Co I        &	0.09	&	0.08	&	11	&	0.41	&	15	&	0.28	&	1	\\
Ni I        &	0.47	&	0.13	&	10	&	0.22	&	15	&	\nodata	&	\nodata	\\
Zn I        &	0.16	&	0.05	&	11	&	0.19	&	18	&	\nodata	&	\nodata	\\
Sr II       &	0.37	&	0.25	&	11	&	0.23	&	36	&	$–$0.30	&	10	\\
Y II       &	0.04	&	0.16	&	11	&	$–$0.12	&	27	&	0.22	&	4	\\
Zr II       &	0.41	&	0.17	&	11	&	0.42	&	23	&	0.61	&	7	\\
Ba II       &	0.23	&	0.29	&	11	&	0.03	&	36	&	0.00	&	7	\\
La II       &	0.19	&	0.30	&	9	&	0.19	&	19	&	\nodata	&	\nodata	\\
Eu II       &	0.31	&	0.21	&	10	&	0.45	&	22	&	\nodata	&	\nodata	
\enddata
\tablenotetext{a}{\citet{For10}}
\end{deluxetable}

\clearpage

\begin{deluxetable}{ccc}
\tablewidth{0pt}
\tablecolumns{3}
\tablecaption{Mean \teff\ Values of RRab Stars.\label{mteff}}
\tablehead{
\colhead{Star} & 
\colhead{$<$\teff$>$\tablenotemark{a}}  & 
\colhead{$\log<$\teff$>$}  \\ 
\colhead{}  &
\colhead{(K)}  &
\colhead{(K)}  
}
\startdata
CD Vel  & 6366 & 3.804 \\
WY Ant  & 6487 & 3.812 \\
DT HYa  & 6460 & 3.810 \\
AS Vir 1 & 6463 & 3.810\\
BS Aps 1 & 6404 & 3.806\\
RV Oct & 6443 & 3.809 \\
XZ Aps & 6438 & 3.809 \\
UV Oct 1 & 6435 & 3.809 \\
Z Mic  & 6246 & 3.796
\enddata
\tablenotetext{a}{Interpolated from Table~\ref{inputmodels}; see text.}
\end{deluxetable}

\clearpage

%% The following command ends your manuscript. LaTeX will ignore any text
%% that appears after it.

\end{document}